\documentclass[traditabstract, longauth]{aa}  %

\usepackage{txfonts}
\usepackage{graphicx}
\usepackage{gensymb} 
\usepackage[dvipsnames]{xcolor}
\usepackage[colorlinks=true, citecolor=blue]{hyperref}
\usepackage{xspace}
\usepackage{siunitx} 
\usepackage{multirow}
\usepackage{hhline}
\usepackage{upgreek} 
\usepackage{pifont}

\newcommand{\neii}{[\ion{Ne}{ii}]\xspace}
\newcommand{\neiii}{[\ion{Ne}{iii}]\xspace}
\newcommand{\nev}{[\ion{Ne}{v}]\xspace}
\newcommand{\oiv}{[\ion{O}{iv}]\xspace}

\newcommand{\nevi}{[\ion{Ne}{vi}]\xspace}
\newcommand{\fevii}{[\ion{Fe}{vii}]\xspace}
\newcommand{\arv}{[\ion{Ar}{v}]\xspace}
\newcommand{\htwo}{H$_2$\xspace}

\newcommand{\Pfa}{{Pf~$\alpha$}\xspace}

\newcommand{\spirit}{{\small SPIRIT}\xspace}
\newcommand{\cafe}{{\small CAFE}\xspace}
\newcommand{\pahfit}{{\small PAHFIT}\xspace}

\newcommand{\upmicron}{$\upmu$m\xspace}
\newcommand{\tabupmicron}{$\upmu$m}

\usepackage{svg}
\newcommand{\orcid}[1]{\href{https://orcid.org/#1}{\includegraphics[width=8pt]{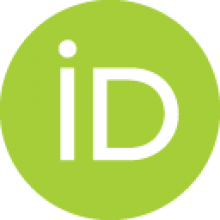}}}

\begin{document} 

   \title{MICONIC: JWST/MIRI-MRS reveals heavily reprocessed PAH emission in the circum-nuclear disc of Centaurus A}
   
  \author{L. Pantoni\orcid{0000-0003-2666-5759}\inst{\ref{Gent}}\fnmsep\thanks{E-mail: \url{lara.pantoni@ugent.be}}
  \and
  M. Baes\inst{\ref{Gent}} 
  \and
  L. Decin\inst{\ref{Leuven}} 
  \and
  P. Guillard\inst{\ref{Sorbonne},\ref{IAP}} 
  \and
  A. Alonso Herrero\inst{\ref{CAB}}
  \and
  L. Hermosa Muñoz\inst{\ref{CAB}}
  \and
  L. Evangelista\inst{\ref{Sorbonne}} 
  \and
  I. García-Bernete\inst{\ref{CAB},\ref{OXFORD}}
  \and
  F. Donnan\inst{\ref{SanDiego}}
  \and
  V. Buiten\inst{\ref{Leiden}}
  \and
  S. Garcia-Burillo\inst{\ref{OAN-Madrid}}
  \and
  G. Wright\inst{\ref{Edinburgh}}
  \and
  L. Colina\inst{}
  \and
  T. Böker\inst{\ref{ESA-USA}}
  \and
  G. Östlin\inst{\ref{Sweden}}
  \and
  D. Dicken\inst{\ref{Edinburgh}}
  \and
  A. Labiano\inst{\ref{ESA-Madrid}}
  \and
  D. Rouan\inst{\ref{Meudon}}
  \and 
  P. van der Werf\inst{\ref{Leiden}}
  \and
  A. Eckart\inst{\ref{MPIfR}}
  \and 
  M. García-Marín\inst{\ref{ESA-USA}}
  \and
  M. Güdel\inst{\ref{Austria},\ref{ETH}}
  \and
  Th. Henning\inst{\ref{Heidelberg}}
  \and
  P.-O. Lagage\inst{\ref{Paris-Saclay}}
  \and
  F. Walter\inst{\ref{Heidelberg}}
  \and
  M. J. Ward\inst{\ref{Durham}}
  }

\institute{
Department of Physics and Astronomy, Universiteit Gent, Proeftuinstraat 86 N3, B-9000 Ghent, Belgium
\label{Gent}
\and
Institute of Astronomy, KU Leuven, Celestijnenlaan 200D, 3001, Leuven, Belgium
\label{Leuven}
\and
Sorbonne Université, CNRS, UMR 7095, Institut d’Astrophysique de Paris, 98bis bd Arago, 75014 Paris, France
\label{Sorbonne}
\and
Institut Universitaire de France, Ministère de l’Education Nationale, de l’Enseignement Supérieur et de la Recherche, 1 rue Descartes, 75231 Paris Cedex 05, France
\label{IAP}
\and
Centro de Astrobiología (CAB), CSIC-INTA, Camino Bajo del Castillo s/n, E-28692 Villanueva de la Cañada, Madrid, Spain
\label{CAB}
\and
Telespazio UK for the European Space Agency (ESA), ESAC, Camino Bajo del Castillo s/n, 28692, Villanueva de la Cañada, Spain
\label{ESA-Madrid}
\and
European Space Agency, c/o Space Telescope Science Institute, 3700 San Martin Drive, Baltimore MD, 21218, USA
\label{ESA-USA}
\and
UK Astronomy Technology Centre, Royal Observatory, Blackford Hill Edinburgh, EH9 3HJ, Scotland, UK
\label{Edinburgh}
\and
Observatorio Astronómico Nacional (OAN-IGN)-Observatorio de Madrid, Alfonso XII, 3, 28014, Madrid, Spain
\label{OAN-Madrid}
\and
Leiden Observatory, Leiden University, PO Box 9513, 2300 RA, Leiden, The Netherlands
\label{Leiden}
\and
Dept. of Astrophysics, University of Vienna, Türkenschanzstr. 17, A-1180, Vienna, Austria
\label{Austria}
\and
ETH Zürich, Institute for Particle Physics and Astrophysics, Wolfgang-Pauli-Str. 27, 8093, Zürich, Switzerland
\label{ETH}
\and
Université Paris-Saclay, Université Paris Cité, CEA, CNRS, AIM, 91191, Gif-Sur-Yvette, France
\label{Paris-Saclay}
\and
Department of Physics, University of Oxford, Keble Road, Oxford, OX1 3RH, UK
\label{OXFORD}
\and
Department of Physics, University of California, San Diego, California 92093, USA
\label{SanDiego}
\and
Physikalisches Institut der Universität zu Köln, Zülpicher Str. 77, D-50937, Köln, Germany; Max-Planck-Institut für Radioastronomie (MPIfR), Auf dem Hügel 69, D-53121, Bonn, Germany
\label{MPIfR}
\and
Department of Astronomy, Stockholm University, The Oskar Klein Centre, AlbaNova, SE-106 91, Stockholm, Sweden
\label{Sweden}
\and
LIRA, Observatoire de Paris, Université PSL, Sorbonne Université, Université Paris Cité, CY Cergy Paris Université, CNRS, 92190, Meudon, France
\label{Meudon}
\and
Max Planck Institute for Astronomy, Königstuhl 17, 69117, Heidelberg, Germany
\label{Heidelberg}
\and
Centre for Extragalactic Astronomy, Durham University, South Road, Durham, DH1 3LE, UK
\label{Durham}
}

   \date{Received 31 December 2025 / Accepted 23 March 2026}
 
  \abstract 
   {Polycyclic Aromatic Hydrocarbons (PAHs) constitute essential components of dust in galaxies and have a fundamental role in the physics of the interstellar medium (ISM). The impact of AGN feedback on these molecules is still under debate.}
   {We present a detailed analysis of the spatially-resolved properties of PAHs in the central $7^{\prime\prime}\times12^{\prime\prime}$ ($\sim100\times200$ pc$^2$) of Centaurus~A (Cen~A). We use the JWST/MIRI-MRS observations at $\lambda\sim5-28$~\upmicron taken as part of the MIRI European consortium’s Guaranteed Time Observation (GTO) program MICONIC, with angular resolution between $0.35^{\prime\prime}$ and $1^{\prime\prime}$ ($\sim6-17$ pc).}
   {We derive PAH moment-0 maps through local continuum subtraction and extract 1-dimensional spectra in five regions of interest, including the nucleus, the circum-nuclear disc, and a region characterized by a prominent deficiency in PAH emission. We decompose the spectra into continuum, emission lines and PAHs, from which we extract PAH intensities and EWs.}
   {PAH emission is predominantly distributed in a ring-like structure with localized intensity enhancements, at a radius of $\sim40$ pc from the active nucleus. Toward the North-West we observe a distinct PAH-deficient area, roughly perpendicular to the jet axis and coincident with enhanced ionized-gas velocity dispersion as well as inflowing warm and cold molecular streamers. The PAH 11.3/7.7~\upmicron and 6.2/7.7~\upmicron intensity ratios exceed model predictions for pericondensed PAHs, suggesting heavily reprocessed populations characterized by more open and irregular molecular structures.
   PAH 11.3/12.7~\upmicron ratios point to a prevalence of solo over duo or trio hydrogen sites, consistent with a non-negligible degree of dehydrogenation, particularly within the PAH-deficient region, where shock-driven erosion may play a major role. We measure the largest EWs in the PAH ring, whereas reduced values in the PAH-deficient region likely reflect partial destruction by shocks; in the nucleus, the small EWs are largely attributable to continuum dilution.} 
   {}


   \titlerunning{MICONIC: JWST/MIRI-MRS reveals heavily reprocessed PAHs in Cen A.}
   \authorrunning{L. Pantoni}
   \maketitle
%

\section{Introduction}
Polycyclic Aromatic Hydrocarbons (PAHs) are compact, planar, carbon-based molecules composed of fused aromatic rings and typically containing between 20 and 500 carbon atoms (up to a few nm in size). These molecules are primarily excited by ultraviolet (UV) photons and subsequently emit in the mid-infrared (MIR) through vibrational relaxation \citep[e.g.,][]{Allamandola1989ApJS...71..733A, Peetersdoi:10.1021/acs.accounts.0c00747}.

PAHs are widely considered the primary carriers of the aromatic infrared bands (AIBs)\footnote{Other proposed carriers include e.g. amorphous hydrogenated carbon (a-C:H; \citealt{Jones2013A&A...558A..62J}). For simplicity, we use `PAHs' to refer collectively to a broader family of carbonaceous species carrying both the 2175~\textup{\AA} extinction bump and the AIBs.}, a set of prominent mid-infrared (MIR) emission features between 3 and 20~\upmicron \citep[e.g.,][]{LegerPuget1984A&A...137L...5L, Allamandola1985ApJ...290L..25A, Verstraet2001A&A...372..981V, Sellgren2010ApJ...722L..54S, Hansen2022CmChe...5...94H}. 
Initially detected with ground-based telescopes \citep{Gillett1973ApJ...183...87G}, AIBs were later observed with space-borne facilities such as ISO \citep[e.g.,][]{Rigopoulou1999AJ....118.2625R,Verstraet2001A&A...372..981V} and Spitzer \citep[e.g.,][]{Werner2004ApJS..154..309W, Sellgren2007ApJ...659.1338S}, and are now known to be ubiquitous across the Universe, with detections out to redshift $z\sim4$ \citep{Riechers2014ApJ...786...31R, Spilker2023Natur.618..708S}.
   \begin{figure*}
   \centering
   \includegraphics[width=1\columnwidth]{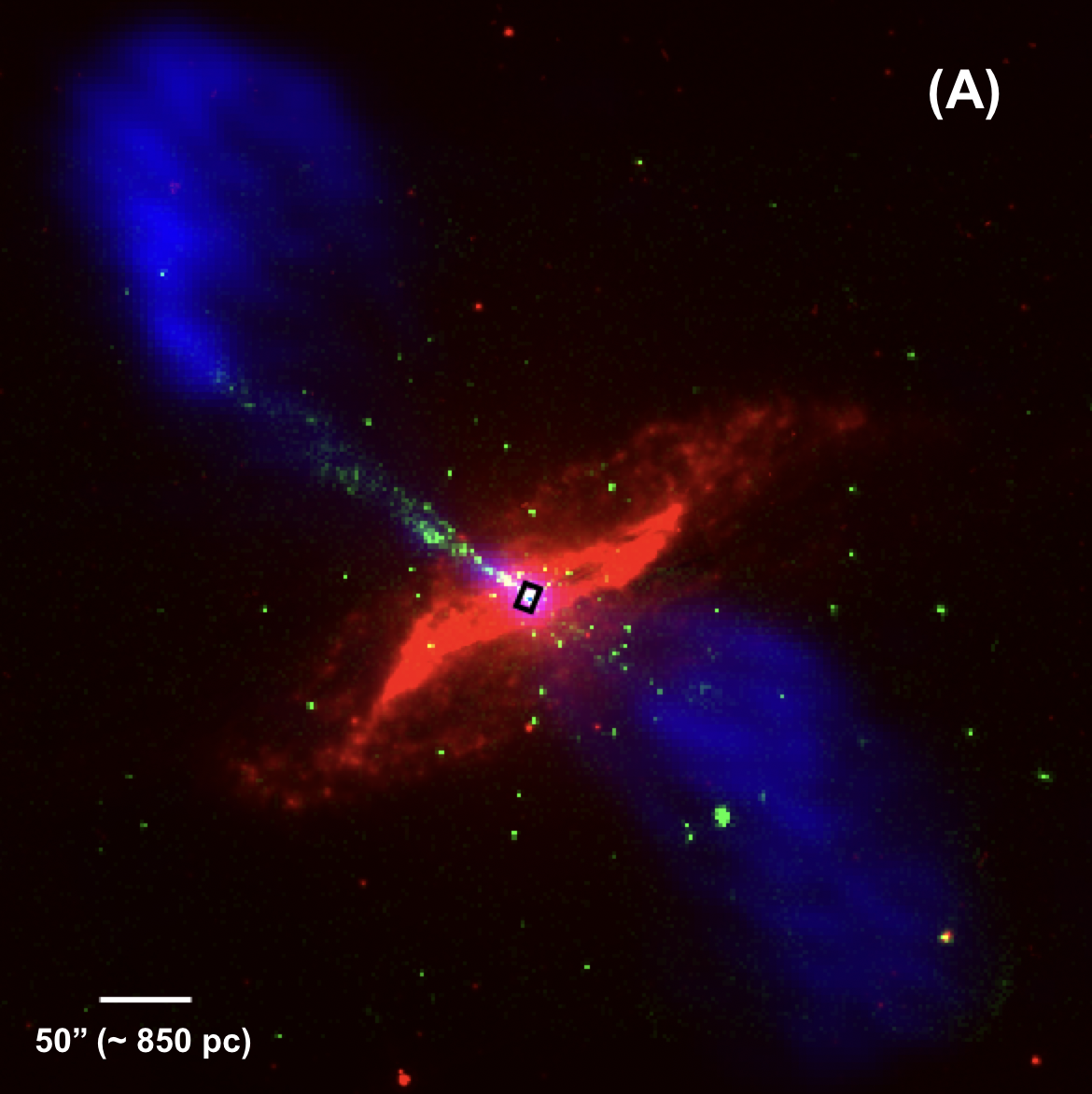}
   \includegraphics[width=0.953\columnwidth]{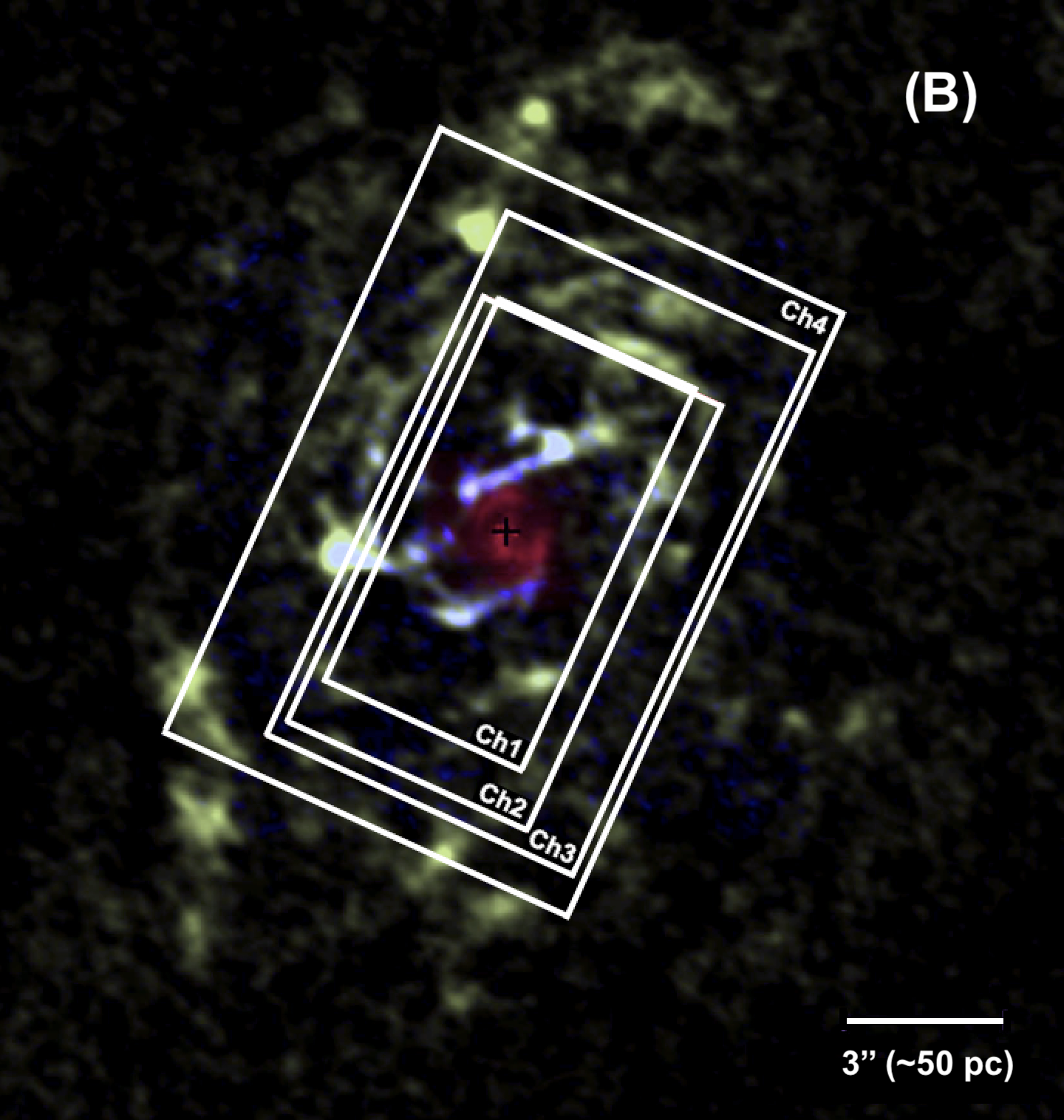}
      \caption{Panel A: RGB composite image of Cen~A. Red colour shows the 5.8~\upmicron emission from Cen~A as observed by Spitzer/IRAC \citep{Quillen2006_disk, Quillen2006_shell}; green colour corresponds to X-ray emission observed with Chandra/ACIS \citep{Hardcastle2007ApJ...670L..81H}; and blue colour traces the radio jet structure imaged with the VLA \citep{Hardcastle2003ApJ...593..169H}. In black we show the footprint of MIRI-MRS observations. Panel B: Zoom-in on the central few hundred parsecs of Cen~A, showing cold molecular gas traced by CO(6--5) and CO(3--2) in blue and green colours, and the VLT/SINFONI \htwo~1-0~S(1) intensity map by \citet{Neumayer2007ApJ...671.1329N} in red \citep[adapted from][]{Espada2017ApJ...843..136E}. White rectangles mark the each MIRI-MRS mosaic of $1\times2$ FoV \citep[adapted from][]{Evangelista2026}. North is up, East is to the left. 
      }
         \label{Fig:CenA_comp}
   \end{figure*}

PAHs play a crucial role in the physics of the interstellar medium \citep[ISM; see][for a review]{Tielens2008ARA&A..46..289T}. 
They contribute significantly to the photoelectric heating and the ionization balance of the ISM \citep[e.g.,][]{Bakes1994ApJ...427..822B,Berne2022A&A...667A.159B} and act as catalysts in the process of forming molecular hydrogen \citep[e.g.,][]{Mennella2012ApJ...745L...2M, Thrower2012ApJ...752....3T, Skov2014FaDi..168..223S, Jones2015A&A...581A..92J, Barrera2023MNRAS.524.3741B}. They contain $\lesssim10-15\%$ of the interstellar carbon \citep[e.g.,][]{Li2001ApJ...554..778L, Zubko2004ApJS..152..211Z,Draine2007ApJ...657..810D, Jones2017A&A...602A..46J, Tielens2021moas.book.....T}, are responsible for the ultraviolet (UV) dust extinction \citep{Weingartner2001ApJ...548..296W}, and can account for up to 20\% of the total infrared (IR) luminosity of star-forming, metal-rich galaxies \citep[e.g.,][]{Smith2007ApJ...656..770S}. As such, PAHs serve as indirect tracers of obscured star formation \citep[e.g.,][]{Farrah2007ApJ...667..149F, Calzetti2011EAS....46..133C, Kim2024ApJ...974..253K, Ujjwal2024A&A...684A..71U}, although the relation may vary depending on the SFR of the system \citep{Mordini2021A&A...653A..36M, Robinson2026arXiv260109810R}.
Extensive observations with ISO and Spitzer, and more recently with JWST, revealed substantial variations in PAH emission across a wide range of Galactic and extragalactic environments, including protoplanetary discs, photodissociation regions (PDRs), HII regions, early-type galaxies, active galactic nuclei (AGNs), and low-metallicity systems \citep[e.g.,][for a review]{Peeters2002A&A...390.1089P,Galliano2008ApJ...679..310G,Smith2007ApJ...656..770S, Garcia-Bernete2022MNRAS.509.4256G, Chastenet2023ApJ...944L..11C, Chastenet2023ApJ...944L..12C, Sandstrom2023ApJ...944L...7S, Rigopoulou2024MNRAS.532.1598R, ZhangCongcong2025ApJS..280....4Z,Li2020NatAs...4..339L}.

Both observational and theoretical studies have demonstrated that the relative strengths of PAH bands, particularly those in the $6-9$~\upmicron range relative to the 11.3~\upmicron feature, serve as effective diagnostics of the local physical conditions in the ISM \citep[e.g.,][]{Galliano2008ApJ...679..310G, ODowd2009ApJ...705..885O, Sales2010ApJ...725..605S, Rigopoulou2021MNRAS.504.5287R, Garcia-Bernete2022MNRAS.509.4256G, Rigopoulou2024MNRAS.532.1598R}.
In the harsh environments surrounding AGNs, especially in Seyfert nuclei, the $6-9$~\upmicron PAH bands are frequently found to be significantly suppressed compared to star-forming regions, whereas the 11.3~\upmicron feature typically remains comparatively strong \citep[e.g.,][]{Smith2007ApJ...656..770S, Garcia-Bernete2022MNRAS.509.4256G, Garcia-Bernete2022A&A...666L...5G, Garcia-Bernete2024A&A...691A.162G, ZhangLulu2023ApJ...953L...9Z, ZhangLulu2024ApJ...975L...2Z}.
The suppression of the $6-9$~\upmicron features is commonly attributed to the preferential destruction of small, ionized PAH molecules and to significant structural modifications of PAHs, while large and neutral species are more likely to survive. These effects are generally interpreted as the consequence of shocks or of the hard radiation fields associated with AGNs \citep[e.g.,][]{Diamond-Stanic2010ApJ...724..140D, Sales2010ApJ...725..605S, ZhangLulu2022ApJ...939...22Z, Li2020NatAs...4..339L}. Reduced PAH equivalent widths (EWs) observed in the immediate vicinity of AGNs, often interpreted as evidence of PAH destruction, may also largely result from dilution by the strong AGN MIR continuum \citep[e.g.,][]{Desai2007ApJ...669..810D,Alonso2014MNRAS.443.2766A, Garcia-Bernete2015MNRAS.449.1309G, RamosAlmeida2023A&A...669L...5R}.

   \begin{figure*}
   \centering
   \includegraphics[width=\hsize]{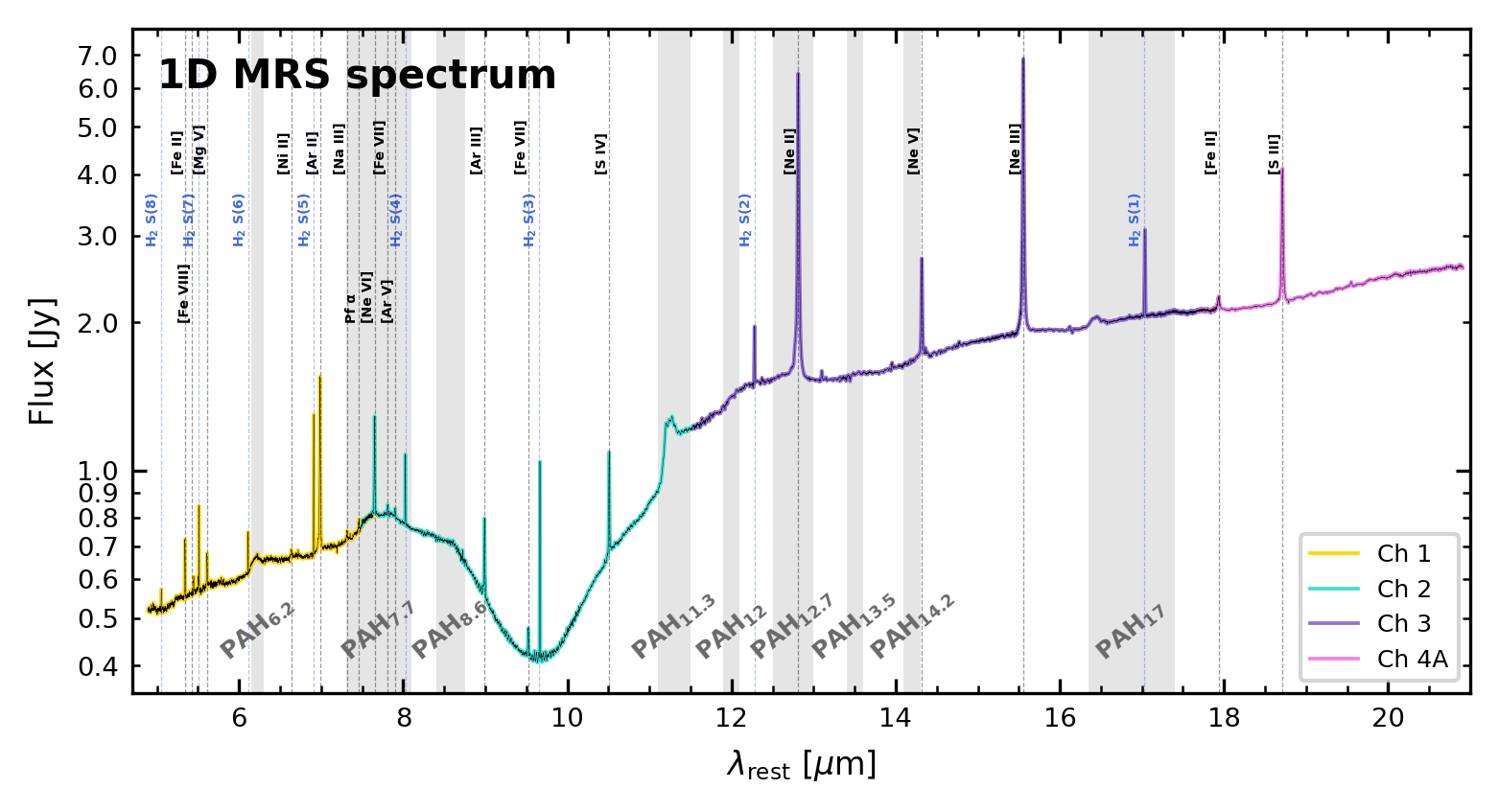}
      \caption{MIRI-MRS spectrum of the central $7.2^{\prime\prime}\times3.6^{\prime\prime}$ region of Cen~A, integrated over the Channel~1A mosaic, shown up to $\lambda_{\rm rest}=21$~\upmicron. The four MRS channels are colour-coded as follows: yellow (Ch~1), cyan (Ch~2), violet (Ch~3), and pink (Ch~4). PAH features commonly observed in galactic and extragalactic environments are highlighted in gray. Ionized gas emission lines and warm molecular hydrogen transitions are marked in black and blue, respectively. Strong silicate absorption feature is evident at 9.8~\upmicron.
      }
    \label{Fig:spectrum_Ch1A}
\end{figure*}
This work presents the first sub-arcsecond, spatially resolved analysis of all major PAH emission features at $\lambda > 4.9$~\upmicron in the innermost circum-nuclear disc (ICND; central $\sim100\times200$~pc$^2$) of Centaurus A (NGC~5128; hereafter Cen~A), based on JWST/MIRI-MRS observations (see footprints in Fig.~\ref{Fig:CenA_comp}).
Cen~A is the nearest active radio galaxy \citep[$D_{\rm L} = 3.5$~Mpc, $1^{\prime\prime}\sim17$ pc;][and $z=0.001825\pm0.000017$; \citealt{BaerWay2024ApJ...964..172B}]{Neumayer2007ApJ...671.1329N}, making it an ideal laboratory for resolving AGN-related processes on scales of a few pc. Its nuclear activity is thought to be triggered by a major merger between an elliptical and a spiral galaxy \citep{Israel1998A&ARv...8..237I}, which also explains the 500 pc expanding shell detected in the MIR \citep{Quillen2006_shell, Quillen2008}. Optical and UV images of Cen~A show a thick dust lane crossing the centre of the galaxy \citep[e.g.,][]{Marconi2000ApJ...528..276M}, while MIR images reveal the presence of a dusty warped disc with a peculiar parallelogram-shape structure \citep[Fig. \ref{Fig:CenA_comp}A, in red; see also][]{Quillen2006_disk}. Classified as a Type~2 Seyfert \citep{Israel1998A&ARv...8..237I}, Cen~A hosts a moderate-luminosity AGN \citep[L$_{\rm bol} \sim 1-4\times10^{43}$~erg~s$^{-1}$;][]{Beckmann2011A&A...531A..70B} embedded within a complex multi-phase circum-nuclear environment, partially shaped by AGN-powered jet and outflows \citep[e.g.,][]{Espada2009ApJ...695..116E, Espada2010ApJ...720..666E,Borkar2021MNRAS.500.3536B, McCoy2017ApJ...851...76M,Alonso2025A&A...699A.334A}. In the central region of Cen A the jet (seen both in radio and in X-ray; Fig. \ref{Fig:CenA_comp}A) has a position angle of $51\deg$, a present-day power $\gtrsim10^{43}$~erg~s$^{-1}$ \citep[e.g.,][]{Neff2015ApJ...802...87N}, and shows apparent subluminal motions \citep{Clarke1992ApJ...395..444C, Hardcastle2003ApJ...593..169H}. It appears to be nearly perpendicular to the circum-nuclear molecular disc detected in CO, extending up to $\sim400$~pc from the central cavity \citep[see Fig. \ref{Fig:CenA_comp}B;][]{Espada2017ApJ...843..136E}. 

Our main goal is to investigate how the AGN and shocks driven by nuclear activity affect the properties of PAHs within the ICND of Cen~A, down to spatial scales of $\sim10$~pc ($\sim 0.6^{\prime\prime}$).
This spatial resolution, comparable to the outer radius of AGN tori, allows us to disentangle nuclear and circum-nuclear emission, whereas previous studies of nearby AGN, limited to $\sim100$ pc scales, probe blended components \citep[see also][]{Lai2023ApJ...957L..26L, ZhangLulu2023ApJ...953L...9Z, Garcia-Bernete2024A&A...691A.162G}.

The observations with JWST/MIRI-MRS were conducted as part of the MIRI European Consortium GTO program `Mid-Infrared Characterization Of Nearby Iconic galaxy Centres' (MICONIC). 
Two companion studies based on MIRI-MRS observations of Cen~A investigate the distribution, temperature, and excitation of warm molecular hydrogen \citep{Evangelista2026}, and the kinematics of the warm and ionized gas \citep{Alonso2025A&A...699A.334A}. Along with Cen~A, the MICONIC program targets several nearby, well-studied galaxies with diverse nuclear environments, including Mrk~231 \citep{AlonsoHerrero2024A&A...690A..95A}, Arp~220 \citep[][van der Werf et al. in prep.]{Buiten2025A&A...699A.312B}, NGC~6240 \citep{HermosaMunoz2025A&A...693A.321H}, SBS~$0335-052$ (Bik et al. in prep.), and the central region surrounding Sgr~A$^*$.

The paper is organized as follows. Section~\ref{sect:data} describes the JWST/MIRI-MRS observations and the data reduction and homogenization across the MRS channels. Section~\ref{sect:methods} outlines the methods used to extract the PAH emission from the MIRI-MRS cubes. In Section~\ref{sect:results}, we present spatially resolved maps of the main PAH features and the corresponding intensity ratios measured in selected regions of the mosaic, which we interpret in the context of recent PAH models. Section~\ref{sect:discussion} discusses the impact of AGN-driven shocks on the PAH population and explores plausible evolutionary pathways. Finally, Section~\ref{sect:conclusions} summarizes our conclusions.

\section{Data}\label{sect:data}

\subsection{MIRI-MRS observations}
This study is based on MIRI-MRS observations obtained as part of JWST Cycle~1's GTO program ID 1269. The data cover the full MRS spectral range of $4.9-27.9$~\upmicron and consist of a $1\times2$ mosaic centred on the nucleus of Cen~A. The spatial coverage of the mosaic (see footprints in Fig.~\ref{Fig:CenA_comp}), spans approximately $7^{\prime\prime}\times12^{\prime\prime}$, corresponding to a physical scale of $\sim100\times200$~pc$^2$. 

MIRI-MRS comprises four integral field units (IFUs; hereafter referred to as channels or Ch), each with distinct spectral coverage, spectral and spatial resolution, and field of view (FoV). Each channel comes with three different grating settings (short, medium, and long; hereafter referred to as band A, B, C) covering adjacent wavelength ranges. The angular resolution varies from $0.3^{\prime\prime}$ in Ch~1 ($\lambda = 4.9-7.65$~\upmicron) to $1^{\prime\prime}$ in Ch~4 ($\lambda = 17.7-27.9$~\upmicron), corresponding to physical scales of approximately $5-17$~pc at the distance of Cen~A \citep{Argyriou2023A&A...675A.111A}. This resolution enables us to resolve the complex interplay between AGN activity and PAH emission in Cen~A on spatial scales previously inaccessible. The spectral resolving power also varies across channels, with $R = \lambda/\Delta\lambda \sim 3700$ in Ch~1 and $\sim 1300$ in Ch~4 \citep{Labiano2021A&A...656A..57L}. 
The FoV (1 pointing) increases with wavelength, ranging from the compact coverage of Ch~1 ($3.2^{\prime\prime} \times 3.7^{\prime\prime}$; Fig.~\ref{Fig:CenA_comp}B) to the nearly quadrupled area of Ch~4 ($6.6^{\prime\prime} \times 7.7^{\prime\prime}$). 

Observations employed the 4-point dither pattern optimized for extended sources \citep{Law2023AJ....166...45L}, with 10 groups per integration and 5 integrations per exposure, in FASTR1 readout mode. The full MRS spectral range was covered in three exposures (one per band), yielding 600 seconds of on-source integration time per setting.  
Free-of-source background observations were obtained west of the galaxy using a 2-point extended source dither pattern with identical exposure settings.

\begin{table*}
\caption{Setting for the mom0 extraction of PAH features.}
\label{tab:PAH2D}
\centering
\begin{tabular}{ccccc}
\hline\hline
\noalign{\smallskip}
$\mathbf{\lambda_{PAH}}$& $\mathbf{\Delta\lambda_{PAH}}$ & $\mathbf{\Delta\lambda_{cont}^{blue}}$ & $\mathbf{\Delta\lambda_{cont}^{red}}$ & \textbf{Other lines} \\ 
\noalign{\smallskip}
[\tabupmicron] & [\tabupmicron] & [\tabupmicron] & [\tabupmicron] & \\ 
\noalign{\smallskip}
\hline     
\noalign{\smallskip}
6.2  & $6.15-6.3$ & $6.13-6.15$ & $6.30-6.31$ & $-$\\ 
7.7 & $7.3-8.1$ & $7.2-7.3$ & $8.10-8.15$& \Pfa, \nevi, \fevii, \arv, \htwo~S(4)\\
8.6 & $8.4-8.75$ & $8.38-8.4$ & $8.75-8.77$ & \\
11.3 & $11.1-11.5$ & $11.0-11.1$ & $11.50-11.52$ & $-$ \\
12.7 & $12.48-13.03$ & $12.46-12.48$ & $13.03-13.05$ & [Ne II] \\
16.5 & $16.35-16.65$ & $16.2-16.35$ & $16.65-16.8$ & $-$ \\
\noalign{\smallskip}
\hline
\end{tabular}
\tablefoot{In the order we list: the observed central wavelength of each PAH complex ($\lambda_{\rm PAH}$), the wavelength interval used for its extraction ($\Delta\lambda_{\rm PAH}$), the blue and red continuum windows ($\Delta\lambda_{\rm cont}^{\rm blue}$, $\Delta\lambda_{\rm cont}^{\rm red}$), and any overlapping gas-phase emission lines. PAH 16.5~\upmicron is usually included in the 17~\upmicron complex.}
\end{table*}

\subsection{Data reduction}

We reduced the MRS data using the JWST Science Calibration Pipeline v1.14.1 \citep[][]{Bushouse2024zndo..12556702B}, within the context 1240 of the Calibration Reference Data System (CRDS), following standard procedures \citep[see e.g.,][]{Alvarez-Marquez2023A&A...672A.108A}. 

We applied detector-level corrections (Stage~1) with default parameters, except for lowering the cosmic-ray jump detection threshold to 3.5$\sigma$ and enabling cosmic ray shower flagging \citep{Morrison2023PASP..135g5004M}.
In Stage~2, we enabled 2D pixel-by-pixel background subtraction and residual fringe correction, and applied the JWST pipeline bad-pixel self-calibration, which statistically identifies outliers in the background frames, resulting in $\sim0.5$\% of pixels being flagged
\citep{Argyriou2023A&A...675A.111A,Gasman2023A&A...673A.102G, Patapis2024A&A...682A..53P}. In Stage~3, we disabled master background subtraction and sky matching due to the introduction of artefacts \citep{Law2023AJ....166...45L}.

The final reconstructed cubes were astrometrically aligned using the MIRI imager in parallel (i.e., off-target) and registered to the Gaia DR3 catalogue, achieving a positional accuracy better than 0.1$^{\prime\prime}$. They were reoriented to standard sky coordinates (North up, East left), resulting in final mosaic areas ranging approximately from $7.5^{\prime\prime} \times 3.8^{\prime\prime}$ in Ch~1 to $11.7^{\prime\prime} \times 7.2^{\prime\prime}$ in Ch~4.

In Fig.~\ref{Fig:spectrum_Ch1A} we present the spectrum extracted over the full Ch~1A mosaic (cf. Fig.~\ref{Fig:CenA_comp}B) covering an area of $7.2^{\prime\prime} \times 3.6^{\prime\prime}$, up to $\lambda_{\rm rest}= 21$~\upmicron. Owing to the improved spectrophotometric calibration \citep{Gasman2023A&A...673A.102G, Law2025AJ....169...67L}, no band stitching was necessary, except for a minor additive adjustment of 0.025 Jy to align Ch~1A and Ch~2A-B. 

The spectrum exhibits prominent PAH emission features at 6.2, 7.7, 8.6, 11.2, 12.0, and 16.5~\upmicron (the latter usually included in the 17~\upmicron complex). Intrinsically fainter features at 12.7, 13.5, and 14.2~\upmicron are more difficult to detect, given the strong MIR continuum from the AGN dusty torus. Alongside PAHs, the spectrum reveals a rich set of ionized gas emission lines and the S$(1)-$S(8) warm \htwo rotational transitions \citep[see][]{Alonso2025A&A...699A.334A, Evangelista2026}.

\section{Extraction of PAH features}\label{sect:methods}
\begin{figure*}
   \centering
   \includegraphics[width=\columnwidth]{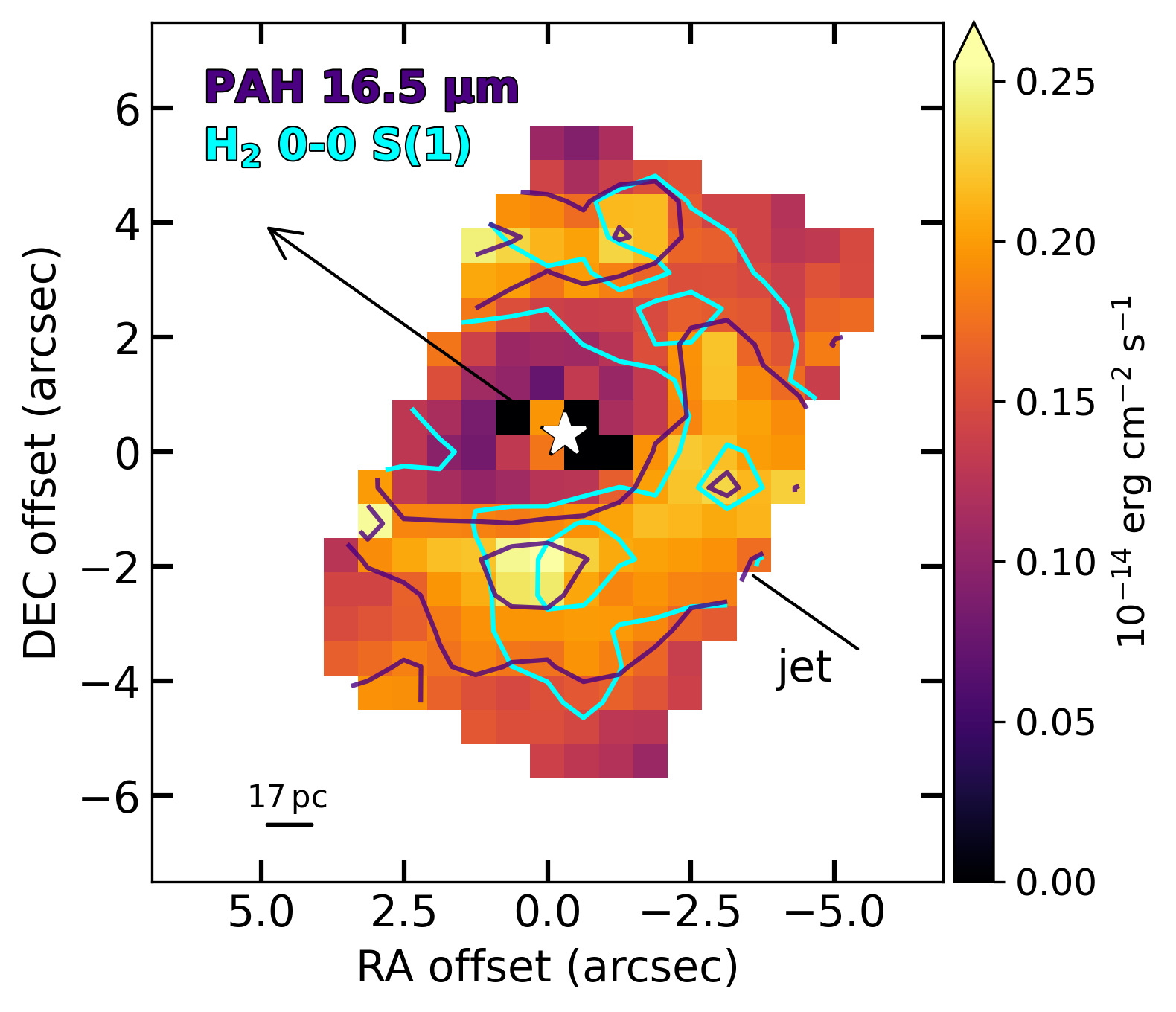}
   \includegraphics[width=\columnwidth]{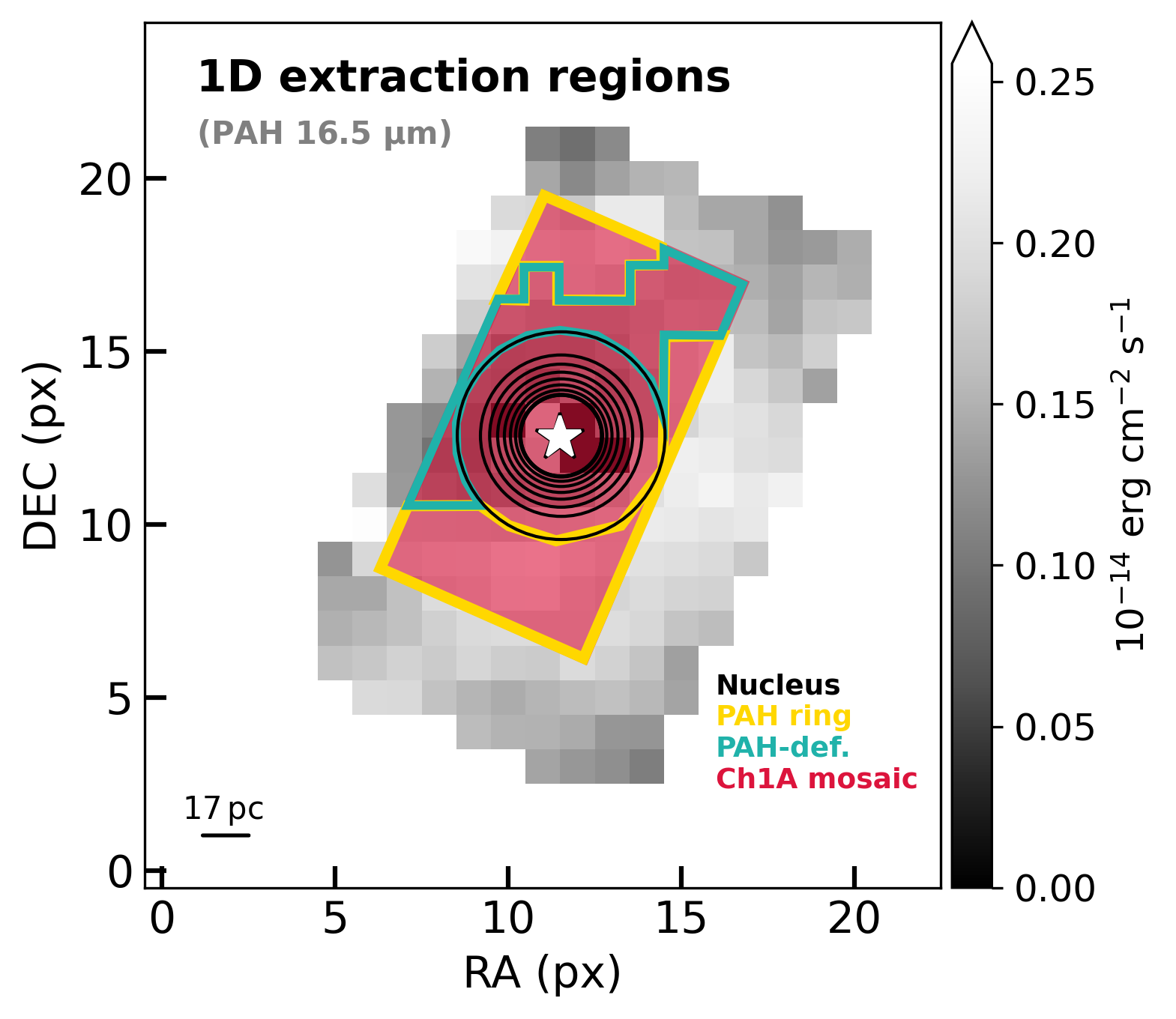}
      \caption{Left: mom0 map of the PAH 16.5~\upmicron feature in the nuclear region of Cen~A at $0.65^{\prime\prime}$ resolution ($\sim11$ pc), obtained after subtraction of the local continuum. Violet contours show the PAH emission at levels of $(0.2,0.25)\times10^{-14}$~erg~cm$^{-2}$~s$^{-1}$. Cyan contours trace the \htwo 0-0 S(1) 17.03~\upmicron rotational line at levels of $(0.19,0.23)\times10^{-14}$~erg~cm$^{-2}$~s$^{-1}$. The black arrows indicate the position angle of the jet (i.e., PA$_{\rm jet}=51\,$deg). Bright pixels near the AGN position are artifacts resulting from the local continuum subtraction. Right: 1D extraction regions defined within the Ch~1A mosaic (red filled area), overlaid on the PAH 16.5~\upmicron surface density map (greyscale): nucleus (concentric circular areas in black), PAH ring (yellow), PAH-deficient region (green). The circum-nuclear region is defined as the full Ch~1A mosaic minus the nuclear emission. In both panels, the star marks the peak of the continuum, corresponding to the position of the AGN. The [0,0] point on the axes (pixel [11,12]) denotes the centre of the sub-channel array (Ch~3C). North is up, and East is to the left. 
      }
    \label{Fig:PAH16p5}
\end{figure*}
To characterise the PAH features in the ICND of Cen~A, we first derived their two-dimensional (2D) spatial distribution. Then, we extracted one-dimensional (1D) spectra from five regions of interest identified in the maps, and measured the PAH feature fluxes and EWs.
In the following, we describe the methods used for the 2D and 1D extractions and subsequent analysis.

\subsection{2D extraction}\label{sect:2D-ext-method}

We constructed the moment-0 (mom0 or 2D) maps of the most prominent PAH emission features (listed in Table \ref{tab:PAH2D}) by locally subtracting the underlying continuum emission. 

Before extracting the mom0 maps, we merged the four individual channels into a single data cube (hereafter `super-cube'), as several of the broader PAH complexes (e.g. the 7.7~\upmicron and 11.3~\upmicron features) straddle adjacent channels. We then resampled all data to a common spaxel scale of $0.6^{\prime\prime}$ \citep[$\sim$10 pc; i.e. Ch~3B angular resolution,][]{Law2023AJ....166...45L} and set the spectral sampling to $\Delta\lambda = 0.005$~\upmicron. We carefully inspected the resulting cube and found no evidence of artifacts introduced by the resampling. Relative to PAHs extracted at the native spaxel scale, the resampled maps exhibit a similar intensity distribution but with an improved signal-to-noise ratio.

We modelled the underlying continuum with a linear fit over narrow wavelength intervals of $\sim 0.01-0.15$~\upmicron to minimize continuum variability, and selected to lie immediately adjacent to each PAH band, as reported in Table \ref{tab:PAH2D}. Alternative choices, such as broader windows or higher-order models, do not give any significant improvement.
We then subtracted the fitted continuum from the super-cube and integrated the PAH flux densities within the corresponding wavelength intervals (Table \ref{tab:PAH2D}). We caution that local continuum subtraction may partly suppress emission in the broad PAH wings that are blended with the strong MIR continuum and, thus, underestimate the total flux. For this reason, we do not use our mom0 maps to compute PAH intensity ratios. 

Several PAH bands overlap with one or more emission lines from warm \htwo and ionized gas (Table \ref{tab:PAH2D}). To minimize contamination, we excluded these lines from the integration intervals\footnote{The excluded wavelength window is centred on the gas emission line and set to $4\times\Delta\lambda\sim0.02$~\upmicron.}. Nevertheless, we cannot fully rule out residual line contributions, which may still influence the resulting PAH mom0 maps. 

\subsection{1D extraction}\label{sect:1Dextraction}

We extracted 1D spectra from five regions of interest: the full Ch~1A mosaic, the nucleus, the circum-nuclear region (defined as the full Ch~1A mosaic minus the nuclear emission), the PAH ring, and a PAH-deficient region located perpendicular to the jet axis. A schematic representation of the extraction regions is shown in the right panel of Fig.~\ref{Fig:PAH16p5}, with additional details on the aperture definitions provided in Appendix~\ref{App:regions}.

Since the PAH emission features are confined to wavelengths $\lambda \leq 17$~\upmicron, we limited our analysis to the $5-21$~\upmicron range (up to Ch~4A), which still ensures adequate sampling of the underlying continuum at longer wavelengths. This choice reduces instrumental systematics, as data at $\lambda \gtrsim 25$~\upmicron are increasingly affected by artifacts and reduced sensitivity, rendering them unreliable for quantitative analysis.
Furthermore, by excluding Ch~4B, we were able to define extraction regions large enough to avoid convolution across channels. Given the non-Gaussian MIRI-MRS point-spread function and the strong diffraction features induced by the nuclear point source in Cen~A, such convolution would likely introduce artifacts and bias the results. For the same reason, we did not perform a pixel-by-pixel spectral decomposition to derive the PAH mom0 maps. Instead, we adopted a local continuum subtraction approach, as described in Sect.~\ref{sect:2D-ext-method}.

\subsection{1D spectral decomposition with \spirit}

We derived the integrated PAH fluxes and EWs using the SPectral InfraRed Inference Tool \citep[\spirit\footnote{\url{https://github.com/FergusDonnan/SPIRIT}};][]{Donnan2024MNRAS.529.1386D}, a spectral-decomposition framework adapted for application to JWST IFU spectroscopy.
\spirit performs a simultaneous fit to the MIR dust continuum, PAH features, and gas emission lines, while accounting for differential extinction. 

The continuum is described by a non-parametric distribution of modified blackbodies (MBBs; $T=35-1500$ K) with wavelength-dependent emissivity from \citet[][their Table 6]{Li2001ApJ...554..778L} and the empirical extinction law of \citet[][their Appendix A]{Donnan2023A&A...669A..87D}. Stellar emission is included through two Flexible Stellar Population Synthesis (FSPS) templates \citep{Conroy2009ApJ...699..486C, Conroy2010ascl.soft10043C} representing 100 Myr and 10 Gyr populations, each subject to an independent foreground extinction screen.

PAH bands are modelled with Drude profiles, incorporating asymmetric components to capture the substructure revealed by MIRI–MRS \citep[e.g.][]{Chown2024A&A...685A..75C}. Emission lines are masked during the continuum and PAH fitting; their fluxes are subsequently recovered from the residual spectrum by integrating each line over its local continuum.

By incorporating differential extinction, \spirit recovers physically plausible dust geometries in which the hottest, most deeply embedded components suffer strong attenuation, while cooler dust remains only mildly obscured. This makes the method particularly well suited for decomposing the spectra of compact obscured nuclei (CONs), frequently found in Seyfert 2 and (ultra-)luminous infrared galaxies (ULIRGs), where the MIR continuum is often dominated by the AGN torus emission \citep[e.g.][]{Donnan2024MNRAS.529.1386D, Garcia-Bernete2024A&A...691A.162G,HermosaMunoz2025A&A...693A.321H}.

\section{Results}\label{sect:results}
\begin{table*}
\caption{PAH, warm molecular hydrogen and ionized gas emission line intensities.}
\label{tab:PAH_1Dfluxes}
\centering
\begin{tabular}{lcccccc}
\hline\hline
\noalign{\smallskip}
\textbf{Line~/~feature}& $\mathbf{\lambda_c}$ & \textbf{Ch~1A mosaic} & \textbf{Nucleus} & \textbf{Circumn-nucl. reg.} & \textbf{PAH ring} & \textbf{PAH def. reg.}\\
\noalign{\smallskip}
& [\tabupmicron] & $[10^{-17}$ W/m$^{2}]$ & $[10^{-17}$ W/m$^{2}]$ & $[10^{-17}$ W/m$^{2}]$ & $[10^{-17}$ W/m$^{2}]$ & $[10^{-17}$ W/m$^{2}]$\\
\noalign{\smallskip}
\hline     
\noalign{\smallskip}
PAH & 6.2 & $108\pm20$ & $8.44\pm2.8$ & $125\pm15$ & $67\pm9$ & $31\pm4$\\
PAH complex & 7.7 &  $251\pm39$ & $70\pm14$ & $268\pm36$ & $132\pm17$& $46\pm6$ \\
PAH complex & 8.6 & $65\pm18$ & $0.12\pm0.05$ & $72\pm17$ & $47\pm9$ & $16\pm5$ \\
PAH complex & 11.3 & $473\pm103$& $164\pm45$ & $224\pm16$& $115\pm8$& $42\pm3$ \\
PAH & 12 & $377\pm71$& $370\pm100$& $92\pm8$ & $28\pm3$  & $8.2\pm0.7$ \\
PAH complex & 12.7 & $233\pm44$& $177\pm49$& $97\pm9$& $48\pm4$& $13\pm1$ \\
PAH complex & 17 & $164\pm32$& $57\pm13$& $113\pm9$ & $71\pm5$ &  $30\pm2$ \\
\htwo S(3) & 9.664 & $15.87\pm0.05$ & $1.36\pm0.03$ & $14.5\pm0.06$ &$5.09\pm0.01$ &  $3.789\pm0.005$ \\
\htwo S(2) & 12.278 & $8.27\pm0.05$ & $0.51\pm0.15$ & $8.19\pm0.05$ & $3.39\pm0.02$ & $2.13\pm0.03$  \\
\htwo S(1) & 17.035 & $14.1\pm0.2$ & $2.63\pm0.09$ & $11.71\pm0.02$ & $6.11\pm0.01$ & $3.455\pm0.005$  \\
\neii & 12.814 & $163.8\pm0.06$ & $132.62\pm0.06$ & $31.05\pm0.03$ & $7.17\pm0.04$ &  $5.34\pm0.02$ \\
\nev & 14.3217 & $23.4\pm0.1$ & $19.4\pm0.1$ & $4.29\pm0.02$ & $0.68\pm0.02$  & $0.857\pm0.008$ \\
\neiii & 15.555& $124.43\pm0.04$ & $103.5\pm0.1$ & $20.609\pm0.005$ & $6.023\pm0.004$ &  $5.71\pm0.01$ \\
\noalign{\smallskip}
\hline 
\end{tabular}
\tablefoot{Intensities are derived using \spirit. The second column lists the central wavelength of each feature. PAH complexes include the following subcomponents: 7.42, 7.55, 7.61, and 7.82~\upmicron features (PAH 7.7); 8.50 and 8.61~\upmicron (PAH 8.6);  11.20 and 11.26~\upmicron (PAH 11.3); 12.60 and 12.77~\upmicron (PAH 12.7); 16.45, 17.04 and 17.375~\upmicron (PAH 17). We list only the lines used in this work.}
\end{table*}
\begin{table*}
\caption{PAH equivalent widths (EWs).}
\label{tab:PAH_1DEW}
\centering
\begin{tabular}{lccccc}
\hline\hline
\noalign{\smallskip}
\textbf{PAH}& \textbf{Ch~1A mosaic} & \textbf{Nucleus} & \textbf{Circumn-nucl. reg.} & \textbf{PAH ring} & \textbf{PAH def. reg.}\\
\noalign{\smallskip}
$\lambda_c$ [\tabupmicron] & [nm] & [nm]  & [nm]  & [nm]  & [nm]  \\ 
\noalign{\smallskip}
\hline     
\noalign{\smallskip}
6.2 & $24\pm2$& $2.3\pm0.2$& $140\pm10$ & $310\pm30$ & $290\pm30$\\   
7.7 complex& $69\pm7$& $24\pm2$& $370\pm40$ & $700\pm70$ & $460\pm50$ \\ 
8.6 complex& $25\pm2$& $0.055\pm0.005$& $130\pm10$ & $380\pm40$ & $240\pm20$ \\
11.3 complex & $220\pm20$& $89\pm9$& $630\pm60$ & $1400\pm100$ & $910\pm90$\\ 
12 & $16\pm2$& $190\pm20$& $240\pm20$ & $320\pm30$ & $160\pm20$ \\
12.7 complex & $93\pm9$& $83\pm8$& $250\pm30$ & $500\pm50$ & $220\pm20$ \\
17 complex & $78\pm8$& $31\pm3$& $400\pm40$ & $1100\pm100$ & $790\pm80$ \\
\noalign{\smallskip}
\hline
\end{tabular}
\tablefoot{Same as Table \ref{tab:PAH_1Dfluxes}. We assume an uncertainty of 10\%.
}
\end{table*}
\begin{table*}
\caption{Intensity ratios.}  
\label{tab:PAH_1Dratios}
\centering 
\begin{tabular}{lccccc}
\hline\hline
\noalign{\smallskip}
\textbf{PAH ratios}& \textbf{Ch~1A mosaic} & \textbf{Nucleus} & \textbf{Circumn-nucl. reg.} & \textbf{PAH ring} & \textbf{PAH def. reg.}\\
\noalign{\smallskip}
\hline     
\noalign{\smallskip}
$6.2/7.7$ & $0.43\pm0.04$& $0.12\pm0.03$& $0.47\pm0.04$& $0.51\pm0.05$& $0.67\pm0.05$ \\
$6.2/11.3$ & $0.23\pm0.02$& $0.052\pm0.004$& $0.56\pm0.04$& $0.58\pm0.04$& $0.74\pm0.05$ \\
$7.7/11.3$ & $0.53\pm0.05$& $0.47\pm0.05$& $1.2\pm0.1$& $1.15\pm0.10$& $1.095\pm0.100$ \\
$11.3/6.2$ & $4.4\pm0.3$& $19\pm1$& $1.8\pm0.1$& $1.7\pm0.1$& $1.4\pm0.1$ \\
$11.3/7.7$ & $1.9\pm0.2$& $2.3\pm0.2$ & $0.8\pm0.2$& $0.9\pm0.3$& $0.9\pm0.3$ \\
$11.3/12.7$ & $2.0\pm0.2$& $0.9\pm0.1$& $2.3\pm0.3$& $2.4\pm0.3$& $3.2\pm0.4$ \\
$L_{\rm H2}/L_{\rm PAH\,7.7}$& 0.11& 0.06 & 0.13& 0.11& 0.20 \\
\noalign{\smallskip}
\hline
\end{tabular}
\tablefoot{Same as Table \ref{tab:PAH_1Dfluxes}. The quoted uncertainties are fully propagated through each ratio. \htwo luminosity (L$_{\rm H2}$) is obtained as the sum of \htwo~S(1), \htwo~S(2), and \htwo~S(3) intensities (Tab. \ref{tab:PAH_1Dfluxes}).}
\end{table*} 

\subsection{PAH spatial distribution}

All PAH features display remarkably similar spatially resolved morphologies, with the exception of the 7.7~\upmicron complex (see Appendix~\ref{App:PAH2Dmaps}). We adopted the 16.5~\upmicron PAH feature as our morphological reference, as it lies within MIRI-MRS Ch~3C, which offers the largest FoV and enables a direct comparison with the least energetic \htwo pure rotational line accessible with MIRI-MRS, namely the \htwo~0-0~S(1) line at 17.03~\upmicron (Fig.~\ref{Fig:PAH16p5}, left panel; see Appendix~\ref{App:PAH2Dmaps} for details on the \htwo mom0 extraction).

Within the ICND, the PAH emission is distributed in a ring-like structure and shows a pronounced deficit relative to the local continuum within the inner $\sim40$~pc (in radius), corresponding to the region from which the jet is launched. However, we cannot rule out the presence of PAH features at $r<40$~pc, which may be buried within the strong MIR continuum.
The PAH emission is not spatially uniform, but instead appears clumpy, with localized intensity enhancements that are offset by several tens of parsecs from the brightest \htwo clumps. 
We also identify a distinct PAH-deficient region toward the North-West, oriented approximately perpendicular to the jet axis and spatially coincident with one of the brightest \htwo peaks. This region further coincides with both an area of enhanced velocity dispersion in the ionized gas \citep{Alonso2025A&A...699A.334A} and the innermost inflowing branch of molecular gas traced by CO(3--2) \citep[][]{Espada2017ApJ...843..136E}, as shown in Fig.~\ref{Fig:multi-phase}.

\subsection{Integrated PAH features}
\begin{figure*}
   \centering
   \includegraphics[width=\hsize]{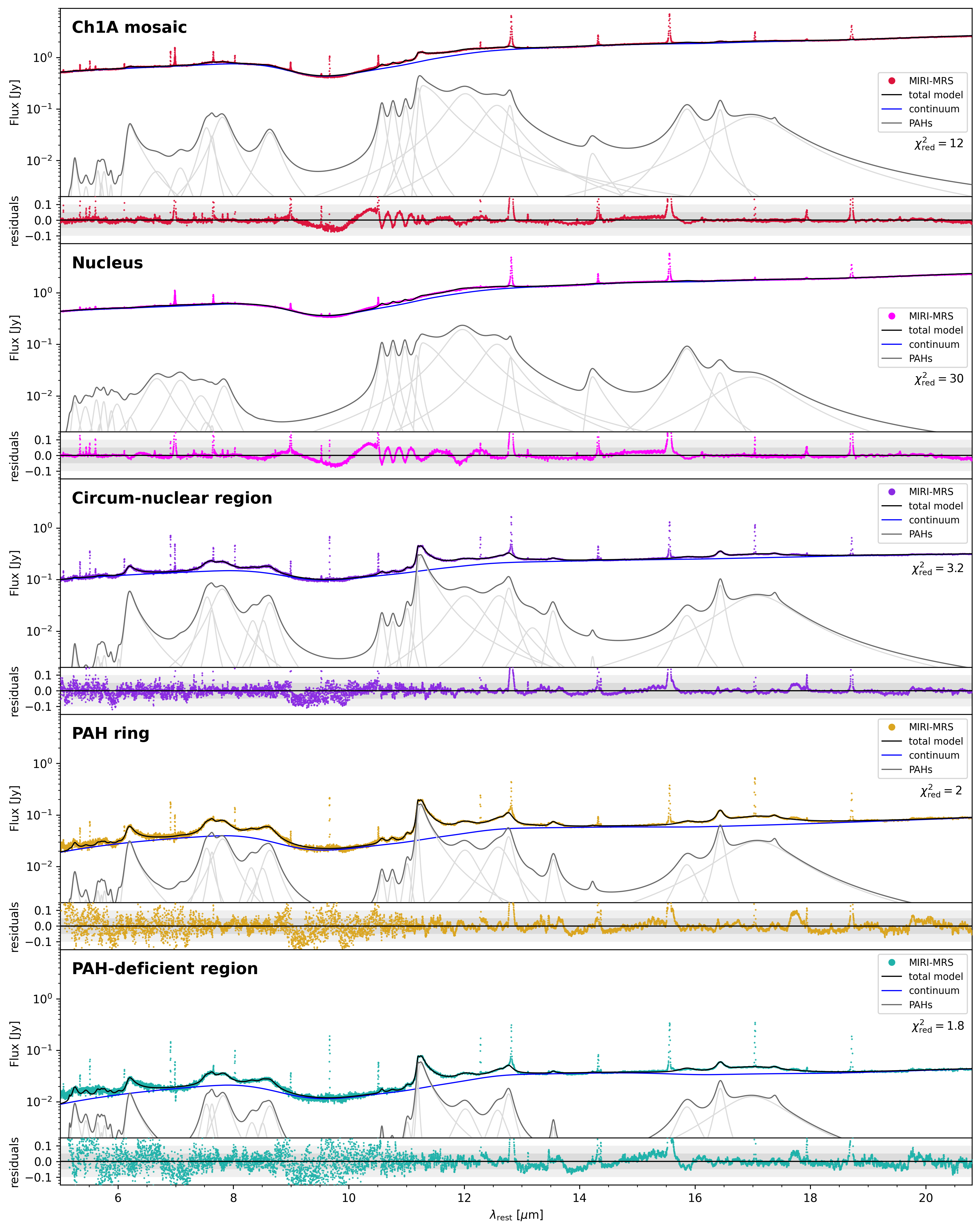}
      \caption{MIRI–MRS spectra extracted from five regions (top to bottom): full Ch~1A mosaic (red), nucleus (magenta), circum-nuclear region (violet), PAH ring (yellow), and PAH-deficient region (green). Extraction regions are shown in the right panel of Fig.~\ref{Fig:PAH16p5}. Black lines indicate the total best-fit models from the MIR decomposition tool \citep{Donnan2024MNRAS.529.1386D}; blue and gray lines show the fitted continua (stellar + AGN) and PAH components, respectively. Fit residuals are displayed below each panel; the reduced $\chi^2$ is reported next to the legend.
      }
    \label{Fig:PAH-SED}
\end{figure*}

We fitted the five spectra extracted from the regions in Fig. \ref{Fig:PAH16p5} (right panel) with \spirit \citep{Donnan2024MNRAS.529.1386D}, as described in Sect.~\ref{sect:1Dextraction}.
Fig.~\ref{Fig:PAH-SED} presents the best-fitting models for the five regions together with their residuals, which remain within $10-15$\%.
The nuclear spectrum yields the poorest fit, with $\chi_{\rm red}^{2}=30$, and shows the largest residuals, particularly across the $9-12$~\upmicron range. The intense AGN continuum makes it difficult to constrain the PAH features, which are strongly diluted. As such, the nuclear PAH intensities are possibly biased and potentially underestimated. In Appendix~\ref{App:nucleus}, we assess the effect of excluding the PAH features from the spectral fit.
The goodness of fit improves markedly when the nucleus is excluded: the reduced $\chi^{2}$ decreases from 12 for the full Ch~1A extraction to 3.2 for the circum-nuclear region (cf. the top and third panels of Fig.~\ref{Fig:PAH-SED}). The PAH ring and the PAH-deficient region spectra are reproduced to a very high accuracy, with $\chi_{\rm red}^{2}$ values of 2 and 1.8, respectively.

The typical uncertainties on the integrated PAH intensities computed by \spirit via bootstrap resampling are below a few percent, reflecting the small errors reported by the JWST pipeline and likely underestimated. We therefore adopt a more realistic error estimate based on the Median Absolute Deviation (MAD) of the PAH flux densities within each extraction region. The MAD is less sensitive to outliers and remains robust even for small samples. In Table~\ref{tab:PAH_1Dfluxes} we list the PAH band intensities and other relevant spectral lines, while in Table~\ref{tab:PAH_1DEW} we report the PAH EWs, and in Table~\ref{tab:PAH_1Dratios} the PAH intensity ratios. 
\begin{figure*}
   \centering
   \includegraphics[width=\hsize]{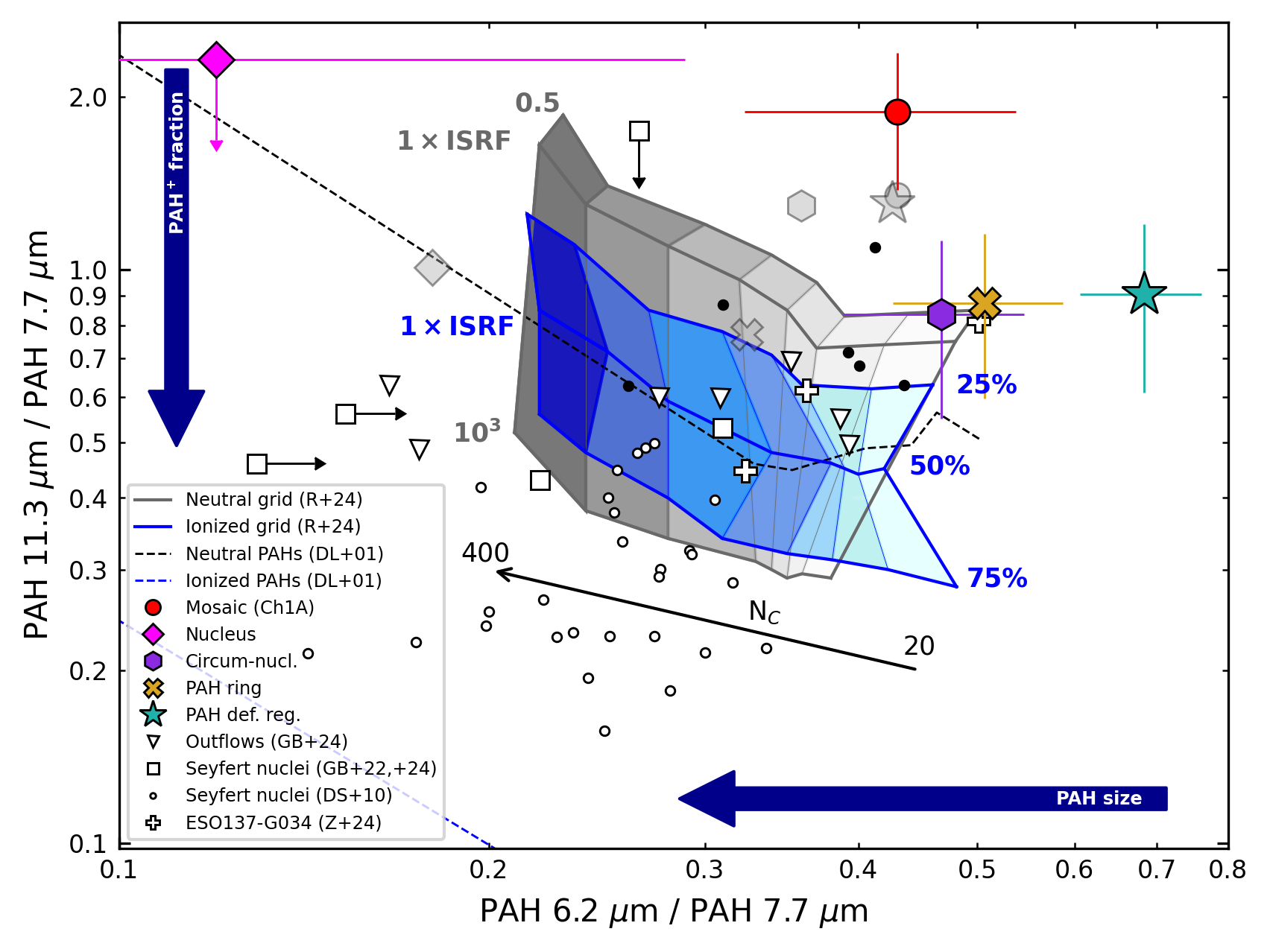}
      \caption{PAH 11.3/7.7~\upmicron versus PAH 6.2/7.7~\upmicron diagnostic plot for the five regions of interest: full Ch~1A mosaic (red circle), nucleus (magenta diamond), circum-nuclear region (violet hexagon), PAH ring (yellow cross), and PAH-deficient region (green star). 
      The large uncertainty on the nuclear PAH 6.2/7.7~\upmicron ratio and the and the upper limit on 11.3/7.7~\upmicron result from partial detection of the 6.2 and 7.7~\upmicron features (Appendix~\ref{App:nucleus}). Grey symbols show \cafe-derived ratios (Appendix \ref{App:cafe}). Uncertainties on the \cafe ratios, which span nearly the full x-range of the figure while remaining close to the \spirit y-axis values, were omitted for clarity. We overlay the grids of \citet{Rigopoulou2021MNRAS.504.5287R} for neutral PAHs (grey; illuminated by $0.5-10^3$ ISRF), and partially ionized PAHs ($25-75$\%; blue; 1~ISRF). PAH size increases along the x-axis (right to left) with the number of carbon atoms (N$_C$) while the ionized fraction increases along the y-axis (top to bottom). The dashed black and blue curves mark the neutral and ionized PAH limits from \citet{DraineLi2001ApJ...551..807D}, for 1~ISRF. 
      For comparison, we include PAH ratios for Seyfert nuclei from \citet[][circles; filled symbols indicate high \htwo/PAH luminosity ratios]{Diamond-Stanic2010ApJ...724..140D}; and from \citet{Garcia-Bernete2022A&A...666L...5G,Garcia-Bernete2024A&A...691A.162G}, shown as squares.
      Triangles denote outflow regions in the same systems. Plus symbols represent the three regions in the central kpc of the Seyfert 2 ESO~137-G034 \citep{ZhangLulu2024ApJ...975L...2Z}. For clarity, we omit the uncertainties associated with literature ratios.
      }
    \label{Fig:PAHratio_d1}
\end{figure*}

Because individual PAH bands arise from distinct vibrational modes, PAH intensity ratios provide powerful diagnostics of PAH size, charge state, and molecular structure \citep[e.g.][]{DraineLi2001ApJ...551..807D, Galliano2008ApJ...679..310G, Rigopoulou2021MNRAS.504.5287R, Rigopoulou2024MNRAS.532.1598R}. In the following sections, we investigate the properties of PAH molecules in the ICND of Cen~A by comparing the measured ratios with theoretical predictions and with results from previous studies of Seyfert nuclei.

\subsubsection{PAH size and charge}
Ratios between bands arising from the same vibrational modes (e.g. the C--C stretching ratio 6.2/7.7) are sensitive to the PAH size distribution, as first noted by \citet{DraineLi2001ApJ...551..807D} and more recently confirmed by e.g., \citet[][see their Fig. 5]{Draine2021ApJ...917....3D,Rigopoulou2021MNRAS.504.5287R}. 
Ratios between the 11.3~\upmicron C--H out-of-plane bending feature, enhanced in neutral species, and the $6-9$~\upmicron C--C stretching features, intrinsically stronger in ionized PAHs, directly trace the PAH charge state and, by extension, the physical conditions of the emitting region \citep{Allamandola1999ApJ...511L.115A, Hudgins1999ApJ...513L..69H, Hony2001A&A...370.1030H, Kim2002ApJS..143..455K, Galliano2008ApJ...679..310G}. 

We used the 6.2/7.7 and 11.3/7.7 PAH band ratios (Table~\ref{tab:PAH_1Dratios}) to constrain the PAH size and charge in the five regions of interest. Both the 6.2 and 7.7~\upmicron features predominantly arise from ionized PAHs, rendering their ratio relatively robust in star-forming environments but less reliable in AGN-dominated regions, where recent evidence points to a predominantly neutral PAH population \citep[e.g.,][]{Garcia-Bernete2024A&A...691A.162G}. In this context, the ratio between the 17 (or 11.3) and 3.3~\upmicron features, emitted mainly by neutral PAHs, has been shown to provide a reliable alternative tracer of PAH size \citep[e.g.][]{Rigopoulou2024MNRAS.532.1598R}. That said, we are compelled to rely on the 6.2/7.7 ratio, as the MIRI-MRS data do not extend to wavelengths shorter than 4.9~\upmicron and the available NIRSpec observations of Cen~A \citep[][]{Dumont2025A&A...703A..54D}
cover only the central $\sim3^{\prime\prime}\times3^{\prime\prime}$ ($\sim50\times50$~pc$^2$).

In Fig.~\ref{Fig:PAHratio_d1} we compare our measurements with the theoretical grids of \citet{Rigopoulou2021MNRAS.504.5287R}, and the model predictions by \citet{DraineLi2001ApJ...551..807D}.
Since theoretical models depend on the intensity of the radiation field exciting the PAH molecules, we estimated an upper limit to the AGN radiation in the centre of Cen~A (Appendix~\ref{App:ISRF}), i.e. $\sim10^{3}\,{\rm ISRF}$, where ${\rm ISRF}=2.2\times10^{-5}\ {\rm W\,m^{-2}} = 220\ {\rm erg\,s^{-1}\,m^{-2}}$ is the interstellar radiation field in the solar neighbourhood \citep{Mathis1983A&A...128..212M, Galliano2022HabT.........1G}.
We restricted accordingly the theoretical predictions in Fig. \ref{Fig:PAHratio_d1} ($0.5$–$10^{3}$ ISRF).

All the five regions (cf. Table~\ref{tab:PAH_1Dratios}) lie in the upper part of the plot, at $11.3/7.7\ge0.7$.
Spectral fitting performed with \cafe \citep{Marshall2007ApJ...670..129M,CAFE2025ascl.soft01001D}, another tool widely used for decomposing MIR spectra of AGNs \citep[e.g.,][]{Armus2023ApJ...942L..37A,Lai2023ApJ...957L..26L}, yields PAH ratios consistent with our
measurements (grey filled symbols in Fig.~\ref{Fig:PAHratio_d1}), albeit with large uncertainties (see Appendix~\ref{App:cafe}). This agreement indicates that our results are robust against the choice of spectral decomposition method.

The nuclear region, lying in the upper-left corner of the plot, is likely biased toward high 11.3~\upmicron values, due to the combined effects of the strong 9.8~\upmicron silicate absorption feature and the intense MIR continuum, which leads to only partial detection of the 6.2 and 7.7~\upmicron features. Consequently, we treat the 11.3/7.7 ratio as an upper limit, while the 6.2/7.7 ratio remains poorly constrained.

The PAH ratios in the other four regions lie at $6.2/7.7 \ge 0.4$ and $11.3/7.7 \ge 0.7$. While large neutral PAHs ($>100$ C atoms) could in principle produce high 11.3/7.7 ratios, they would be expected to show low 6.2/7.7 ratios, contrary to our observations. Only the PAH ring and circum-nuclear region are consistent with small (N$_C\sim 20$) neutral PAHs \citep{Rigopoulou2021MNRAS.504.5287R}, yet they are offset from the predictions by \citet{DraineLi2001ApJ...551..807D}.

The predictions of both \citet{DraineLi2001ApJ...551..807D} and \citet{Rigopoulou2021MNRAS.504.5287R} are based on models of pericondensed PAHs (e.g., pyrene, coronene, ovalene). Our ratios could be naturally explained, instead, if the PAH population is dominated by catacondensed species (e.g., naphthalene, pentaphene, pentacene), which have more open and irregular structures \citep[see][and references therein]{Li2020NatAs...4..339L}. Catacondensed PAHs contain more H atoms per C atom than pericondensed PAHs, resulting in intrinsically higher 11.3/7.7 ratios at a given 6.2/7.7 ratio.

Computational studies of catacondensed PAHs \citep[e.g.,][]{Pathak2005CP....313..133P} indicate that neutral species show enhanced C--H stretching (3.3~\upmicron), while in cations the C--C stretch (6.2 and 7.7~\upmicron) and C--H in-plane modes (8.6~\upmicron) dominate.
Since the PAH population in the five Cen~A's regions appears to be dominated by neutral species, diagnostics based on the 6.2/7.7 ratio may bias our conclusions. Direct access to the 3.3~\upmicron feature would provide an independent verification of our findings. Ultimately, theoretical models that incorporate catacondensed PAH species are required to further test and confirm this scenario.

Alternatively, the extreme 11.3/7.7 ratios could result from poor PAH fitting. This is likely the case for the nucleus and the full Ch~1A mosaic, where the AGN MIR continuum dominates.
\begin{figure}
   \centering
   \includegraphics[width=\hsize]{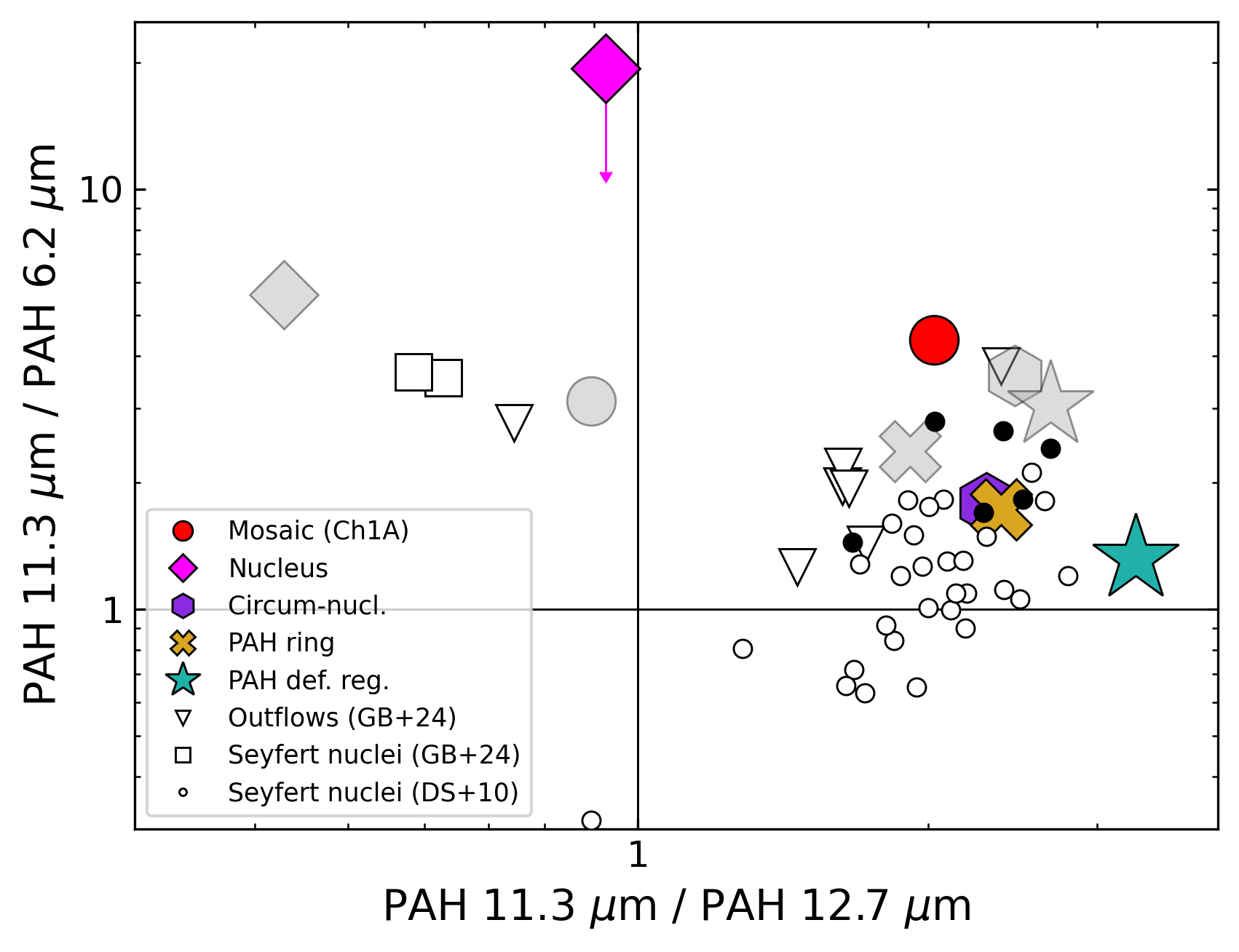}
      \caption{PAH 11.3/6.2~\upmicron versus PAH 11.3/12.7~\upmicron diagnostic plot. Symbols are the same as Fig. \ref{Fig:PAHratio_d1}. Uncertainties, between $10-50$\%, are omitted for clarity. The nuclear upper limit results from partial detection of the 6.2~\upmicron feature (Appendix~\ref{App:nucleus}).
      }
    \label{Fig:PAHratio_d12.7}
\end{figure}

\citet{Diamond-Stanic2010ApJ...724..140D} reported similar PAH ratios for the central regions ($3.6^{\prime\prime}\times7.2^{\prime\prime}$, median distance 22 Mpc) of six Seyferts (NGC~5194, NGC~4501\footnote{Comparable ratios for NGC~4501 were found by \citet{Garcia-Bernete2022MNRAS.509.4256G}.}, NGC~4639, NGC~1433, NGC~2639, NGC~5005), shown in Fig.~\ref{Fig:PAHratio_d1} as black filled circles. These objects lie beyond the predictions of \citet{DraineLi2001ApJ...551..807D}, but generally agree with the neutral PAH grid of \citet{Rigopoulou2021MNRAS.504.5287R} for $N_C \lesssim 200$, except for NGC~5194 ($11.3/7.7 \sim 1.09$, $6.2/11.3 \sim 0.41$), which is also classified as LINER. The authors reported for these objects enhanced \htwo emission, which might be a signature of the presence of AGN-driven shocks that collisionally excite the \htwo. In this case, shocks might play a role in setting the observed PAH ratios (for further discussion see Sect. \ref{sect:shocks}).

In Fig.~\ref{Fig:PAHratio_d1} we also include recent MIRI–MRS measurements of local Seyfert nuclei. These comprise nuclear and outflow regions of NGC~5728, NGC~5506, and NGC~7172 \citep[$\sim75-100$ pc;][]{Garcia-Bernete2024A&A...691A.162G}, and the nuclear regions of NGC~6552, NGC~7319, and NGC~7469 \citep[$\sim142-245$ pc][]{Garcia-Bernete2022A&A...666L...5G}, comparable in size to our Cen~A mosaic. Additionally, we include three 500 pc regions sampling the central kiloparsec of ESO~137-G034 \citep{ZhangLulu2024ApJ...975L...2Z}, dominated by collisional shock heating.

Most of these regions lie beyond the predictions of \citet{DraineLi2001ApJ...551..807D} but remain consistent with the neutral PAH grid of \citet{Rigopoulou2021MNRAS.504.5287R}, with ionized PAH fractions below 25\%. The nuclei of NGC~5728, NGC~5506, and NGC~6652, and the outflow regions of NGC~5506 and NGC~7172, are consistent with emission dominated by large, neutral PAHs. Overall, these comparisons show that AGN-driven shocks (e.g. by nuclear outflows) preferentially destroys ionized PAHs, which is consistent with the elevated 11.3/7.7 ratios observed in the central regions of Cen~A.

\subsubsection{PAH hydrogenation state}

PAH intensity ratios can also provide insights on the hydrogenation state of the dominant PAH species. 
The 11.3~\upmicron feature is due to out-of-plane C--H bending of solo hydrogen atoms \citep{Hony2001A&A...370.1030H}, while the 12.7~\upmicron, 13.6~\upmicron and 14.2~\upmicron PAH bands originate from out-of-plane bending vibrations of PAHs with duo and trio/quartet peripheral hydrogens \citep{Hudgins1999ApJ...513L..69H, Hony2001A&A...370.1030H, Tielens2008ARA&A..46..289T}; as such, the ratio 11.3~/~12.7 is frequently used to infer the geometry or edge structure of interstellar PAHs. Lower 11.3~/~12.7 ratios imply more irregular PAHs (i.e., more duo sites), while higher 11.3~/~12.7 ratios point towards a dominant population of more compact PAHs (fewer duo sites).

In Fig.~\ref{Fig:PAHratio_d12.7} we plot the 11.3/6.2 ratio against 11.3/12.7, the latter tracing the relative contribution of duo and trio C--H groups compared to solo groups. Seyfert nuclei and outflows span a broad range of 11.3/12.7 values, although clustering at values $>1$, while their 11.3/6.2 ratios range from $\sim$1 to $\sim4$. The circum-nuclear regions of Cen~A exhibit 11.3/12.7 ratios of $\gtrsim 2$, consistent with those measured in Seyfert nuclei \citep[e.g.][]{Diamond-Stanic2010ApJ...724..140D}. Such high ratios indicate a dominance of solo over duo/trio C--H groups, suggesting partial PAH dehydrogenation. This effect is strongest in Cen~A's PAH-deficient region, which shows the largest 11.3/12.7 ratio.

\subsubsection{PAH equivalent widths}
\begin{figure}
   \centering
   \includegraphics[width=0.96\columnwidth]{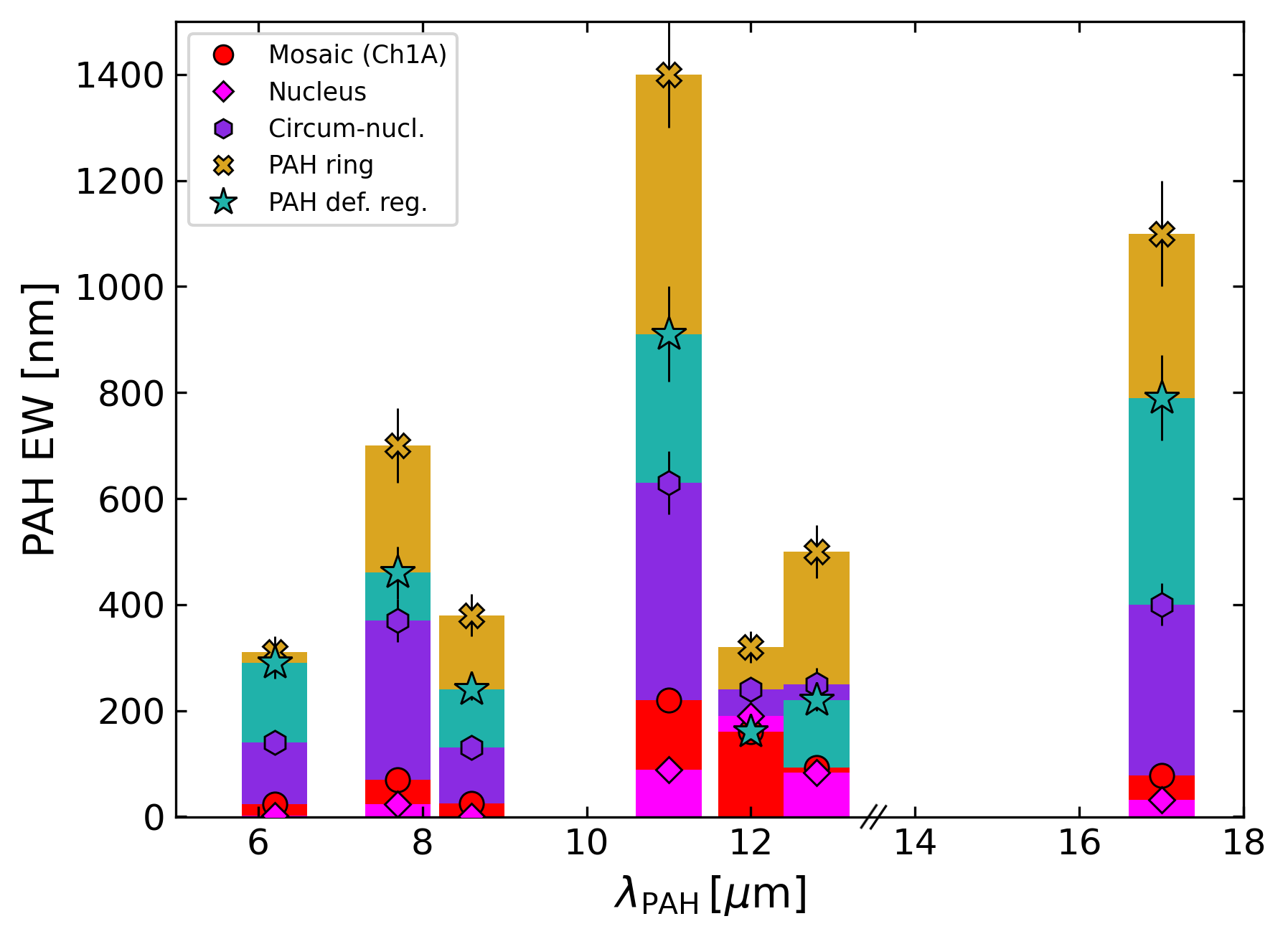}\\
   \includegraphics[width=0.96\columnwidth]{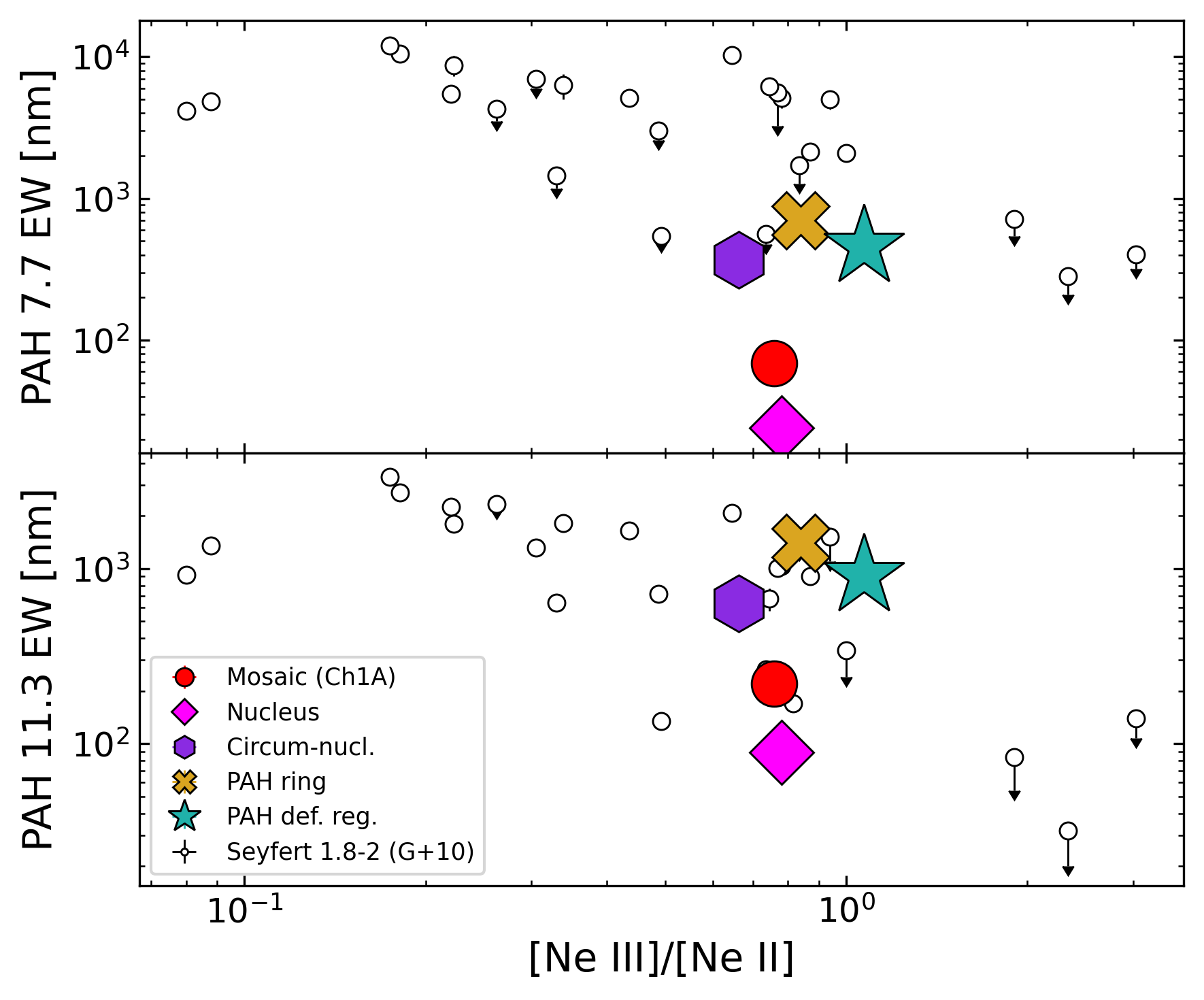}\\
   \includegraphics[width=0.96\columnwidth]{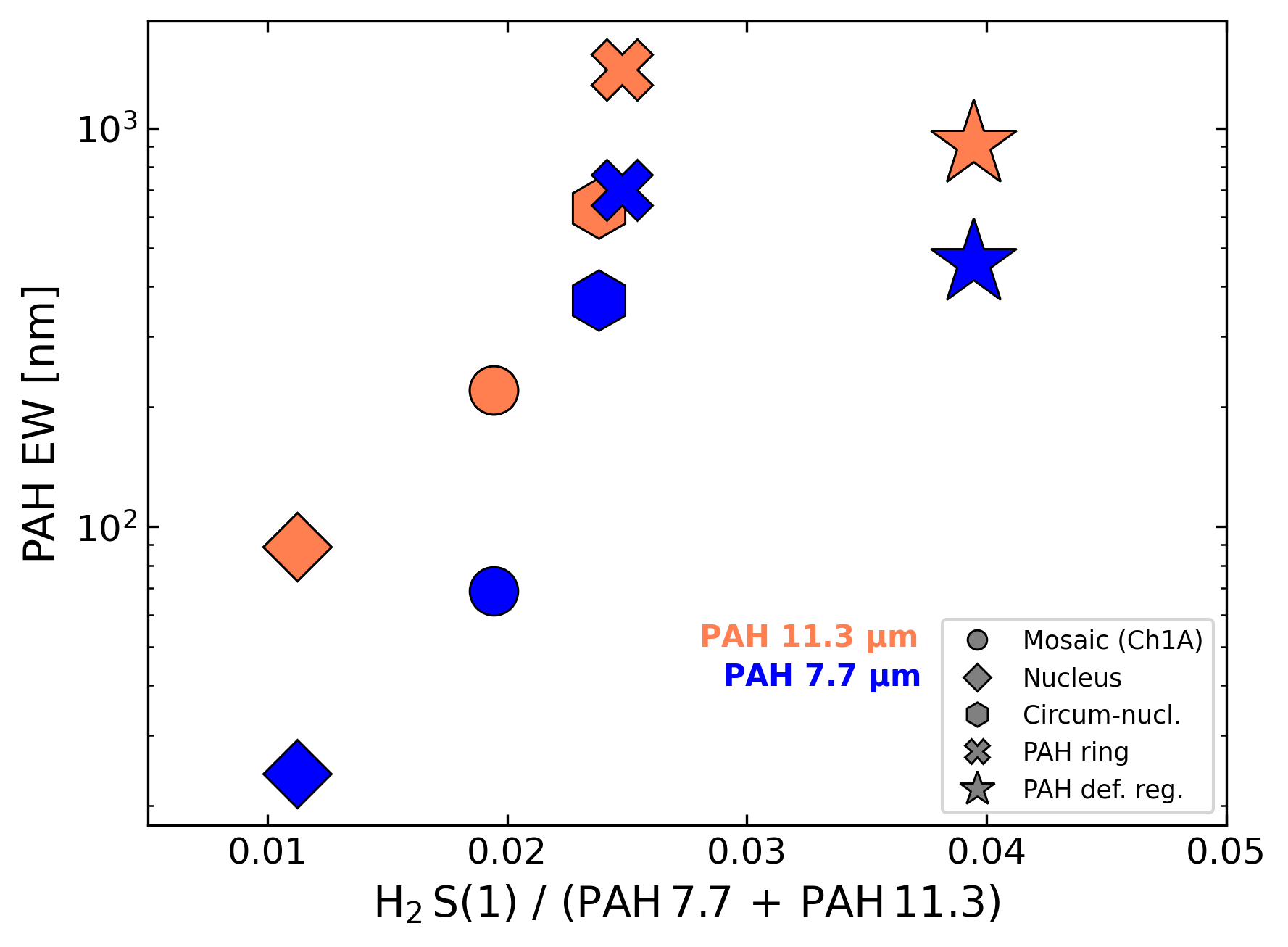}
      \caption{Top: Bar plot of PAH EWs for the full Ch~1A mosaic (red), nucleus (magenta), circum-nuclear region (violet), PAH ring (yellow), and PAH-deficient region (green). Middle: EWs of the PAH 7.7~\upmicron (top) and 11.3~\upmicron (bottom) complexes versus \neiii/\neii, tracing the radiation-field hardness. Points correspond to the five regions defined above, colour-coded as in the top panel. For comparison, we include Seyfert 1.8–2 nuclei from \citet{Gallimore2010ApJS..187..172G} and \citet{Sales2010ApJ...725..605S}. Bottom: EWs of the PAH 7.7~\upmicron (blue) and 11.3~\upmicron (orange) complexes versus \htwo~S(1) intensity relative to the PAH bands, tracing the role of shock-driven collisional excitation. Symbols denote region type: circle (Ch~1A mosaic), diamond (nucleus), hexagon (circum-nuclear), cross (PAH ring), and star (PAH-deficient). EW uncertainties are smaller than or comparable to the symbol size.
      }
    \label{Fig:PAH_EW}
\end{figure}

PAH EWs quantify the strength of PAH features relative to the local MIR continuum. They provide a useful diagnostic of PAH abundance and destruction \citep[e.g.,][]{Roche1991MNRAS.248..606R, Voit1992MNRAS.258..841V}, though they are also strongly affected by dilution from hot-dust continuum emission \citep[e.g.,][]{Alonso2014MNRAS.443.2766A, Robinson2026arXiv260109810R}. As a result, PAH EWs are widely used to probe the resilience of PAHs in harsh environments, to differentiate star-forming from AGN-dominated regions \citep[e.g.,][]{Treyer2010arXiv1005.1316T, Garcia-Bernete2022MNRAS.509.4256G}, and to trace how PAH excitation responds to variations in the intensity and hardness of the radiation field \citep[e.g.,][]{Sales2010ApJ...725..605S, Sajina2022Univ....8..356S}.

The top panel of Fig.~\ref{Fig:PAH_EW} shows the PAH EWs in the five regions of interest. The EWs vary substantially: they are lowest in the nucleus and in the full Ch~1A mosaic, and highest in the PAH ring. The 7.7, 11.3, and 17~\upmicron complexes exhibit the strongest variations, whereas the 12~\upmicron feature is comparatively uniform. In the following, we focus on the 7.7~\upmicron (primarily ionized PAHs) and 11.3~\upmicron (primarily neutral PAHs) complexes.

In the middle panel of Fig.~\ref{Fig:PAH_EW} we examine the dependence of PAH EWs on the \neiii/\neii ratio, a standard tracer of radiation-field hardness \citep[e.g.,][]{Thornley2000ApJ...539..641T, Dale2006ApJ...646..161D, Snijders2007ApJ...669..269S, Pereira-Santaella2010ApJ...725.2270P, Inami2013ApJ...777..156I}. We found a pronounced decrease in both 7.7~\upmicron and 11.3~\upmicron EWs with increasing \neiii/\neii, consistent with previous studies \citep[e.g.,][]{Gallimore2010ApJS..187..172G, Sales2010ApJ...725..605S}. Two regimes are evident:
(i) a steep decline for the nucleus, the Ch~1A mosaic, and the circum-nuclear region, and
(ii) a shallower trend for the PAH ring and the PAH-deficient region, both of which display higher \neiii/\neii.
For the former regions, the steep decline is most naturally explained by AGN continuum dilution, consistent with their prominent MIR continua and high $\chi^2_{\rm red}$ values (cf. Fig.~\ref{Fig:PAH-SED}). However, we can not exclude contribution from PAH photo-erosion by the AGN radiation field, particularly in the nucleus. The high \neiii/\neii ratios in the PAH ring and PAH-deficient region may indicate an extra contribution from shocks to the gas excitation and ionization \citep{Alonso2025A&A...699A.334A}, as further supported by Neon line ratio diagnostics (Appendix~\ref{App:AGNmodels}). Given their weaker MIR continua and low $\chi^2_{\rm red}$ continuum dilution appears negligible here, and shocks are likely the dominant mechanism driving the reduced EWs in the PAH-deficient region compared to the PAH ring.

Finally, in Fig.~\ref{Fig:PAH_EW} (bottom) we further investigate the influence of shocks by plotting PAH EWs as a function of the \htwo~S(1) line intensity relative to PAHs (cf. Sect.~\ref{sect:shocks}).
For the nucleus, the Ch~1A mosaic, and the circum-nuclear region, PAH EWs increase with \htwo~S(1)/PAH luminosity ratio, suggesting that the steep decrease in EW with \neiii/\neii is not shock-driven. In contrast, the PAH-deficient region shows both the highest \htwo/PAH ratio and significantly reduced EWs compared to the PAH ring, reinforcing the interpretation that shocks lead to the partial destruction of both neutral and ionized PAHs in this region.

\section{Discussion}\label{sect:discussion}

In this section, we explore the impact of AGN-driven shocks on the PAH emission features in the ICND of Cen~A. Then, we assess possible evolutionary pathways of the PAH population, by combining our results with complementary sub-arcsecond-resolution observations of the molecular and ionized gas.

\subsection{Impact of AGN-driven shocks}\label{sect:shocks}

Previous studies found that collisional heating by AGN-driven shocks is associated to enhanced warm \htwo emission in the MIR \citep[e.g.,][]{Roussel2007ApJ...669..959R, Ogle_2010, Guillard2012ApJ...747...95G}.
Within the MIRI-MRS mosaic of Cen~A, \citet{Evangelista2026} measured an excess in \htwo relatively to the upper limit predicted by the PDR/XDR models and required by Cosmic Ray heating, which indicates that \htwo is largely collisionally-excited. This is confirmed by the \htwo/7.7~\upmicron luminosity ratios (Table \ref{tab:PAH_1Dratios}), which always exceed the threshold for the presence of significant non-radiative heating component \citep[i.e., $L_{\rm H2}/L_{\rm PAH\,7.7} = 0.04$;][]{Ogle_2007,Ogle_2010,Guillard2012ApJ...747...95G}. 
Between the five regions analysed in this work, the PAH deficient region shows the highest $L_{\rm H2}/L_{\rm PAH\,7.7}$ ratio, i.e. 0.20.

\begin{figure}
   \centering
   \includegraphics[width=\columnwidth]{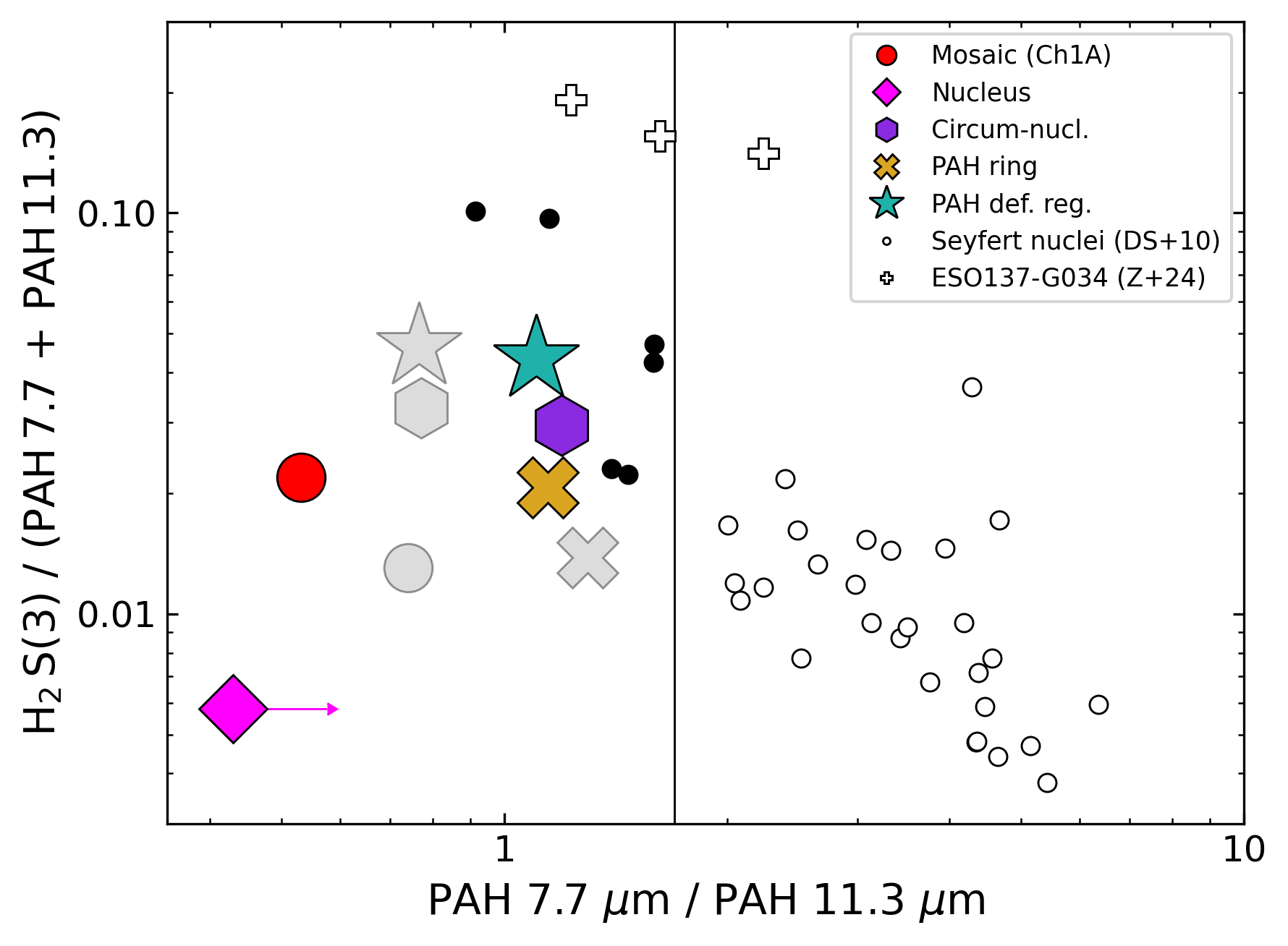}
      \caption{\htwo~S(3) intensity relative to 7.7 and 11.3~\upmicron PAH bands diagnostic plot, quantifying shock-driven collisional excitation of \htwo. Symbols are as in Fig.~\ref{Fig:PAHratio_d1}. Uncertainties, between $10-50$\%, are omitted for clarity. The nuclear lower limit results from partial detection of the 7.7~\upmicron feature (Appendix~\ref{App:nucleus}).
      }
    \label{Fig:PAHratio_dH2}
\end{figure}

Regions with enhanced, shock-heated \htwo emission are known to exhibit systematically lower 7.7/11.3 ratios \citep[e.g.,][]{Ogle2007ApJ...668..699O, Smith2007ApJ...656..770S, Guillard2010A&A...518A..59G, Vega2010ApJ...721.1090V}. As shown in Fig.~\ref{Fig:PAHratio_dH2}, the strength of the \htwo S(3) line, normalized to the 7.7 and 11.3~\upmicron PAH features, anti-correlates with the 7.7/11.3 ratio. The five regions in Cen~A's MIRI-MRS mosaic follow the same trend. Notably, they occupy the same locus as the six Seyfert nuclei from \citet{Diamond-Stanic2010ApJ...724..140D} and the two shock-dominated regions in the central kpc of ESO~137-G034 \citep{ZhangLulu2024ApJ...975L...2Z} that lie beyond the \citet{DraineLi2001ApJ...551..807D} predictions for pericondensed PAHs (see Fig.~\ref{Fig:PAHratio_d1}).

\citet{Micelotta2010A&A...510A..36M} showed that PAHs with $N_{\rm C}<200$ experience severe structural modification in shocks of $75-100$ km s$^{-1}$ and are fully destroyed for velocities $\gtrsim$125 km s$^{-1}$. Although the impact of such processing on the resulting PAH band ratios remains uncertain, one plausible interpretation of the observed association between altered PAH ratios and strong H$_2$ emission is that shocks may leave the surviving PAHs with more open or irregular structures (i.e., catacondensed PAHs). 

\begin{figure}
   \centering
   \includegraphics[width=\columnwidth]{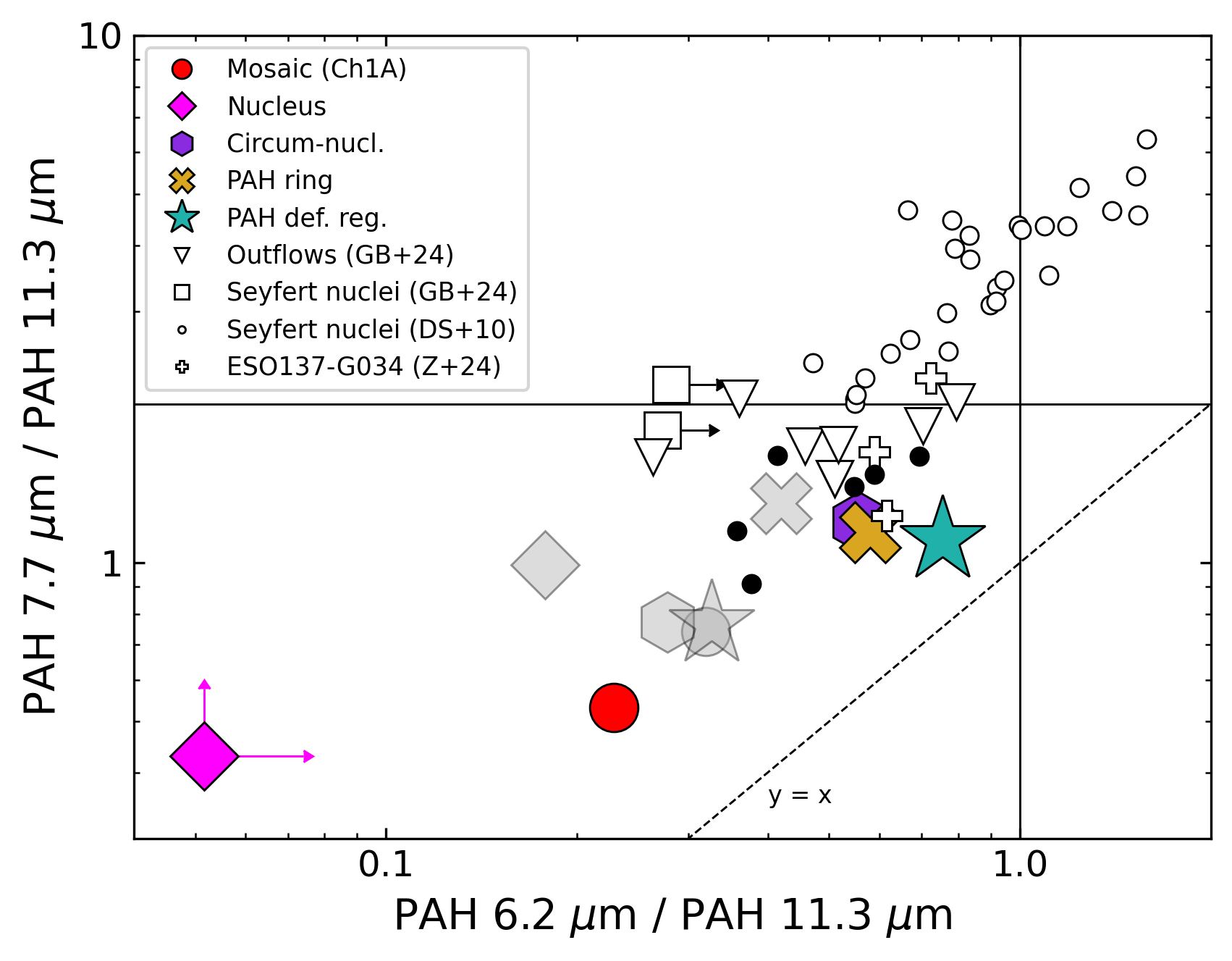}
      \caption{PAH 7.7/11.3~\upmicron versus PAH 6.2/11.3~\upmicron diagnostic plot. Symbols are the same as Fig. \ref{Fig:PAHratio_d1}. Uncertainties, between $10-40$\%, are omitted for clarity. The nuclear lower limits result from partial detection of the 6.2 and 7.7~\upmicron features (Appendix~\ref{App:nucleus}).
      }
    \label{Fig:PAHratio_d11.3}
\end{figure}

Notably, if collisional shocks are responsible for the observed behaviour, they appear to preferentially suppress the 6.2 and 7.7~\upmicron features (ionized PAHs), while the 11.3~\upmicron feature (neutral PAHs) remains the most intense. This is evident from the direct relation between 7.7/11.3 and 6.2/11.3 ratios in Fig. \ref{Fig:PAHratio_d11.3}, where Seyfert nuclear regions and outflows populate the bottom left corner at low 7.7/11.3 ($\lesssim2$) and 6.2/11.3 ($<1$), along with the five regions of Cen~A studied in this work. This is consistent with the recent results by \citet[][]{Garcia-Bernete2024A&A...691A.162G, ZhangLulu2022ApJ...939...22Z, ZhangLulu2024ApJ...975L...2Z}, who found neutral PAHs to be the most resilient in Seyfert nuclei with low/intermediate-ionization emission lines, typical of LINERs.
Overall, the 7.7/11.3 ratio varies more steeply than 6.2/11.3 (i.e., the relation is shifted from the bisector to higher 7.7/11.3 values; Fig. \ref{Fig:PAHratio_d11.3}), indicating stronger suppression of the 7.7~\upmicron feature. 

\subsection{PAH evolutionary pathways in the centre of Cen~A}
\begin{figure}
   \centering
   \includegraphics[width=\hsize]{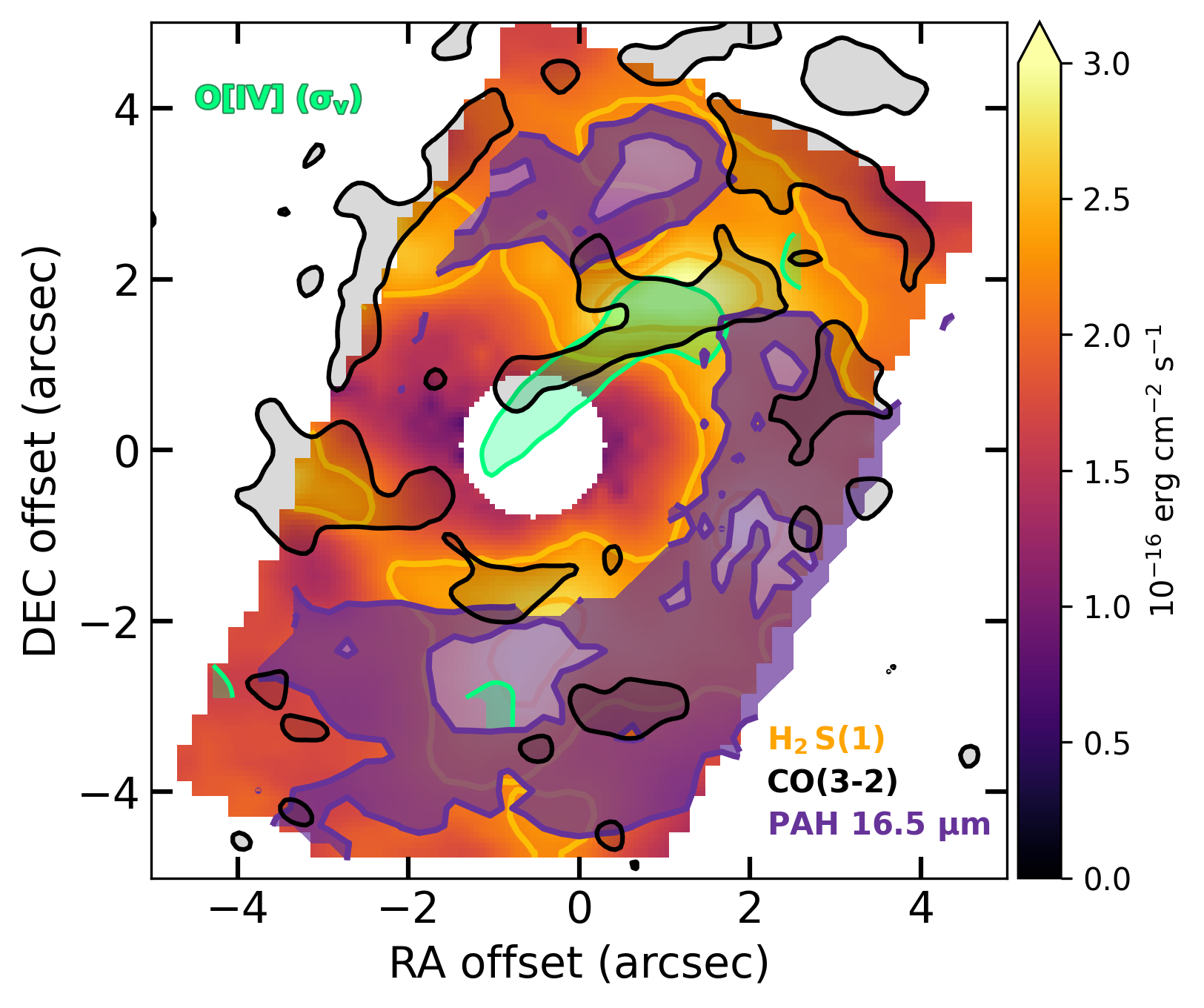}
      \caption{Relative spatial distribution within the central $\sim$200 pc of warm molecular hydrogen (\htwo~0-0~S(1); coloured map with yellow contours), CO(3--2) emission (black contours; \citealt{Espada2017ApJ...843..136E}), and PAH emission (PAH 16.5~\upmicron feature; purple-filled contours). In green we show the \oiv velocity dispersion from \citet{Alonso2025A&A...699A.334A} for values exceeding 130 km s$^{-1}$. All maps are resampled to match the CO(3--2) mom0 map ($\Delta\rm{xy}=0.35^{\prime\prime}$).
      }
    \label{Fig:multi-phase}
\end{figure}

The synergy between sub-arcsecond maps of PAHs, ionised gas, and cold and warm molecular gas is key to probing the spatially resolved ISM physics and the impact of AGN feedback.
In Fig. \ref{Fig:multi-phase} we present an overview of the multi-phase ISM in the ICND of Cen~A. The \htwo~0-0~S(1) mom0 map \citep[see also][]{Evangelista2026} is overlaid with the CO(3--2) line emission by \citet{Espada2017ApJ...843..136E}, the PAH 16.5~\upmicron feature emission, and the velocity dispersion map of \oiv\ by \citet{Alonso2025A&A...699A.334A}. 

The central area of the mosaic, from where the jet is launched, shows a remarkable deficiency in CO \citep[for $r\le20-30$~pc;][]{Espada2017ApJ...843..136E}\footnote{Similar evidence is found by \citet{GarciaBurillo2024A&A...689A.347G} for a sample of 45 AGNs.}, \htwo~0-0~S(1) and PAHs ($r\le40$ pc). 
The latter are distributed on a ring-like structure, showing local regions of intensity enhancement. In the direction perpendicular to the jet, high velocity dispersion is observed in the ionized gas phase, which is consistent with the presence of an expanding jet-inflated bubble, as predicted by hydrodynamical simulations \citep[][]{Alonso2025A&A...699A.334A}. The bubble likely interacts with the circum-nuclear disc, potentially inducing gas inflows and/or outflows \citep[][]{Alonso2025A&A...699A.334A}.
A pronounced deficit of PAH emission is observed in the direction of the jet-inflated bubble (Fig. \ref{Fig:multi-phase}), as well as deviations from circular motion in the molecular gas and inflow streamers \citep[][]{Espada2017ApJ...843..136E, Alonso2025A&A...699A.334A, Evangelista2026}. The resulting shocks can strongly affect interstellar dust through grain-grain shattering and ion-grain sputtering \citep[e.g.][]{Jones1994AAS...185.2605J, Jones1994ApJ...433..797J, Jones1996ApJ...469..740J}. Indeed, this region (i.e. the PAH deficient region) shows the highest $L_{\rm H2}/L_{\rm PAH}$ ratio, the most reduced PAH EWs and the highest dehydrogenation. 
Although the gas within the MIRI-MRS mosaic is partly collisionally excited by shocks \citep[][]{Alonso2025A&A...699A.334A, Evangelista2026}, some of the prominent PAH emission in the ICND may arise from entrained clumps shielded by cold \htwo and not fully exposed to shocks or the hot post-shock gas \citep{Micelotta2010A&A...510A..36M}, which may explain the largest PAH EWs.
While the column density of warm \htwo \citep[$\sim 8\times10^{20}$~cm$^{-2}$;][]{Evangelista2026} is too low to afford significant protection, the cold \htwo column density reported by \citet{Israel2014A&A...562A..96I}, i.e. $N_{\rm H_2}^{\rm cold} = (1.7\pm0.1)\times10^{22}$~cm$^{-2}$, is sufficient to shield PAHs, although this value is averaged over the central $22^{\prime\prime}$ ($\sim 400$~pc in diameter).
Overall, PAH intensity ratios across the Cen~A MIRI-MRS mosaic point to a strongly reprocessed PAH population dominated by neutral species with solo hydrogen sites, potentially exhibiting open and irregular structures.

\section{Summary and conclusions}\label{sect:conclusions}

We presented a detailed analysis of the PAH emission in the inner $7^{\prime\prime}\times12^{\prime\prime}$ ($\sim100\times200$ pc$^{2}$) of Cen~A, observed with JWST/MIRI-MRS as part of the Cycle~1's GTO program ID 1269. The spatial resolution across the MRS channels ranges from $0.35^{\prime\prime}$ to $1^{\prime\prime}$ ($6-17$ pc).
Our study focuses on five regions of interest:
the full Ch~1A mosaic, the nuclear region, the circum-nuclear region (i.e. the mosaic excluding the nucleus), the PAH ring, and the PAH-deficient region. 
In the following we summarize our results.
\begin{itemize}
    \item The PAH emission is mostly distributed in a ring-like structure with localized clumps, located at a radius of $\sim40$ pc from the AGN. A distinct PAH-deficient area is detected toward the North-West, approximately perpendicular to the jet axis. 
    \item The 1D spectra extracted from the five regions reveal a strong AGN MIR continuum that complicates the spectral decomposition, particularly in the nucleus, where $\chi^2_{\rm red}=30$. Our analysis indicates that PAH emission is still present in the nuclear region, although it cannot be reliably isolated or quantified. The PAH ring and PAH-deficient regions are reproduced best by our fits, yielding $\chi^2_{\rm red}=2$ and $1.8$, respectively.
    \item PAH 11.3/7.7~\upmicron and 6.2/7.7~\upmicron intensity ratios we measured fall beyond model predictions by \citet{DraineLi2001ApJ...551..807D}, although the PAH ring and the circum-nuclear region are consistent with small, neutral PAHs by \citet{Rigopoulou2021MNRAS.504.5287R}. These extreme ratio can be naturally explained if the PAH population is dominated by more open and irregular species. Alternatively, they could reflect an imperfect subtraction of the underlying continuum beneath the PAH bands.
    \item The PAH 11.3/12.7~\upmicron intensity ratios for the five regions indicate a dominance of solo over duo or trio hydrogen sites, consistent with previous findings for other Seyfert galaxies. The most extreme case occurs in the PAH-deficient region, where we measured the lowest 11.3/12.7 ratio.
    \item We measured the lowest EWs in the three regions whose spectra have the highest MIR continuum contribution, i.e. the nucleus, the Ch~1A mosaic and the circum-nuclear region. We interpret this as evidence that the low PAH EWs in these regions are mainly driven by AGN continuum dilution. We find the largest EWs in the PAH ring, and a marked decrease in the PAH-deficient region.
    \item All the five regions show $L_{\rm H2}/L_{\rm PAH\,7.7}$ ratio exceeding the limit set by PDR/XDR models, with the most extreme value reported for the PAH-deficient region ($L_{\rm H2}/L_{\rm PAH\,7.7}=0.20$), suggesting that the molecular gas is at least partially collisional excited. Comparable results for the ionized gas further suggest that the ISM in the ICND of Cen~A is heavily affected by AGN-driven shocks. When exposed to shocks, PAHs may experience severe structural modification, supporting the hypothesis of small, neutral, catacondensed PAHs as the dominant PAH species in Cen~A MIRI-MRS mosaic. In the most extreme cases, shocks may cause erosion or destruction of PAHs, as we claim for the PAH-deficient region, which shows both the lowest hydrogenation state and reduced EWs relative to the PAH ring. Nonetheless, the PAH-deficient region overlaps with the zone of enhanced ionized-gas velocity dispersion \citep{Alonso2025A&A...699A.334A} and with inflowing warm and cold molecular streamers \citep[][]{Espada2017ApJ...843..136E, Evangelista2026}.
\end{itemize}
The synergy between sub-arcsecond-resolution maps of PAH emission, ionized gas, and cold and warm molecular gas obtained with ALMA and JWST/MIRI-MRS was crucial to get insights on the spatially-resolved ISM physics and the role of AGN feedback in the nuclear and circum-nuclear regions of Cen~A. Future observations with access to the 3.3~\upmicron PAH feature will be essential to independently test and refine our interpretation of the PAH properties.

On the theoretical side, further progress will rely on extending current models to encompass a broader range of PAH species, including catacondensed molecules, enabling a more comprehensive interpretation of PAH processing in AGN-dominated environments.

\begin{acknowledgements}
      LP acknowledges K. Matsumoto for valuable discussions. LP and MB acknowledge funding from the Belgian Science Policy Office (BELSPO) through the PRODEX project ``JWST/MIRI Science exploitation'' (C4000142239). AAH and LHM acknowledge support from grant PID2021-124665NB-I00 funded by MCIN/AEI/10.13039/501100011033 and by ERDF A way of making Europe. SGB acknowledges support from the Spanish grant PID2022-138560NB-I00, funded by MCIN/AEI/10.13039/501100011033/FEDER, EU.  
\end{acknowledgements}

\bibliographystyle{aa.bst}
\bibliography{bibliography.bib} 

@ARTICLE{Allamandola1985ApJ...290L..25A,
       author = {{Allamandola}, L.~J. and {Tielens}, A.~G.~G.~M. and {Barker}, J.~R.},
        title = "{Polycyclic aromatic hydrocarbons and the unidentified infrared emission bands: auto exhaust along the milky way.}",
      journal = {\apjl},
         year = 1985,
        month = mar,
       volume = {290},
        pages = {L25-L28},
          doi = {10.1086/184435},
       adsurl = {https://ui.adsabs.harvard.edu/abs/1985ApJ...290L..25A}
}

@ARTICLE{Allamandola1989ApJS...71..733A,
       author = {{Allamandola}, L.~J. and {Tielens}, A.~G.~G.~M. and {Barker}, J.~R.},
        title = "{Interstellar Polycyclic Aromatic Hydrocarbons: The Infrared Emission Bands, the Excitation/Emission Mechanism, and the Astrophysical Implications}",
      journal = {\apjs},
         year = 1989,
        month = dec,
       volume = {71},
        pages = {733},
          doi = {10.1086/191396},
       adsurl = {https://ui.adsabs.harvard.edu/abs/1989ApJS...71..733A}
}

@ARTICLE{Allamandola1999ApJ...511L.115A,
       author = {{Allamandola}, L.~J. and {Hudgins}, D.~M. and {Sandford}, S.~A.},
        title = "{Modeling the Unidentified Infrared Emission with Combinations of Polycyclic Aromatic Hydrocarbons}",
      journal = {\apjl},
         year = 1999,
        month = feb,
       volume = {511},
       number = {2},
        pages = {L115-L119},
          doi = {10.1086/311843},
       adsurl = {https://ui.adsabs.harvard.edu/abs/1999ApJ...511L.115A}
}

@ARTICLE{Alonso2014MNRAS.443.2766A,
       author = {{Alonso-Herrero}, A. and {Ramos Almeida}, C. and {Esquej}, P. and {Roche}, P.~F. and {Hern{\'a}n-Caballero}, A. and {H{\"o}nig}, S.~F. and {Gonz{\'a}lez-Mart{\'\i}n}, O. and {Aretxaga}, I. and {Mason}, R.~E. and {Packham}, C. and {Levenson}, N.~A. and {Rodr{\'\i}guez Espinosa}, J.~M. and {Siebenmorgen}, R. and {Pereira-Santaella}, M. and {D{\'\i}az-Santos}, T. and {Colina}, L. and {Alvarez}, C. and {Telesco}, C.~M.},
        title = "{Nuclear 11.3 {\ensuremath{\mu}}m PAH emission in local active galactic nuclei}",
      journal = {\mnras},
         year = 2014,
        month = sep,
       volume = {443},
       number = {3},
        pages = {2766-2782},
          doi = {10.1093/mnras/stu1293},
archivePrefix = {arXiv},
       eprint = {1407.1154},
 primaryClass = {astro-ph.GA},
       adsurl = {https://ui.adsabs.harvard.edu/abs/2014MNRAS.443.2766A}
}

@ARTICLE{AlonsoHerrero2024A&A...690A..95A,
       author = {{Alonso Herrero}, A. and {Hermosa Mu{\~n}oz}, L. and {Labiano}, A. and {Guillard}, P. and {Buiten}, V.~A. and {Dicken}, D. and {van der Werf}, P. and {{\'A}lvarez-M{\'a}rquez}, J. and {B{\"o}ker}, T. and {Colina}, L. and {Eckart}, A. and {Garc{\'\i}a-Mar{\'\i}n}, M. and {Jones}, O.~C. and {Pantoni}, L. and {P{\'e}rez-Gonz{\'a}lez}, P.~G. and {Rouan}, D. and {Ward}, M.~J. and {Baes}, M. and {{\"O}stlin}, G. and {Royer}, P. and {Wright}, G.~S. and {G{\"u}del}, M. and {Henning}, Th. and {Lagage}, P. -O. and {van Dishoeck}, E.~F.},
        title = "{MICONIC: JWST/MIRI MRS observations of the nuclear and circumnuclear regions of Mrk 231}",
      journal = {\aap},
         year = 2024,
        month = oct,
       volume = {690},
          eid = {A95},
        pages = {A95},
          doi = {10.1051/0004-6361/202450071},
archivePrefix = {arXiv},
       eprint = {2407.02180},
 primaryClass = {astro-ph.GA},
       adsurl = {https://ui.adsabs.harvard.edu/abs/2024A&A...690A..95A}
}

@ARTICLE{Alonso2025A&A...699A.334A,
       author = {{Alonso Herrero}, A. and {Hermosa Mu{\~n}oz}, L. and {Labiano}, A. and {Guillard}, P. and {Garc{\'\i}a-Mar{\'\i}n}, M. and {Dicken}, D. and {Garc{\'\i}a-Burillo}, S. and {Pantoni}, L. and {Buiten}, V. and {Colina}, L. and {B{\"o}ker}, T. and {Baes}, M. and {Eckart}, A. and {Evangelista}, L. and {{\"O}stlin}, G. and {Rouan}, D. and {van der Werf}, P. and {Walter}, F. and {Ward}, M.~J. and {Wright}, G. and {G{\"u}del}, M. and {Henning}, Th. and {Lagage}, P. -O.},
        title = "{MICONIC: JWST/MIRI MRS reveals a fast ionized gas outflow in the central region of Centaurus A}",
      journal = {\aap},
         year = 2025,
        month = jul,
       volume = {699},
          eid = {A334},
        pages = {A334},
          doi = {10.1051/0004-6361/202554823},
archivePrefix = {arXiv},
       eprint = {2506.15286},
 primaryClass = {astro-ph.GA},
       adsurl = {https://ui.adsabs.harvard.edu/abs/2025A&A...699A.334A}
}

@ARTICLE{Alvarez-Marquez2023A&A...672A.108A,
       author = {{{\'A}lvarez-M{\'a}rquez}, J. and {Labiano}, A. and {Guillard}, P. and {Dicken}, D. and {Argyriou}, I. and {Patapis}, P. and {Law}, D.~R. and {Kavanagh}, P.~J. and {Larson}, K.~L. and {Gasman}, D. and {Mueller}, M. and {Alberts}, S. and {Brandl}, B.~R. and {Colina}, L. and {Garc{\'\i}a-Mar{\'\i}n}, M. and {Jones}, O.~C. and {Noriega-Crespo}, A. and {Shivaei}, I. and {Temim}, T. and {Wright}, G.~S.},
        title = "{Nuclear high-ionisation outflow in the Compton-thick AGN NGC 6552 as seen by the JWST mid-infrared instrument}",
      journal = {\aap},
         year = 2023,
        month = apr,
       volume = {672},
          eid = {A108},
        pages = {A108},
          doi = {10.1051/0004-6361/202244880},
archivePrefix = {arXiv},
       eprint = {2209.01695},
 primaryClass = {astro-ph.GA},
       adsurl = {https://ui.adsabs.harvard.edu/abs/2023A&A...672A.108A}
}

@ARTICLE{Argyriou2023A&A...675A.111A,
       author = {{Argyriou}, Ioannis and {Glasse}, Alistair and {Law}, David R. and {Labiano}, Alvaro and {{\'A}lvarez-M{\'a}rquez}, Javier and {Patapis}, Polychronis and {Kavanagh}, Patrick J. and {Gasman}, Danny and {Mueller}, Michael and {Larson}, Kirsten and {Vandenbussche}, Bart and {Glauser}, Adrian M. and {Royer}, Pierre and {Dicken}, Daniel and {Harkett}, Jake and {Sargent}, Beth A. and {Engesser}, Michael and {Jones}, Olivia C. and {Kendrew}, Sarah and {Noriega-Crespo}, Alberto and {Brandl}, Bernhard and {Rieke}, George H. and {Wright}, Gillian S. and {Lee}, David and {Wells}, Martyn},
        title = "{JWST MIRI flight performance: The Medium-Resolution Spectrometer}",
      journal = {\aap},
         year = 2023,
        month = jul,
       volume = {675},
          eid = {A111},
        pages = {A111},
          doi = {10.1051/0004-6361/202346489},
archivePrefix = {arXiv},
       eprint = {2303.13469},
 primaryClass = {astro-ph.IM},
       adsurl = {https://ui.adsabs.harvard.edu/abs/2023A&A...675A.111A}
}

@ARTICLE{Armus2023ApJ...942L..37A,
       author = {{Armus}, L. and {Lai}, T. and {U}, V. and {Larson}, K.~L. and {Diaz-Santos}, T. and {Evans}, A.~S. and {Malkan}, M.~A. and {Rich}, J. and {Medling}, A.~M. and {Law}, D.~R. and {Inami}, H. and {Muller-Sanchez}, F. and {Charmandaris}, V. and {van der Werf}, P. and {Stierwalt}, S. and {Linden}, S. and {Privon}, G.~C. and {Barcos-Mu{\~n}oz}, L. and {Hayward}, C. and {Song}, Y. and {Appleton}, P. and {Aalto}, S. and {Bohn}, T. and {B{\"o}ker}, T. and {Brown}, M.~J.~I. and {Finnerty}, L. and {Howell}, J. and {Iwasawa}, K. and {Kemper}, F. and {Marshall}, J. and {Mazzarella}, J.~M. and {McKinney}, J. and {Murphy}, E.~J. and {Sanders}, D. and {Surace}, J.},
        title = "{GOALS-JWST: Mid-infrared Spectroscopy of the Nucleus of NGC 7469}",
      journal = {\apjl},
         year = 2023,
        month = jan,
       volume = {942},
       number = {2},
          eid = {L37},
        pages = {L37},
          doi = {10.3847/2041-8213/acac66},
archivePrefix = {arXiv},
       eprint = {2209.13125},
 primaryClass = {astro-ph.GA},
       adsurl = {https://ui.adsabs.harvard.edu/abs/2023ApJ...942L..37A}
}

@ARTICLE{BaerWay2024ApJ...964..172B,
       author = {{Baer-Way}, Raphael and {DeGraw}, Asia and {Zheng}, WeiKang and {Van Dyk}, Schuyler D. and {Filippenko}, Alexei V. and {Fox}, Ori D. and {Brink}, Thomas G. and {Kelly}, Patrick L. and {Smith}, Nathan and {Vasylyev}, Sergiy S. and {de Jaeger}, Thomas and {Zhang}, Keto and {Stegman}, Samantha and {Ross}, Timothy and {Yunus}, Sameen},
        title = "{A Snapshot Survey of Nearby Supernovae with the Hubble Space Telescope}",
      journal = {\apj},
         year = 2024,
        month = apr,
       volume = {964},
       number = {2},
          eid = {172},
        pages = {172},
          doi = {10.3847/1538-4357/ad2175},
archivePrefix = {arXiv},
       eprint = {2401.12185},
 primaryClass = {astro-ph.HE},
       adsurl = {https://ui.adsabs.harvard.edu/abs/2024ApJ...964..172B}
}

@ARTICLE{Bakes1994ApJ...427..822B,
       author = {{Bakes}, E.~L.~O. and {Tielens}, A.~G.~G.~M.},
        title = "{The Photoelectric Heating Mechanism for Very Small Graphitic Grains and Polycyclic Aromatic Hydrocarbons}",
      journal = {\apj},
         year = 1994,
        month = jun,
       volume = {427},
        pages = {822},
          doi = {10.1086/174188},
       adsurl = {https://ui.adsabs.harvard.edu/abs/1994ApJ...427..822B}
}

@ARTICLE{Beckmann2011A&A...531A..70B,
       author = {{Beckmann}, V. and {Jean}, P. and {Lubi{\'n}ski}, P. and {Soldi}, S. and {Terrier}, R.},
        title = "{The hard X-ray emission of Centaurus A}",
      journal = {\aap},
         year = 2011,
        month = jul,
       volume = {531},
          eid = {A70},
        pages = {A70},
          doi = {10.1051/0004-6361/201016020},
archivePrefix = {arXiv},
       eprint = {1104.4253},
 primaryClass = {astro-ph.CO},
       adsurl = {https://ui.adsabs.harvard.edu/abs/2011A&A...531A..70B}
}

@ARTICLE{Berne2022A&A...667A.159B,
       author = {{Bern{\'e}}, O. and {Foschino}, S. and {Jalabert}, F. and {Joblin}, C.},
        title = "{Contribution of polycyclic aromatic hydrocarbon ionization to neutral gas heating in galaxies: model versus observations}",
      journal = {\aap},
         year = 2022,
        month = nov,
       volume = {667},
          eid = {A159},
        pages = {A159},
          doi = {10.1051/0004-6361/202243171},
archivePrefix = {arXiv},
       eprint = {2208.08762},
 primaryClass = {astro-ph.GA},
       adsurl = {https://ui.adsabs.harvard.edu/abs/2022A&A...667A.159B}
}

@ARTICLE{Barrera2023MNRAS.524.3741B,
       author = {{Barrera}, Nicol{\'a}s F. and {Fuentealba}, Patricio and {Mu{\~n}oz}, Francisco and {G{\'o}mez}, Tatiana and {C{\'a}rdenas}, Carlos},
        title = "{Formation of H$_{2}$ on polycyclic aromatic hydrocarbons under conditions of the ISM: an ab initio molecular dynamics study}",
      journal = {\mnras},
         year = 2023,
        month = sep,
       volume = {524},
       number = {3},
        pages = {3741-3748},
          doi = {10.1093/mnras/stad2106},
archivePrefix = {arXiv},
       eprint = {2305.14206},
 primaryClass = {astro-ph.GA},
       adsurl = {https://ui.adsabs.harvard.edu/abs/2023MNRAS.524.3741B}
}

@ARTICLE{Bohn2024ApJ...977...36B,
       author = {{Bohn}, Thomas and {Inami}, Hanae and {Togi}, Aditya and {Armus}, Lee and {Lai}, Thomas S.-Y. and {Barcos-Munoz}, Loreto and {Song}, Yiqing and {Linden}, S.~T. and {Surace}, Jason and {Bianchin}, Marina and {U}, Vivian and {Evans}, Aaron S. and {B{\"o}ker}, Torsten and {Malkan}, Matthew A. and {Larson}, Kirsten L. and {Stierwalt}, Sabrina and {Buiten}, Victorine A. and {Charmandaris}, Vassilis and {Diaz-Santos}, Tanio and {Howell}, Justin H. and {Privon}, George C. and {Ricci}, Claudio and {van der Werf}, Paul P. and {Aalto}, Susanne and {Hayward}, Christopher C. and {Kader}, Justin A. and {Mazzarella}, Joseph M. and {Muller-Sanchez}, Francisco and {Sanders}, David B.},
        title = "{GOALS-JWST: The Warm Molecular Outflows of the Merging Starburst Galaxy NGC 3256}",
      journal = {\apj},
         year = 2024,
        month = dec,
       volume = {977},
       number = {1},
          eid = {36},
        pages = {36},
          doi = {10.3847/1538-4357/ad87d3},
archivePrefix = {arXiv},
       eprint = {2403.14751},
 primaryClass = {astro-ph.GA},
       adsurl = {https://ui.adsabs.harvard.edu/abs/2024ApJ...977...36B}
}

@ARTICLE{Borkar2021MNRAS.500.3536B,
       author = {{Borkar}, A. and {Adhikari}, T.~P. and {R{\'o}{\.z}a{\'n}ska}, A. and {Markowitz}, A.~G. and {Boorman}, P. and {Czerny}, B. and {Migliori}, G. and {De Marco}, B. and {Karas}, V.},
        title = "{The multiphase environment in the centre of Centaurus A}",
      journal = {\mnras},
         year = 2021,
        month = jan,
       volume = {500},
       number = {3},
        pages = {3536-3551},
          doi = {10.1093/mnras/staa3515},
archivePrefix = {arXiv},
       eprint = {2006.01099},
 primaryClass = {astro-ph.GA},
       adsurl = {https://ui.adsabs.harvard.edu/abs/2021MNRAS.500.3536B}
}

@ARTICLE{Buiten2025A&A...699A.312B,
       author = {{Buiten}, Victorine A. and {van der Werf}, Paul P. and {Viti}, Serena and {Dicken}, Daniel and {Alonso Herrero}, Almudena and {Wright}, Gillian S. and {Baes}, Maarten and {B{\"o}ker}, Torsten and {Brandl}, Bernhard R. and {Colina}, Luis and {Garc{\'\i}a Mar{\'\i}n}, Macarena and {Greve}, Thomas R. and {Guillard}, Pierre and {Jones}, Olivia C. and {Hermosa Mu{\~n}oz}, Laura and {Labiano}, {\'A}lvaro and {{\"O}stlin}, G{\"o}ran and {Pantoni}, Lara and {Walter}, Fabian and {Ward}, Martin J. and {Perna}, Michele and {van Dishoeck}, Ewine F. and {Henning}, Thomas and {G{\"u}del}, Manuel and {Ray}, Thomas P.},
        title = "{The rich JWST spectrum of the western nucleus of Arp 220: Shocked hot core chemistry dominates the inner disc}",
      journal = {\aap},
         year = 2025,
        month = jul,
       volume = {699},
          eid = {A312},
        pages = {A312},
          doi = {10.1051/0004-6361/202554141},
archivePrefix = {arXiv},
       eprint = {2502.10271},
 primaryClass = {astro-ph.GA},
       adsurl = {https://ui.adsabs.harvard.edu/abs/2025A&A...699A.312B}
}

@article{Bushouse2024zndo..12556702B,
       author = {{Bushouse}, Howard and {Eisenhamer}, Jonathan and {Dencheva}, Nadia and {Davies}, James and {Greenfield}, Perry and {Morrison}, Jane and {Hodge}, Phil and {Simon}, Bernie and {Grumm}, David and {Droettboom}, Michael and {Slavich}, Edward and {Sosey}, Megan and {Pauly}, Tyler and {Miller}, Todd and {Jedrzejewski}, Robert and {Hack}, Warren and {Davis}, David and {Crawford}, Steven and {Law}, David and {Gordon}, Karl and {Regan}, Michael and {Cara}, Mihai and {MacDonald}, Ken and {Bradley}, Larry and {Shanahan}, Clare and {Jamieson}, William and {Teodoro}, Mairan and {Williams}, Thomas and {Pena-Guerrero}, Maria},
        title = "{JWST Calibration Pipeline}",
         year = 2024,
        month = jun,
          eid = {10.5281/zenodo.12556702},
          doi = {10.5281/zenodo.12556702},
      version = {1.15.0},
    publisher = {Zenodo},
       adsurl = {https://ui.adsabs.harvard.edu/abs/2024zndo..12556702B}
}

@article{CAFE2025ascl.soft01001D,
       author = {{Diaz-Santos}, Tanio and {Lai}, Thomas S.-Y. and {Finnerty}, Luke and {Privon}, George and {Bonfini}, Paolo and {Larson}, Kirsten and {Marshall}, Jason and {Armus}, Lee and {Charmandaris}, Vassilis},
        title = "{CAFE: Continuum And Feature Extraction tool}",
 howpublished = {Astrophysics Source Code Library, record ascl:2501.001},
         year = 2025,
        month = jan,
          eid = {ascl:2501.001},
archivePrefix = {ascl},
       eprint = {2501.001},
       adsurl = {https://ui.adsabs.harvard.edu/abs/2025ascl.soft01001D}
}

@INPROCEEDINGS{Calzetti2011EAS....46..133C,
       author = {{Calzetti}, D.},
        title = "{Polycyclic Aromatic Hydrocarbons as Star Formation Rate Indicators}",
    booktitle = {EAS Publications Series},
         year = 2011,
       editor = {{Joblin}, C. and {Tielens}, A.~G.~G.~M.},
       series = {EAS Publications Series},
       volume = {46},
        month = mar,
        pages = {133-141},
          doi = {10.1051/eas/1146014},
archivePrefix = {arXiv},
       eprint = {1010.4996},
 primaryClass = {astro-ph.CO},
       adsurl = {https://ui.adsabs.harvard.edu/abs/2011EAS....46..133C}
}

@ARTICLE{Chastenet2023ApJ...944L..11C,
       author = {{Chastenet}, J{\'e}r{\'e}my and {Sutter}, Jessica and {Sandstrom}, Karin and {Belfiore}, Francesco and {Egorov}, Oleg V. and {Larson}, Kirsten L. and {Leroy}, Adam K. and {Liu}, Daizhong and {Rosolowsky}, Erik and {Thilker}, David A. and {Watkins}, Elizabeth J. and {Williams}, Thomas G. and {Barnes}, Ashley. T. and {Bigiel}, Frank and {Boquien}, M{\'e}d{\'e}ric and {Chevance}, M{\'e}lanie and {Chiang}, I-Da and {Dale}, Daniel A. and {Kruijssen}, J.~M. Diederik and {Emsellem}, Eric and {Grasha}, Kathryn and {Groves}, Brent and {Hassani}, Hamid and {Hughes}, Annie and {Kreckel}, Kathryn and {Meidt}, Sharon E. and {Rickards Vaught}, Ryan J. and {Sardone}, Amy and {Schinnerer}, Eva},
        title = "{PHANGS-JWST First Results: Variations in PAH Fraction as a Function of ISM Phase and Metallicity}",
      journal = {\apjl},
         year = 2023,
        month = feb,
       volume = {944},
       number = {2},
          eid = {L11},
        pages = {L11},
          doi = {10.3847/2041-8213/acadd7},
archivePrefix = {arXiv},
       eprint = {2301.00578},
 primaryClass = {astro-ph.GA},
       adsurl = {https://ui.adsabs.harvard.edu/abs/2023ApJ...944L..11C}
}

@ARTICLE{Chastenet2023ApJ...944L..12C,
       author = {{Chastenet}, J{\'e}r{\'e}my and {Sutter}, Jessica and {Sandstrom}, Karin and {Belfiore}, Francesco and {Egorov}, Oleg V. and {Larson}, Kirsten L. and {Leroy}, Adam K. and {Liu}, Daizhong and {Rosolowsky}, Erik and {Thilker}, David A. and {Watkins}, Elizabeth J. and {Williams}, Thomas G. and {Barnes}, Ashley. T. and {Bigiel}, F. and {Boquien}, M{\'e}d{\'e}ric and {Chevance}, M{\'e}lanie and {Dale}, Daniel A. and {Kruijssen}, J.~M. Diederik and {Emsellem}, Eric and {Grasha}, Kathryn and {Groves}, Brent and {Hassani}, Hamid and {Hughes}, Annie and {Kreckel}, Kathryn and {Meidt}, Sharon E. and {Pan}, Hsi-An and {Querejeta}, Miguel and {Schinnerer}, Eva and {Whitcomb}, Cory M.},
        title = "{PHANGS-JWST First Results: Measuring Polycyclic Aromatic Hydrocarbon Properties across the Multiphase Interstellar Medium}",
      journal = {\apjl},
         year = 2023,
        month = feb,
       volume = {944},
       number = {2},
          eid = {L12},
        pages = {L12},
          doi = {10.3847/2041-8213/acac94},
       adsurl = {https://ui.adsabs.harvard.edu/abs/2023ApJ...944L..12C}
}

@ARTICLE{Chastenet2026,
       author = {{Chastenet}, J. and {De Looze}, I. and {Gordon}, K.},
        title = "{Survival of very small carbonaceous dust grain in the inner-CGM of NGC 891 from JWST/MIRI MRS}",
      journal = {\aap},
         year = 2026,
        note = {submitted}
}

@ARTICLE{Chown2024A&A...685A..75C,
       author = {{Chown}, Ryan and {Sidhu}, Ameek and {Peeters}, Els and {Tielens}, Alexander G.~G.~M. and {Cami}, Jan and {Bern{\'e}}, Olivier and {Habart}, Emilie and {Alarc{\'o}n}, Felipe and {Canin}, Am{\'e}lie and {Schroetter}, Ilane and {Trahin}, Boris and {Van De Putte}, Dries and {Abergel}, Alain and {Bergin}, Edwin A. and {Bernard-Salas}, Jeronimo and {Boersma}, Christiaan and {Bron}, Emeric and {Cuadrado}, Sara and {Dartois}, Emmanuel and {Dicken}, Daniel and {El-Yajouri}, Meriem and {Fuente}, Asunci{\'o}n and {Goicoechea}, Javier R. and {Gordon}, Karl D. and {Issa}, Lina and {Joblin}, Christine and {Kannavou}, Olga and {Khan}, Baria and {Lacinbala}, Ozan and {Languignon}, David and {Le Gal}, Romane and {Maragkoudakis}, Alexandros and {Meshaka}, Raphael and {Okada}, Yoko and {Onaka}, Takashi and {Pasquini}, Sofia and {Pound}, Marc W. and {Robberto}, Massimo and {R{\"o}llig}, Markus and {Schefter}, Bethany and {Schirmer}, Thi{\'e}baut and {Vicente}, S{\'\i}lvia and {Wolfire}, Mark G. and {Zannese}, Marion and {Aleman}, Isabel and {Allamandola}, Louis and {Auchettl}, Rebecca and {Baratta}, Giuseppe Antonio and {Bejaoui}, Salma and {Bera}, Partha P. and {Black}, John H. and {Boulanger}, Fran{\c{c}}ois and {Bouwman}, Jordy and {Brandl}, Bernhard and {Brechignac}, Philippe and {Br{\"u}nken}, Sandra and {Buragohain}, Mridusmita and {Burkhardt}, Andrew and {Candian}, Alessandra and {Cazaux}, St{\'e}phanie and {Cernicharo}, Jose and {Chabot}, Marin and {Chakraborty}, Shubhadip and {Champion}, Jason and {Colgan}, Sean W.~J. and {Cooke}, Ilsa R. and {Coutens}, Audrey and {Cox}, Nick L.~J. and {Demyk}, Karine and {Meyer}, Jennifer Donovan and {Foschino}, Sacha and {Garc{\'\i}a-Lario}, Pedro and {Gavilan}, Lisseth and {Gerin}, Maryvonne and {Gottlieb}, Carl A. and {Guillard}, Pierre and {Gusdorf}, Antoine and {Hartigan}, Patrick and {He}, Jinhua and {Herbst}, Eric and {Hornekaer}, Liv and {J{\"a}ger}, Cornelia and {Janot-Pacheco}, Eduardo and {Kaufman}, Michael and {Kemper}, Francisca and {Kendrew}, Sarah and {Kirsanova}, Maria S. and {Klaassen}, Pamela and {Kwok}, Sun and {Labiano}, {\'A}lvaro and {Lai}, Thomas S.-Y. and {Lee}, Timothy J. and {Lefloch}, Bertrand and {Le Petit}, Franck and {Li}, Aigen and {Linz}, Hendrik and {Mackie}, Cameron J. and {Madden}, Suzanne C. and {Mascetti}, Jo{\"e}lle and {McGuire}, Brett A. and {Merino}, Pablo and {Micelotta}, Elisabetta R. and {Misselt}, Karl and {Morse}, Jon A. and {Mulas}, Giacomo and {Neelamkodan}, Naslim and {Ohsawa}, Ryou and {Omont}, Alain and {Paladini}, Roberta and {Palumbo}, Maria Elisabetta and {Pathak}, Amit and {Pendleton}, Yvonne J. and {Petrignani}, Annemieke and {Pino}, Thomas and {Puga}, Elena and {Rangwala}, Naseem and {Rapacioli}, Mathias and {Ricca}, Alessandra and {Roman-Duval}, Julia and {Roser}, Joseph and {Roueff}, Evelyne and {Rouill{\'e}}, Ga{\"e}l and {Salama}, Farid and {Sales}, Dinalva A. and {Sandstrom}, Karin and {Sarre}, Peter and {Sciamma-O'Brien}, Ella and {Sellgren}, Kris and {Shenoy}, Sachindev S. and {Teyssier}, David and {Thomas}, Richard D. and {Togi}, Aditya and {Verstraete}, Laurent and {Witt}, Adolf N. and {Wootten}, Alwyn and {Zettergren}, Henning and {Zhang}, Yong and {Zhang}, Ziwei E. and {Zhen}, Junfeng},
        title = "{PDRs4All. IV. An embarrassment of riches: Aromatic infrared bands in the Orion Bar}",
      journal = {\aap},
         year = 2024,
        month = may,
       volume = {685},
          eid = {A75},
        pages = {A75},
          doi = {10.1051/0004-6361/202346662},
archivePrefix = {arXiv},
       eprint = {2308.16733},
 primaryClass = {astro-ph.GA},
       adsurl = {https://ui.adsabs.harvard.edu/abs/2024A&A...685A..75C}
}

@ARTICLE{Clarke1992ApJ...395..444C,
       author = {{Clarke}, David A. and {Burns}, Jack O. and {Norman}, Michael L.},
        title = "{VLA Observations of the Inner Lobes of Centaurus A}",
      journal = {\apj},
         year = 1992,
        month = aug,
       volume = {395},
        pages = {444},
          doi = {10.1086/171663},
       adsurl = {https://ui.adsabs.harvard.edu/abs/1992ApJ...395..444C}
}

@ARTICLE{Conroy2009ApJ...699..486C,
       author = {{Conroy}, Charlie and {Gunn}, James E. and {White}, Martin},
        title = "{The Propagation of Uncertainties in Stellar Population Synthesis Modeling. I. The Relevance of Uncertain Aspects of Stellar Evolution and the Initial Mass Function to the Derived Physical Properties of Galaxies}",
      journal = {\apj},
         year = 2009,
        month = jul,
       volume = {699},
       number = {1},
        pages = {486-506},
          doi = {10.1088/0004-637X/699/1/486},
archivePrefix = {arXiv},
       eprint = {0809.4261},
 primaryClass = {astro-ph},
       adsurl = {https://ui.adsabs.harvard.edu/abs/2009ApJ...699..486C}
}

@article{Conroy2010ascl.soft10043C,
       author = {{Conroy}, Charlie and {Gunn}, James E.},
        title = "{FSPS: Flexible Stellar Population Synthesis}",
 howpublished = {Astrophysics Source Code Library, record ascl:1010.043},
         year = 2010,
        month = oct,
          eid = {ascl:1010.043},
archivePrefix = {ascl},
       eprint = {1010.043},
       adsurl = {https://ui.adsabs.harvard.edu/abs/2010ascl.soft10043C}
}

@ARTICLE{Dale2006ApJ...646..161D,
       author = {{Dale}, D.~A. and {Smith}, J.~D.~T. and {Armus}, L. and {Buckalew}, B.~A. and {Helou}, G. and {Kennicutt}, Jr., R.~C. and {Moustakas}, J. and {Roussel}, H. and {Sheth}, K. and {Bendo}, G.~J. and {Calzetti}, D. and {Draine}, B.~T. and {Engelbracht}, C.~W. and {Gordon}, K.~D. and {Hollenbach}, D.~J. and {Jarrett}, T.~H. and {Kewley}, L.~J. and {Leitherer}, C. and {Li}, A. and {Malhotra}, S. and {Murphy}, E.~J. and {Walter}, F.},
        title = "{Mid-Infrared Spectral Diagnostics of Nuclear and Extranuclear Regions in Nearby Galaxies}",
      journal = {\apj},
         year = 2006,
        month = jul,
       volume = {646},
       number = {1},
        pages = {161-173},
          doi = {10.1086/504835},
archivePrefix = {arXiv},
       eprint = {astro-ph/0604007},
 primaryClass = {astro-ph},
       adsurl = {https://ui.adsabs.harvard.edu/abs/2006ApJ...646..161D}
}

@ARTICLE{Desai2007ApJ...669..810D,
       author = {{Desai}, V. and {Armus}, L. and {Spoon}, H.~W.~W. and {Charmandaris}, V. and {Bernard-Salas}, J. and {Brandl}, B.~R. and {Farrah}, D. and {Soifer}, B.~T. and {Teplitz}, H.~I. and {Ogle}, P.~M. and {Devost}, D. and {Higdon}, S.~J.~U. and {Marshall}, J.~A. and {Houck}, J.~R.},
        title = "{PAH Emission from Ultraluminous Infrared Galaxies}",
      journal = {\apj},
         year = 2007,
        month = nov,
       volume = {669},
       number = {2},
        pages = {810-820},
          doi = {10.1086/522104},
archivePrefix = {arXiv},
       eprint = {0707.4190},
 primaryClass = {astro-ph},
       adsurl = {https://ui.adsabs.harvard.edu/abs/2007ApJ...669..810D}
}

@ARTICLE{Diamond-Stanic2010ApJ...724..140D,
       author = {{Diamond-Stanic}, Aleksandar M. and {Rieke}, George H.},
        title = "{The Effect of Active Galactic Nuclei on the Mid-infrared Aromatic Features}",
      journal = {\apj},
         year = 2010,
        month = nov,
       volume = {724},
       number = {1},
        pages = {140-153},
          doi = {10.1088/0004-637X/724/1/140},
archivePrefix = {arXiv},
       eprint = {1009.2752},
 primaryClass = {astro-ph.CO},
       adsurl = {https://ui.adsabs.harvard.edu/abs/2010ApJ...724..140D}
}

@ARTICLE{Donnan2023A&A...669A..87D,
       author = {{Donnan}, F.~R. and {Rigopoulou}, D. and {Garc{\'\i}a-Bernete}, I. and {Pereira-Santaella}, M. and {Alonso-Herrero}, A. and {Roche}, P.~F. and {Aalto}, S. and {Hern{\'a}n-Caballero}, A. and {Spoon}, H.~W.~W.},
        title = "{A detailed look at the most obscured galactic nuclei in the mid-infrared}",
      journal = {\aap},
         year = 2023,
        month = jan,
       volume = {669},
          eid = {A87},
        pages = {A87},
          doi = {10.1051/0004-6361/202244937},
archivePrefix = {arXiv},
       eprint = {2211.09628},
 primaryClass = {astro-ph.GA},
       adsurl = {https://ui.adsabs.harvard.edu/abs/2023A&A...669A..87D}
}

@ARTICLE{Donnan2024MNRAS.529.1386D,
       author = {{Donnan}, F.~R. and {Garc{\'\i}a-Bernete}, I. and {Rigopoulou}, D. and {Pereira-Santaella}, M. and {Roche}, P.~F. and {Alonso-Herrero}, A.},
        title = "{Peeling back the layers of extinction of dusty galaxies in the era of JWST: modelling joint NIRSpec + MIRI spectra at rest-frame 1.5-28 {\ensuremath{\mu}}m}",
      journal = {\mnras},
         year = 2024,
        month = apr,
       volume = {529},
       number = {2},
        pages = {1386-1404},
          doi = {10.1093/mnras/stae612},
archivePrefix = {arXiv},
       eprint = {2402.17479},
 primaryClass = {astro-ph.GA},
       adsurl = {https://ui.adsabs.harvard.edu/abs/2024MNRAS.529.1386D}
}

@ARTICLE{DraineLi2001ApJ...551..807D,
       author = {{Draine}, B.~T. and {Li}, Aigen},
        title = "{Infrared Emission from Interstellar Dust. I. Stochastic Heating of Small Grains}",
      journal = {\apj},
         year = 2001,
        month = apr,
       volume = {551},
       number = {2},
        pages = {807-824},
          doi = {10.1086/320227},
archivePrefix = {arXiv},
       eprint = {astro-ph/0011318},
 primaryClass = {astro-ph},
       adsurl = {https://ui.adsabs.harvard.edu/abs/2001ApJ...551..807D}
}

@ARTICLE{Draine2007ApJ...657..810D,
       author = {{Draine}, B.~T. and {Li}, Aigen},
        title = "{Infrared Emission from Interstellar Dust. IV. The Silicate-Graphite-PAH Model in the Post-Spitzer Era}",
      journal = {\apj},
         year = 2007,
        month = mar,
       volume = {657},
       number = {2},
        pages = {810-837},
          doi = {10.1086/511055},
archivePrefix = {arXiv},
       eprint = {astro-ph/0608003},
 primaryClass = {astro-ph},
       adsurl = {https://ui.adsabs.harvard.edu/abs/2007ApJ...657..810D}
}

@ARTICLE{Draine2021ApJ...917....3D,
       author = {{Draine}, B.~T. and {Li}, Aigen and {Hensley}, Brandon S. and {Hunt}, L.~K. and {Sandstrom}, K. and {Smith}, J.-D.~T.},
        title = "{Excitation of Polycyclic Aromatic Hydrocarbon Emission: Dependence on Size Distribution, Ionization, and Starlight Spectrum and Intensity}",
      journal = {\apj},
         year = 2021,
        month = aug,
       volume = {917},
       number = {1},
          eid = {3},
        pages = {3},
          doi = {10.3847/1538-4357/abff51},
archivePrefix = {arXiv},
       eprint = {2011.07046},
 primaryClass = {astro-ph.GA},
       adsurl = {https://ui.adsabs.harvard.edu/abs/2021ApJ...917....3D}
}

@ARTICLE{Dumont2025A&A...703A..54D,
       author = {{Dumont}, Antoine and {Neumayer}, Nadine and {Seth}, Anil C. and {B{\"o}ker}, Torsten and {Eracleous}, Michael and {Goold}, Kameron and {Greene}, Jenny E. and {G{\"u}ltekin}, Kayhan and {Ho}, Luis C. and {Walsh}, Jonelle L. and {L{\"u}tzgendorf}, Nora},
        title = "{WIggle Corrector Kit for NIRSpEc Data: WICKED}",
      journal = {\aap},
         year = 2025,
        month = nov,
       volume = {703},
          eid = {A54},
        pages = {A54},
          doi = {10.1051/0004-6361/202554494},
archivePrefix = {arXiv},
       eprint = {2503.09697},
 primaryClass = {astro-ph.IM},
       adsurl = {https://ui.adsabs.harvard.edu/abs/2025A&A...703A..54D}
}

@ARTICLE{Espada2009ApJ...695..116E,
       author = {{Espada}, D. and {Matsushita}, S. and {Peck}, A. and {Henkel}, C. and {Iono}, D. and {Israel}, F.~P. and {Muller}, S. and {Petitpas}, G. and {Pihlstr{\"o}m}, Y. and {Taylor}, G.~B. and {Dinh-V-Trung}},
        title = "{Disentangling the Circumnuclear Environs of Centaurus A. I. High-Resolution Molecular Gas Imaging}",
      journal = {\apj},
         year = 2009,
        month = apr,
       volume = {695},
       number = {1},
        pages = {116-134},
          doi = {10.1088/0004-637X/695/1/116},
archivePrefix = {arXiv},
       eprint = {0901.1656},
 primaryClass = {astro-ph.GA},
       adsurl = {https://ui.adsabs.harvard.edu/abs/2009ApJ...695..116E}
}

@ARTICLE{Espada2010ApJ...720..666E,
       author = {{Espada}, D. and {Peck}, A.~B. and {Matsushita}, S. and {Sakamoto}, K. and {Henkel}, C. and {Iono}, D. and {Israel}, F.~P. and {Muller}, S. and {Petitpas}, G. and {Pihlstr{\"o}m}, Y. and {Taylor}, G.~B. and {Trung}, D.~V.},
        title = "{Disentangling the Circumnuclear Environs of Centaurus A. II. On the Nature of the Broad Absorption Line}",
      journal = {\apj},
         year = 2010,
        month = sep,
       volume = {720},
       number = {1},
        pages = {666-678},
          doi = {10.1088/0004-637X/720/1/666},
archivePrefix = {arXiv},
       eprint = {1007.2061},
 primaryClass = {astro-ph.CO},
       adsurl = {https://ui.adsabs.harvard.edu/abs/2010ApJ...720..666E}
}

@ARTICLE{Espada2017ApJ...843..136E,
       author = {{Espada}, D. and {Matsushita}, S. and {Miura}, R.~E. and {Israel}, F.~P. and {Neumayer}, N. and {Martin}, S. and {Henkel}, C. and {Izumi}, T. and {Iono}, D. and {Aalto}, S. and {Ott}, J. and {Peck}, A.~B. and {Quillen}, A.~C. and {Kohno}, K.},
        title = "{Disentangling the Circumnuclear Environs of Centaurus A. III. An Inner Molecular Ring, Nuclear Shocks, and the CO to Warm H$_{2}$ Interface}",
      journal = {\apj},
         year = 2017,
        month = jul,
       volume = {843},
       number = {2},
          eid = {136},
        pages = {136},
          doi = {10.3847/1538-4357/aa78a9},
archivePrefix = {arXiv},
       eprint = {1706.05762},
 primaryClass = {astro-ph.GA},
       adsurl = {https://ui.adsabs.harvard.edu/abs/2017ApJ...843..136E}
}

@ARTICLE{Evangelista2026,
       author = {{Evangelista}, L. and {Guillard}, P. and {Martin}, J.},
        title = "{MICONIC: The multiphase circumnuclear region of Centaurus A as seen with JWST/MIRI MRS observations I. Spectral inventory and properties of the warm molecular disk}",
      journal = {\aap},
         year = 2026,
        note = {submitted}
}

@ARTICLE{Farrah2007ApJ...667..149F,
       author = {{Farrah}, D. and {Bernard-Salas}, J. and {Spoon}, H.~W.~W. and {Soifer}, B.~T. and {Armus}, L. and {Brandl}, B. and {Charmandaris}, V. and {Desai}, V. and {Higdon}, S. and {Devost}, D. and {Houck}, J.},
        title = "{High-Resolution Mid-Infrared Spectroscopy of Ultraluminous Infrared Galaxies}",
      journal = {\apj},
         year = 2007,
        month = sep,
       volume = {667},
       number = {1},
        pages = {149-169},
          doi = {10.1086/520834},
archivePrefix = {arXiv},
       eprint = {0706.0513},
 primaryClass = {astro-ph},
       adsurl = {https://ui.adsabs.harvard.edu/abs/2007ApJ...667..149F}
}

@ARTICLE{Feltre2016MNRAS.456.3354F,
       author = {{Feltre}, A. and {Charlot}, S. and {Gutkin}, J.},
        title = "{Nuclear activity versus star formation: emission-line diagnostics at ultraviolet and optical wavelengths}",
      journal = {\mnras},
         year = 2016,
        month = mar,
       volume = {456},
       number = {3},
        pages = {3354-3374},
          doi = {10.1093/mnras/stv2794},
archivePrefix = {arXiv},
       eprint = {1511.08217},
 primaryClass = {astro-ph.GA},
       adsurl = {https://ui.adsabs.harvard.edu/abs/2016MNRAS.456.3354F}
}

@ARTICLE{Feltre2023A&A...675A..74F,
       author = {{Feltre}, A. and {Gruppioni}, C. and {Marchetti}, L. and {Mahoro}, A. and {Salvestrini}, F. and {Mignoli}, M. and {Bisigello}, L. and {Calura}, F. and {Charlot}, S. and {Chevallard}, J. and {Romero-Colmenero}, E. and {Curtis-Lake}, E. and {Delvecchio}, I. and {Dors}, O.~L. and {Hirschmann}, M. and {Jarrett}, T. and {Marchesi}, S. and {Moloko}, M.~E. and {Plat}, A. and {Pozzi}, F. and {Sefako}, R. and {Traina}, A. and {Vaccari}, M. and {V{\"a}is{\"a}nen}, P. and {Vallini}, L. and {Vidal-Garc{\'\i}a}, A. and {Vignali}, C.},
        title = "{Optical and mid-infrared line emission in nearby Seyfert galaxies}",
      journal = {\aap},
         year = 2023,
        month = jul,
       volume = {675},
          eid = {A74},
        pages = {A74},
          doi = {10.1051/0004-6361/202245516},
archivePrefix = {arXiv},
       eprint = {2301.02252},
 primaryClass = {astro-ph.GA},
       adsurl = {https://ui.adsabs.harvard.edu/abs/2023A&A...675A..74F}
}

@ARTICLE{Galliano2008ApJ...679..310G,
       author = {{Galliano}, Fr{\'e}d{\'e}ric and {Madden}, Suzanne C. and {Tielens}, Alexander G.~G.~M. and {Peeters}, Els and {Jones}, Anthony P.},
        title = "{Variations of the Mid-IR Aromatic Features inside and among Galaxies}",
      journal = {\apj},
         year = 2008,
        month = may,
       volume = {679},
       number = {1},
        pages = {310-345},
          doi = {10.1086/587051},
archivePrefix = {arXiv},
       eprint = {0801.4955},
 primaryClass = {astro-ph},
       adsurl = {https://ui.adsabs.harvard.edu/abs/2008ApJ...679..310G}
}

@ARTICLE{Galliano2022HabT.........1G,
       author = {{Galliano}, Fr{\'e}d{\'e}ric},
        title = "{A Nearby Galaxy Perspective on Interstellar Dust Properties and their Evolution}",
      journal = {Habilitation Thesis},
         year = 2022,
        month = feb,
        pages = {1},
          doi = {10.48550/arXiv.2202.01868},
archivePrefix = {arXiv},
       eprint = {2202.01868},
 primaryClass = {astro-ph.GA},
       adsurl = {https://ui.adsabs.harvard.edu/abs/2022HabT.........1G}
}

@ARTICLE{Garcia-Bernete2015MNRAS.449.1309G,
       author = {{Garc{\'\i}a-Bernete}, I. and {Ramos Almeida}, C. and {Acosta-Pulido}, J.~A. and {Alonso-Herrero}, A. and {S{\'a}nchez-Portal}, M. and {Castillo}, M. and {Pereira-Santaella}, M. and {Esquej}, P. and {Gonz{\'a}lez-Mart{\'\i}n}, O. and {D{\'\i}az-Santos}, T. and {Roche}, P. and {Fisher}, S. and {Povi{\'c}}, M. and {P{\'e}rez Garc{\'\i}a}, A.~M. and {Valtchanov}, I. and {Packham}, C. and {Levenson}, N.~A.},
        title = "{The nuclear and extended infrared emission of the Seyfert galaxy NGC 2992 and the interacting system Arp 245}",
      journal = {\mnras},
         year = 2015,
        month = may,
       volume = {449},
       number = {2},
        pages = {1309-1326},
          doi = {10.1093/mnras/stv338},
archivePrefix = {arXiv},
       eprint = {1502.04501},
 primaryClass = {astro-ph.GA},
       adsurl = {https://ui.adsabs.harvard.edu/abs/2015MNRAS.449.1309G}
}

@ARTICLE{Gallimore2010ApJS..187..172G,
       author = {{Gallimore}, J.~F. and {Yzaguirre}, A. and {Jakoboski}, J. and {Stevenosky}, M.~J. and {Axon}, D.~J. and {Baum}, S.~A. and {Buchanan}, C.~L. and {Elitzur}, M. and {Elvis}, M. and {O'Dea}, C.~P. and {Robinson}, A.},
        title = "{Infrared Spectral Energy Distributions of Seyfert Galaxies: Spitzer Space Telescope Observations of the 12 {\ensuremath{\mu}}m Sample of Active Galaxies}",
      journal = {\apjs},
         year = 2010,
        month = mar,
       volume = {187},
       number = {1},
        pages = {172-211},
          doi = {10.1088/0067-0049/187/1/172},
archivePrefix = {arXiv},
       eprint = {1001.4974},
 primaryClass = {astro-ph.CO},
       adsurl = {https://ui.adsabs.harvard.edu/abs/2010ApJS..187..172G}
}

@ARTICLE{Garcia-Bernete2022MNRAS.509.4256G,
       author = {{Garc{\'\i}a-Bernete}, I. and {Rigopoulou}, D. and {Alonso-Herrero}, A. and {Pereira-Santaella}, M. and {Roche}, P.~F. and {Kerkeni}, B.},
        title = "{Polycyclic aromatic hydrocarbons in Seyfert and star-forming galaxies}",
      journal = {\mnras},
         year = 2022,
        month = jan,
       volume = {509},
       number = {3},
        pages = {4256-4275},
          doi = {10.1093/mnras/stab3127},
archivePrefix = {arXiv},
       eprint = {2011.10882},
 primaryClass = {astro-ph.GA},
       adsurl = {https://ui.adsabs.harvard.edu/abs/2022MNRAS.509.4256G}
}

@ARTICLE{Garcia-Bernete2022A&A...666L...5G,
       author = {{Garc{\'\i}a-Bernete}, I. and {Rigopoulou}, D. and {Alonso-Herrero}, A. and {Donnan}, F.~R. and {Roche}, P.~F. and {Pereira-Santaella}, M. and {Labiano}, A. and {Peralta de Arriba}, L. and {Izumi}, T. and {Ramos Almeida}, C. and {Shimizu}, T. and {H{\"o}nig}, S. and {Garc{\'\i}a-Burillo}, S. and {Rosario}, D.~J. and {Ward}, M.~J. and {Bellocchi}, E. and {Hicks}, E.~K.~S. and {Fuller}, L. and {Packham}, C.},
        title = "{A high angular resolution view of the PAH emission in Seyfert galaxies using JWST/MRS data}",
      journal = {\aap},
         year = 2022,
        month = oct,
       volume = {666},
          eid = {L5},
        pages = {L5},
          doi = {10.1051/0004-6361/202244806},
archivePrefix = {arXiv},
       eprint = {2208.11620},
 primaryClass = {astro-ph.GA},
       adsurl = {https://ui.adsabs.harvard.edu/abs/2022A&A...666L...5G}
}

@ARTICLE{Garcia-Bernete2024A&A...691A.162G,
       author = {{Garc{\'\i}a-Bernete}, I. and {Rigopoulou}, D. and {Donnan}, F.~R. and {Alonso-Herrero}, A. and {Pereira-Santaella}, M. and {Shimizu}, T. and {Davies}, R. and {Roche}, P.~F. and {Garc{\'\i}a-Burillo}, S. and {Labiano}, A. and {Hermosa Mu{\~n}oz}, L. and {Zhang}, L. and {Audibert}, A. and {Bellocchi}, E. and {Bunker}, A. and {Combes}, F. and {Delaney}, D. and {Esparza-Arredondo}, D. and {Gandhi}, P. and {Gonz{\'a}lez-Mart{\'\i}n}, O. and {H{\"o}nig}, S.~F. and {Imanishi}, M. and {Hicks}, E.~K.~S. and {Fuller}, L. and {Leist}, M. and {Levenson}, N.~A. and {Lopez-Rodriguez}, E. and {Packham}, C. and {Ramos Almeida}, C. and {Ricci}, C. and {Stalevski}, M. and {Villar Mart{\'\i}n}, M. and {Ward}, M.~J.},
        title = "{The Galaxy Activity, Torus, and Outflow Survey (GATOS): V. Unveiling PAH survival and resilience in the circumnuclear regions of AGNs with JWST}",
      journal = {\aap},
         year = 2024,
        month = nov,
       volume = {691},
          eid = {A162},
        pages = {A162},
          doi = {10.1051/0004-6361/202450086},
archivePrefix = {arXiv},
       eprint = {2409.05686},
 primaryClass = {astro-ph.GA},
       adsurl = {https://ui.adsabs.harvard.edu/abs/2024A&A...691A.162G}
}

@ARTICLE{GarciaBurillo2024A&A...689A.347G,
       author = {{Garc{\'\i}a-Burillo}, S. and {Hicks}, E.~K.~S. and {Alonso-Herrero}, A. and {Pereira-Santaella}, M. and {Usero}, A. and {Querejeta}, M. and {Gonz{\'a}lez-Mart{\'\i}n}, O. and {Delaney}, D. and {Ramos Almeida}, C. and {Combes}, F. and {Angl{\'e}s-Alc{\'a}zar}, D. and {Audibert}, A. and {Bellocchi}, E. and {Davies}, R.~I. and {Davis}, T.~A. and {Elford}, J.~S. and {Garc{\'\i}a-Bernete}, I. and {H{\"o}nig}, S. and {Labiano}, A. and {Leist}, M.~T. and {Levenson}, N.~A. and {L{\'o}pez-Rodr{\'\i}guez}, E. and {Mercedes-Feliz}, J. and {Packham}, C. and {Ricci}, C. and {Rosario}, D.~J. and {Shimizu}, T. and {Stalevski}, M. and {Zhang}, L.},
        title = "{Deciphering the imprint of active galactic nucleus feedback in Seyfert galaxies: Nuclear-scale molecular gas deficits}",
      journal = {\aap},
         year = 2024,
        month = sep,
       volume = {689},
          eid = {A347},
        pages = {A347},
          doi = {10.1051/0004-6361/202450268},
archivePrefix = {arXiv},
       eprint = {2406.11398},
 primaryClass = {astro-ph.GA},
       adsurl = {https://ui.adsabs.harvard.edu/abs/2024A&A...689A.347G}
}

@ARTICLE{Gasman2023A&A...673A.102G,
       author = {{Gasman}, Danny and {Argyriou}, Ioannis and {Sloan}, G.~C. and {Aringer}, Bernhard and {{\'A}lvarez-M{\'a}rquez}, Javier and {Fox}, Ori and {Glasse}, Alistair and {Glauser}, Adrian and {Jones}, Olivia C. and {Justtanont}, Kay and {Kavanagh}, Patrick J. and {Klaassen}, Pamela and {Labiano}, Alvaro and {Larson}, Kirsten and {Law}, David R. and {Mueller}, Michael and {Nayak}, Omnarayani and {Noriega-Crespo}, Alberto and {Patapis}, Polychronis and {Royer}, Pierre and {Vandenbussche}, Bart},
        title = "{JWST MIRI/MRS in-flight absolute flux calibration and tailored fringe correction for unresolved sources}",
      journal = {\aap},
         year = 2023,
        month = may,
       volume = {673},
          eid = {A102},
        pages = {A102},
          doi = {10.1051/0004-6361/202245633},
archivePrefix = {arXiv},
       eprint = {2212.03596},
 primaryClass = {astro-ph.IM},
       adsurl = {https://ui.adsabs.harvard.edu/abs/2023A&A...673A.102G}
}

@ARTICLE{Gillett1973ApJ...183...87G,
       author = {{Gillett}, F.~C. and {Forrest}, W.~J. and {Merrill}, K.~M.},
        title = "{8 - 13-micron spectra of NGC 7027, BD +30 3639, and NGC 6572.}",
      journal = {\apj},
         year = 1973,
        month = jul,
       volume = {183},
        pages = {87},
          doi = {10.1086/152211},
       adsurl = {https://ui.adsabs.harvard.edu/abs/1973ApJ...183...87G}
}

@ARTICLE{Guillard2010A&A...518A..59G,
       author = {{Guillard}, P. and {Boulanger}, F. and {Cluver}, M.~E. and {Appleton}, P.~N. and {Pineau Des For{\^e}ts}, G. and {Ogle}, P.},
        title = "{Observations and modeling of the dust emission from the H$_{2}$-bright galaxy-wide shock in Stephan's Quintet}",
      journal = {\aap},
         year = 2010,
        month = jul,
       volume = {518},
          eid = {A59},
        pages = {A59},
          doi = {10.1051/0004-6361/200913430},
archivePrefix = {arXiv},
       eprint = {1004.0677},
 primaryClass = {astro-ph.CO},
       adsurl = {https://ui.adsabs.harvard.edu/abs/2010A&A...518A..59G}
}

@ARTICLE{Guillard2012ApJ...747...95G,
       author = {{Guillard}, P. and {Ogle}, P.~M. and {Emonts}, B.~H.~C. and {Appleton}, P.~N. and {Morganti}, R. and {Tadhunter}, C. and {Oosterloo}, T. and {Evans}, D.~A. and {Evans}, A.~S.},
        title = "{Strong Molecular Hydrogen Emission and Kinematics of the Multiphase Gas in Radio Galaxies with Fast Jet-driven Outflows}",
      journal = {\apj},
         year = 2012,
        month = mar,
       volume = {747},
       number = {2},
          eid = {95},
        pages = {95},
          doi = {10.1088/0004-637X/747/2/95},
archivePrefix = {arXiv},
       eprint = {1201.1503},
 primaryClass = {astro-ph.CO},
       adsurl = {https://ui.adsabs.harvard.edu/abs/2012ApJ...747...95G}
}

@ARTICLE{Hansen2022CmChe...5...94H,
       author = {{Hansen}, Christopher S. and {Peeters}, Els and {Cami}, Jan and {Schmidt}, Timothy W.},
        title = "{Open questions on carbon-based molecules in space}",
      journal = {Communications Chemistry},
         year = 2022,
        month = aug,
       volume = {5},
       number = {1},
          eid = {94},
        pages = {94},
          doi = {10.1038/s42004-022-00714-3},
       adsurl = {https://ui.adsabs.harvard.edu/abs/2022CmChe...5...94H}
}

@ARTICLE{Hardcastle2003ApJ...593..169H,
       author = {{Hardcastle}, M.~J. and {Worrall}, D.~M. and {Kraft}, R.~P. and {Forman}, W.~R. and {Jones}, C. and {Murray}, S.~S.},
        title = "{Radio and X-Ray Observations of the Jet in Centaurus A}",
      journal = {\apj},
         year = 2003,
        month = aug,
       volume = {593},
       number = {1},
        pages = {169-183},
          doi = {10.1086/376519},
archivePrefix = {arXiv},
       eprint = {astro-ph/0304443},
 primaryClass = {astro-ph},
       adsurl = {https://ui.adsabs.harvard.edu/abs/2003ApJ...593..169H}
}

@ARTICLE{Hardcastle2007ApJ...670L..81H,
       author = {{Hardcastle}, M.~J. and {Kraft}, R.~P. and {Sivakoff}, G.~R. and {Goodger}, J.~L. and {Croston}, J.~H. and {Jord{\'a}n}, A. and {Evans}, D.~A. and {Worrall}, D.~M. and {Birkinshaw}, M. and {Raychaudhury}, S. and {Brassington}, N.~J. and {Forman}, W.~R. and {Harris}, W.~E. and {Jones}, C. and {Juett}, A.~M. and {Murray}, S.~S. and {Nulsen}, P.~E.~J. and {Sarazin}, C.~L. and {Woodley}, K.~A.},
        title = "{New Results on Particle Acceleration in the Centaurus A Jet and Counterjet from a Deep Chandra Observation}",
      journal = {\apjl},
         year = 2007,
        month = dec,
       volume = {670},
       number = {2},
        pages = {L81-L84},
          doi = {10.1086/524197},
archivePrefix = {arXiv},
       eprint = {0710.1277},
 primaryClass = {astro-ph},
       adsurl = {https://ui.adsabs.harvard.edu/abs/2007ApJ...670L..81H}
}

@ARTICLE{HermosaMunoz2025A&A...693A.321H,
       author = {{Hermosa Mu{\~n}oz}, L. and {Alonso-Herrero}, A. and {Labiano}, A. and {Guillard}, P. and {Pantoni}, L. and {Buiten}, V. and {Dicken}, D. and {Baes}, M. and {B{\"o}ker}, T. and {Colina}, L. and {Donnan}, F. and {Garc{\'\i}a-Bernete}, I. and {{\"O}stlin}, G. and {van der Werf}, P. and {Ward}, M.~J. and {Brandl}, B.~R. and {Walter}, F. and {Wright}, G. and {G{\"u}del}, M. and {Henning}, Th. and {Lagage}, P. -O. and {Ray}, T.},
        title = "{MICONIC: Dual active galactic nuclei, star formation, and ionised gas outflows in NGC 6240 seen with MIRI/JWST}",
      journal = {\aap},
         year = 2025,
        month = jan,
       volume = {693},
          eid = {A321},
        pages = {A321},
          doi = {10.1051/0004-6361/202452437},
archivePrefix = {arXiv},
       eprint = {2412.14707},
 primaryClass = {astro-ph.GA},
       adsurl = {https://ui.adsabs.harvard.edu/abs/2025A&A...693A.321H}
}

@article{Hony2001A&A...370.1030H,
  author = {Hony, S. and Van Kerckhoven, C. and Peeters, E. and Tielens, A. G. G. M. and Hudgins, D. M. and Allamandola, L. J.},
  title = {The CH out-of-plane bending modes of PAH molecules in astrophysical environments},
  journal = {Astronomy \& Astrophysics},
  year = {2001},
  volume = {370},
  pages = {1030--1043},
  doi = {10.1051/0004-6361:20010248}
}

@article{Hudgins1999ApJ...513L..69H,
  author = {Hudgins, D. M. and Allamandola, L. J.},
  title = {Infrared spectroscopy of matrix-isolated polycyclic aromatic hydrocarbons. 2. PAHs containing five or more rings},
  journal = {Astrophysical Journal Letters},
  year = {1999},
  volume = {513},
  pages = {L69--L73},
  doi = {10.1086/311912}
}

@ARTICLE{Inami2013ApJ...777..156I,
       author = {{Inami}, H. and {Armus}, L. and {Charmandaris}, V. and {Groves}, B. and {Kewley}, L. and {Petric}, A. and {Stierwalt}, S. and {D{\'\i}az-Santos}, T. and {Surace}, J. and {Rich}, J. and {Haan}, S. and {Howell}, J. and {Evans}, A.~S. and {Mazzarella}, J. and {Marshall}, J. and {Appleton}, P. and {Lord}, S. and {Spoon}, H. and {Frayer}, D. and {Matsuhara}, H. and {Veilleux}, S.},
        title = "{Mid-infrared Atomic Fine-structure Emission-line Spectra of Luminous Infrared Galaxies: Spitzer/IRS Spectra of the GOALS Sample}",
      journal = {\apj},
         year = 2013,
        month = nov,
       volume = {777},
       number = {2},
          eid = {156},
        pages = {156},
          doi = {10.1088/0004-637X/777/2/156},
archivePrefix = {arXiv},
       eprint = {1309.4788},
 primaryClass = {astro-ph.CO},
       adsurl = {https://ui.adsabs.harvard.edu/abs/2013ApJ...777..156I}
}

@ARTICLE{Israel1998A&ARv...8..237I,
       author = {{Israel}, F.~P.},
        title = "{Centaurus A - NGC 5128}",
      journal = {\aapr},
         year = 1998,
        month = jan,
       volume = {8},
       number = {4},
        pages = {237-278},
          doi = {10.1007/s001590050011},
archivePrefix = {arXiv},
       eprint = {astro-ph/9811051},
 primaryClass = {astro-ph},
       adsurl = {https://ui.adsabs.harvard.edu/abs/1998A&ARv...8..237I}
}

@ARTICLE{Israel2014A&A...562A..96I,
       author = {{Israel}, F.~P. and {G{\"u}sten}, R. and {Meijerink}, R. and {Loenen}, A.~F. and {Requena-Torres}, M.~A. and {Stutzki}, J. and {van der Werf}, P. and {Harris}, A. and {Kramer}, C. and {Martin-Pintado}, J. and {Weiss}, A.},
        title = "{The molecular circumnuclear disk (CND) in Centaurus A. A multi-transition CO and [CI] survey with Herschel, APEX, JCMT, and SEST}",
      journal = {\aap},
         year = 2014,
        month = feb,
       volume = {562},
          eid = {A96},
        pages = {A96},
          doi = {10.1051/0004-6361/201322780},
archivePrefix = {arXiv},
       eprint = {1402.0999},
 primaryClass = {astro-ph.GA},
       adsurl = {https://ui.adsabs.harvard.edu/abs/2014A&A...562A..96I}
}

@ARTICLE{Jensen2017MNRAS.470.3071J,
       author = {{Jensen}, J.~J. and {H{\"o}nig}, S.~F. and {Rakshit}, S. and {Alonso-Herrero}, A. and {Asmus}, D. and {Gandhi}, P. and {Kishimoto}, M. and {Smette}, A. and {Tristram}, K.~R.~W.},
        title = "{PAH features within few hundred parsecs of active galactic nuclei}",
      journal = {\mnras},
         year = 2017,
        month = sep,
       volume = {470},
       number = {3},
        pages = {3071-3094},
          doi = {10.1093/mnras/stx1447},
archivePrefix = {arXiv},
       eprint = {1706.04811},
 primaryClass = {astro-ph.GA},
       adsurl = {https://ui.adsabs.harvard.edu/abs/2017MNRAS.470.3071J}
}

@ARTICLE{Jones1994ApJ...433..797J,
       author = {{Jones}, A.~P. and {Tielens}, A.~G.~G.~M. and {Hollenbach}, D.~J. and {McKee}, C.~F.},
        title = "{Grain Destruction in Shocks in the Interstellar Medium}",
      journal = {\apj},
         year = 1994,
        month = oct,
       volume = {433},
        pages = {797},
          doi = {10.1086/174689},
       adsurl = {https://ui.adsabs.harvard.edu/abs/1994ApJ...433..797J}
}

@INPROCEEDINGS{Jones1994AAS...185.2605J,
       author = {{Jones}, A.~P. and {Tielens}, A.~G.~G.~M. and {Hollenbach}, D.~J. and {McKee}, C.~F.},
        title = "{The Shattering of Grains in Interstellar Shock Waves}",
    booktitle = {American Astronomical Society Meeting Abstracts},
         year = 1994,
       series = {American Astronomical Society Meeting Abstracts},
       volume = {185},
        month = dec,
          eid = {26.05},
        pages = {26.05},
       adsurl = {https://ui.adsabs.harvard.edu/abs/1994AAS...185.2605J}
}

@ARTICLE{Jones2013A&A...558A..62J,
       author = {{Jones}, A.~P. and {Fanciullo}, L. and {K{\"o}hler}, M. and {Verstraete}, L. and {Guillet}, V. and {Bocchio}, M. and {Ysard}, N.},
        title = "{The evolution of amorphous hydrocarbons in the ISM: dust modelling from a new vantage point}",
      journal = {\aap},
         year = 2013,
        month = oct,
       volume = {558},
          eid = {A62},
        pages = {A62},
          doi = {10.1051/0004-6361/201321686},
archivePrefix = {arXiv},
       eprint = {1411.6293},
 primaryClass = {astro-ph.GA},
       adsurl = {https://ui.adsabs.harvard.edu/abs/2013A&A...558A..62J}
}

@ARTICLE{Jones1996ApJ...469..740J,
       author = {{Jones}, A.~P. and {Tielens}, A.~G.~G.~M. and {Hollenbach}, D.~J.},
        title = "{Grain Shattering in Shocks: The Interstellar Grain Size Distribution}",
      journal = {\apj},
         year = 1996,
        month = oct,
       volume = {469},
        pages = {740},
          doi = {10.1086/177823},
       adsurl = {https://ui.adsabs.harvard.edu/abs/1996ApJ...469..740J}
}

@ARTICLE{Jones2015A&A...581A..92J,
       author = {{Jones}, A.~P. and {Habart}, E.},
        title = "{H$_{2}$ formation via the UV photo-processing of a-C:H nano-particles}",
      journal = {\aap},
         year = 2015,
        month = sep,
       volume = {581},
          eid = {A92},
        pages = {A92},
          doi = {10.1051/0004-6361/201526487},
archivePrefix = {arXiv},
       eprint = {1507.08534},
 primaryClass = {astro-ph.GA},
       adsurl = {https://ui.adsabs.harvard.edu/abs/2015A&A...581A..92J}
}

@ARTICLE{Jones2017A&A...602A..46J,
       author = {{Jones}, A.~P. and {K{\"o}hler}, M. and {Ysard}, N. and {Bocchio}, M. and {Verstraete}, L.},
        title = "{The global dust modelling framework THEMIS}",
      journal = {\aap},
         year = 2017,
        month = jun,
       volume = {602},
          eid = {A46},
        pages = {A46},
          doi = {10.1051/0004-6361/201630225},
archivePrefix = {arXiv},
       eprint = {1703.00775},
 primaryClass = {astro-ph.GA},
       adsurl = {https://ui.adsabs.harvard.edu/abs/2017A&A...602A..46J}
}

@ARTICLE{Kim2002ApJS..143..455K,
       author = {{Kim}, Hack-Sung and {Saykally}, Richard J.},
        title = "{Single-Photon Infrared Emission Spectroscopy of Gaseous Polycyclic Aromatic Hydrocarbon Cations: A Direct Test for Proposed Carriers of the Unidentified Infrared Emission Bands}",
      journal = {\apjs},
         year = 2002,
        month = dec,
       volume = {143},
       number = {2},
        pages = {455-467},
          doi = {10.1086/343078},
       adsurl = {https://ui.adsabs.harvard.edu/abs/2002ApJS..143..455K}
}

@ARTICLE{Kim2024ApJ...974..253K,
       author = {{Kim}, Helen Kyung and {Malkan}, Matthew A. and {Takagi}, Toshinobu and {Oi}, Nagisa and {Burgarella}, Denis and {Miyaji}, Takamitsu and {Shim}, Hyunjin and {Matsuhara}, Hideo and {Goto}, Tomotsugu and {Ohyama}, Yoichi and {Buat}, Veronique and {Kim}, Seong Jin},
        title = "{The Calibration of Polycyclic Aromatic Hydrocarbon Dust Emission as a Star Formation Rate Indicator in the AKARI NEP Survey}",
      journal = {\apj},
         year = 2024,
        month = oct,
       volume = {974},
       number = {2},
          eid = {253},
        pages = {253},
          doi = {10.3847/1538-4357/ad72e6},
archivePrefix = {arXiv},
       eprint = {2408.14005},
 primaryClass = {astro-ph.GA},
       adsurl = {https://ui.adsabs.harvard.edu/abs/2024ApJ...974..253K}
}

@ARTICLE{Labiano2021A&A...656A..57L,
       author = {{Labiano}, A. and {Argyriou}, I. and {{\'A}lvarez-M{\'a}rquez}, J. and {Glasse}, A. and {Glauser}, A. and {Patapis}, P. and {Law}, D. and {Brandl}, B.~R. and {Justtanont}, K. and {Lahuis}, F. and {Mart{\'\i}nez-Galarza}, J.~R. and {Mueller}, M. and {Noriega-Crespo}, A. and {Royer}, P. and {Shaughnessy}, B. and {Vandenbussche}, B.},
        title = "{Wavelength calibration and resolving power of the JWST MIRI Medium Resolution Spectrometer}",
      journal = {\aap},
         year = 2021,
        month = dec,
       volume = {656},
          eid = {A57},
        pages = {A57},
          doi = {10.1051/0004-6361/202140614},
archivePrefix = {arXiv},
       eprint = {2109.04254},
 primaryClass = {astro-ph.IM},
       adsurl = {https://ui.adsabs.harvard.edu/abs/2021A&A...656A..57L}
}

@ARTICLE{Lai2022ApJ...941L..36L,
       author = {{Lai}, Thomas S.-Y. and {Armus}, Lee and {U}, Vivian and {D{\'\i}az-Santos}, Tanio and {Larson}, Kirsten L. and {Evans}, Aaron and {Malkan}, Matthew A. and {Appleton}, Philip and {Rich}, Jeff and {M{\"u}ller-S{\'a}nchez}, Francisco and {Inami}, Hanae and {Bohn}, Thomas and {McKinney}, Jed and {Finnerty}, Luke and {Law}, David R. and {Linden}, Sean T. and {Medling}, Anne M. and {Privon}, George C. and {Song}, Yiqing and {Stierwalt}, Sabrina and {van der Werf}, Paul P. and {Barcos-Mu{\~n}oz}, Loreto and {Smith}, J.~D.~T. and {Togi}, Aditya and {Aalto}, Susanne and {B{\"o}ker}, Torsten and {Charmandaris}, Vassilis and {Howell}, Justin and {Iwasawa}, Kazushi and {Kemper}, Francisca and {Mazzarella}, Joseph M. and {Murphy}, Eric J. and {Brown}, Michael J.~I. and {Hayward}, Christopher C. and {Marshall}, Jason and {Sanders}, David and {Surace}, Jason},
        title = "{GOALS-JWST: Tracing AGN Feedback on the Star-forming Interstellar Medium in NGC 7469}",
      journal = {\apjl},
         year = 2022,
        month = dec,
       volume = {941},
       number = {2},
          eid = {L36},
        pages = {L36},
          doi = {10.3847/2041-8213/ac9ebf},
archivePrefix = {arXiv},
       eprint = {2209.06741},
 primaryClass = {astro-ph.GA},
       adsurl = {https://ui.adsabs.harvard.edu/abs/2022ApJ...941L..36L}
}

@ARTICLE{Lai2023ApJ...957L..26L,
       author = {{Lai}, Thomas S. -Y. and {Armus}, Lee and {Bianchin}, Marina and {D{\'\i}az-Santos}, Tanio and {Linden}, Sean T. and {Privon}, George C. and {Inami}, Hanae and {U}, Vivian and {Bohn}, Thomas and {Evans}, Aaron S. and {Larson}, Kirsten L. and {Hensley}, Brandon S. and {Smith}, J. -D.~T. and {Malkan}, Matthew A. and {Song}, Yiqing and {Stierwalt}, Sabrina and {van der Werf}, Paul P. and {McKinney}, Jed and {Aalto}, Susanne and {Buiten}, Victorine A. and {Rich}, Jeff and {Charmandaris}, Vassilis and {Appleton}, Philip and {Barcos-Mu{\~n}oz}, Loreto and {B{\"o}ker}, Torsten and {Finnerty}, Luke and {Kader}, Justin A. and {Law}, David R. and {Medling}, Anne M. and {Brown}, Michael J.~I. and {Hayward}, Christopher C. and {Howell}, Justin and {Iwasawa}, Kazushi and {Kemper}, Francisca and {Marshall}, Jason and {Mazzarella}, Joseph M. and {M{\"u}ller-S{\'a}nchez}, Francisco and {Murphy}, Eric J. and {Sanders}, David and {Surace}, Jason},
        title = "{GOALS-JWST: Small Neutral Grains and Enhanced 3.3 {\ensuremath{\mu}}m PAH Emission in the Seyfert Galaxy NGC 7469}",
      journal = {\apjl},
         year = 2023,
        month = nov,
       volume = {957},
       number = {2},
          eid = {L26},
        pages = {L26},
          doi = {10.3847/2041-8213/ad0387},
archivePrefix = {arXiv},
       eprint = {2307.15169},
 primaryClass = {astro-ph.GA},
       adsurl = {https://ui.adsabs.harvard.edu/abs/2023ApJ...957L..26L}
}

@ARTICLE{Law2023AJ....166...45L,
       author = {{Law}, David R. and {E. Morrison}, Jane and {Argyriou}, Ioannis and {Patapis}, Polychronis and {{\'A}lvarez-M{\'a}rquez}, J. and {Labiano}, Alvaro and {Vandenbussche}, Bart},
        title = "{A 3D Drizzle Algorithm for JWST and Practical Application to the MIRI Medium Resolution Spectrometer}",
      journal = {\aj},
         year = 2023,
        month = aug,
       volume = {166},
       number = {2},
          eid = {45},
        pages = {45},
          doi = {10.3847/1538-3881/acdddc},
archivePrefix = {arXiv},
       eprint = {2306.05520},
 primaryClass = {astro-ph.IM},
       adsurl = {https://ui.adsabs.harvard.edu/abs/2023AJ....166...45L}
}

@ARTICLE{Law2025AJ....169...67L,
       author = {{Law}, David R. and {Argyriou}, Ioannis and {Gordon}, Karl D. and {Sloan}, G.~C. and {Gasman}, Danny and {Glasse}, Alistair and {Larson}, Kirsten and {Fletcher}, Leigh N. and {Labiano}, Alvaro and {Noriega-Crespo}, Alberto},
        title = "{The James Webb Space Telescope Absolute Flux Calibration. III. Mid-infrared Instrument Medium Resolution Integral Field Unit Spectrometer}",
      journal = {\aj},
         year = 2025,
        month = feb,
       volume = {169},
       number = {2},
          eid = {67},
        pages = {67},
          doi = {10.3847/1538-3881/ad9685},
archivePrefix = {arXiv},
       eprint = {2409.15435},
 primaryClass = {astro-ph.IM},
       adsurl = {https://ui.adsabs.harvard.edu/abs/2025AJ....169...67L}
}

@ARTICLE{LegerPuget1984A&A...137L...5L,
       author = {{Leger}, A. and {Puget}, J.~L.},
        title = "{Identification of the Unidentified Infrared Emission Features of Interstellar Dust}",
      journal = {\aap},
         year = 1984,
        month = aug,
       volume = {137},
        pages = {L5-L8},
       adsurl = {https://ui.adsabs.harvard.edu/abs/1984A&A...137L...5L}
}

@ARTICLE{Li2001ApJ...554..778L,
       author = {{Li}, Aigen and {Draine}, B.~T.},
        title = "{Infrared Emission from Interstellar Dust. II. The Diffuse Interstellar Medium}",
      journal = {\apj},
         year = 2001,
        month = jun,
       volume = {554},
       number = {2},
        pages = {778-802},
          doi = {10.1086/323147},
archivePrefix = {arXiv},
       eprint = {astro-ph/0011319},
 primaryClass = {astro-ph},
       adsurl = {https://ui.adsabs.harvard.edu/abs/2001ApJ...554..778L}
}

@ARTICLE{Li2020NatAs...4..339L,
       author = {{Li}, Aigen},
        title = "{Spitzer's perspective of polycyclic aromatic hydrocarbons in galaxies}",
      journal = {Nature Astronomy},
         year = 2020,
        month = mar,
       volume = {4},
        pages = {339-351},
          doi = {10.1038/s41550-020-1051-1},
archivePrefix = {arXiv},
       eprint = {2003.10489},
 primaryClass = {astro-ph.GA},
       adsurl = {https://ui.adsabs.harvard.edu/abs/2020NatAs...4..339L}
}

@ARTICLE{Maragkoudakis2025ApJ...979...90M,
       author = {{Maragkoudakis}, A. and {Boersma}, C. and {Temi}, P. and {Bregman}, J.~D. and {Allamandola}, L.~J. and {Esposito}, V.~J. and {Ricca}, A. and {Peeters}, E.},
        title = "{A Sensitivity Analysis of the Modeling of Polycyclic Aromatic Hydrocarbon Emission in Galaxies}",
      journal = {\apj},
         year = 2025,
        month = jan,
       volume = {979},
       number = {1},
          eid = {90},
        pages = {90},
          doi = {10.3847/1538-4357/ad9918},
archivePrefix = {arXiv},
       eprint = {2412.01875},
 primaryClass = {astro-ph.GA},
       adsurl = {https://ui.adsabs.harvard.edu/abs/2025ApJ...979...90M}
}

@ARTICLE{Marconi2000ApJ...528..276M,
       author = {{Marconi}, Alessandro and {Schreier}, Ethan J. and {Koekemoer}, Anton and {Capetti}, Alessandro and {Axon}, David and {Macchetto}, Duccio and {Caon}, Nicola},
        title = "{Unveiling the Active Nucleus of Centaurus A}",
      journal = {\apj},
         year = 2000,
        month = jan,
       volume = {528},
       number = {1},
        pages = {276-291},
          doi = {10.1086/308168},
archivePrefix = {arXiv},
       eprint = {astro-ph/9907378},
 primaryClass = {astro-ph},
       adsurl = {https://ui.adsabs.harvard.edu/abs/2000ApJ...528..276M}
}

@ARTICLE{Marshall2007ApJ...670..129M,
       author = {{Marshall}, J.~A. and {Herter}, T.~L. and {Armus}, L. and {Charmandaris}, V. and {Spoon}, H.~W.~W. and {Bernard-Salas}, J. and {Houck}, J.~R.},
        title = "{Decomposing Dusty Galaxies. I. Multicomponent Spectral Energy Distribution Fitting}",
      journal = {\apj},
         year = 2007,
        month = nov,
       volume = {670},
       number = {1},
        pages = {129-155},
          doi = {10.1086/521588},
archivePrefix = {arXiv},
       eprint = {0707.2962},
 primaryClass = {astro-ph},
       adsurl = {https://ui.adsabs.harvard.edu/abs/2007ApJ...670..129M}
}

@ARTICLE{Mathis1983A&A...128..212M,
       author = {{Mathis}, J.~S. and {Mezger}, P.~G. and {Panagia}, N.},
        title = "{Interstellar radiation field and dust temperatures in the diffuse interstellar medium and in giant molecular clouds}",
      journal = {\aap},
         year = 1983,
        month = nov,
       volume = {128},
        pages = {212-229},
       adsurl = {https://ui.adsabs.harvard.edu/abs/1983A&A...128..212M}
}

@ARTICLE{McCoy2017ApJ...851...76M,
       author = {{McCoy}, Mark and {Ott}, J{\"u}rgen and {Meier}, David S. and {Muller}, S{\'e}bastien and {Espada}, Daniel and {Mart{\'\i}n}, Sergio and {Israel}, Frank P. and {Henkel}, Christian and {Impellizzeri}, Violette and {Aalto}, Susanne and {Edwards}, Philip G. and {Brunthaler}, Andreas and {Neumayer}, Nadine and {Peck}, Alison B. and {van der Werf}, Paul and {Feain}, Ilana},
        title = "{ALMA Observations of the Physical and Chemical Conditions in Centaurus A}",
      journal = {\apj},
         year = 2017,
        month = dec,
       volume = {851},
       number = {2},
          eid = {76},
        pages = {76},
          doi = {10.3847/1538-4357/aa99d6},
archivePrefix = {arXiv},
       eprint = {1712.00360},
 primaryClass = {astro-ph.GA},
       adsurl = {https://ui.adsabs.harvard.edu/abs/2017ApJ...851...76M}
}

@ARTICLE{Mennella2012ApJ...745L...2M,
       author = {{Mennella}, Vito and {Hornek{\ae}r}, Liv and {Thrower}, John and {Accolla}, Mario},
        title = "{The Catalytic Role of Coronene for Molecular Hydrogen Formation}",
      journal = {\apjl},
         year = 2012,
        month = jan,
       volume = {745},
       number = {1},
          eid = {L2},
        pages = {L2},
          doi = {10.1088/2041-8205/745/1/L2},
       adsurl = {https://ui.adsabs.harvard.edu/abs/2012ApJ...745L...2M}
}

@ARTICLE{Micelotta2010A&A...510A..36M,
       author = {{Micelotta}, E.~R. and {Jones}, A.~P. and {Tielens}, A.~G.~G.~M.},
        title = "{Polycyclic aromatic hydrocarbon processing in interstellar shocks}",
      journal = {\aap},
         year = 2010,
        month = feb,
       volume = {510},
          eid = {A36},
        pages = {A36},
          doi = {10.1051/0004-6361/200911682},
archivePrefix = {arXiv},
       eprint = {0910.2461},
 primaryClass = {astro-ph.GA},
       adsurl = {https://ui.adsabs.harvard.edu/abs/2010A&A...510A..36M}
}

@ARTICLE{Mordini2021A&A...653A..36M,
       author = {{Mordini}, Sabrina and {Spinoglio}, Luigi and {Fern{\'a}ndez-Ontiveros}, Juan Antonio},
        title = "{Calibration of mid- to far-infrared spectral lines in galaxies}",
      journal = {\aap},
         year = 2021,
        month = sep,
       volume = {653},
          eid = {A36},
        pages = {A36},
          doi = {10.1051/0004-6361/202140696},
archivePrefix = {arXiv},
       eprint = {2105.04584},
 primaryClass = {astro-ph.GA},
       adsurl = {https://ui.adsabs.harvard.edu/abs/2021A&A...653A..36M}
}

@ARTICLE{Morrison2023PASP..135g5004M,
       author = {{Morrison}, Jane E. and {Dicken}, Daniel and {Argyriou}, Ioannis and {Ressler}, Michael E. and {Gordon}, Karl D. and {Regan}, Michael W. and {Cracraft}, Misty and {Rieke}, George H. and {Engesser}, Michael and {Alberts}, Stacey and {Alvarez-Marquez}, Javier and {Colbert}, James W. and {Fox}, Ori D. and {Gasman}, Danny and {Law}, David R. and {Garcia Marin}, Macarena and {G{\'a}sp{\'a}r}, Andr{\'a}s and {Guillard}, Pierre and {Kendrew}, Sarah and {Labiano}, A. and {Laine}, Seppo and {Noriega-Crespo}, A. and {Shivaei}, Irene and {Sloan}, G.~C.},
        title = "{JWST MIRI Flight Performance: Detector Effects and Data Reduction Algorithms}",
      journal = {\pasp},
         year = 2023,
        month = jul,
       volume = {135},
       number = {1049},
          eid = {075004},
        pages = {075004},
          doi = {10.1088/1538-3873/acdea6},
archivePrefix = {arXiv},
       eprint = {2308.16327},
 primaryClass = {astro-ph.IM},
       adsurl = {https://ui.adsabs.harvard.edu/abs/2023PASP..135g5004M}
}

@ARTICLE{Neff2015ApJ...802...87N,
       author = {{Neff}, Susan G. and {Eilek}, Jean A. and {Owen}, Frazer N.},
        title = "{The Complex North Transition Region of Centaurus A: Radio Structure}",
      journal = {\apj},
         year = 2015,
        month = apr,
       volume = {802},
       number = {2},
          eid = {87},
        pages = {87},
          doi = {10.1088/0004-637X/802/2/87},
archivePrefix = {arXiv},
       eprint = {1502.05066},
 primaryClass = {astro-ph.GA},
       adsurl = {https://ui.adsabs.harvard.edu/abs/2015ApJ...802...87N}
}

@ARTICLE{Neumayer2007ApJ...671.1329N,
       author = {{Neumayer}, N. and {Cappellari}, M. and {Reunanen}, J. and {Rix}, H. -W. and {van der Werf}, P.~P. and {de Zeeuw}, P.~T. and {Davies}, R.~I.},
        title = "{The Central Parsecs of Centaurus A: High-excitation Gas, a Molecular Disk, and the Mass of the Black Hole}",
      journal = {\apj},
         year = 2007,
        month = dec,
       volume = {671},
       number = {2},
        pages = {1329-1344},
          doi = {10.1086/523039},
archivePrefix = {arXiv},
       eprint = {0709.1877},
 primaryClass = {astro-ph},
       adsurl = {https://ui.adsabs.harvard.edu/abs/2007ApJ...671.1329N}
}

@ARTICLE{ODowd2009ApJ...705..885O,
       author = {{O'Dowd}, Matthew J. and {Schiminovich}, David and {Johnson}, Benjamin D. and {Treyer}, Marie A. and {Martin}, Christopher D. and {Wyder}, Ted K. and {Charlot}, S. and {Heckman}, Timothy M. and {Martins}, Lucimara P. and {Seibert}, Mark and {van der Hulst}, J.~M.},
        title = "{Polycyclic Aromatic Hydrocarbons in Galaxies at z \raisebox{-0.5ex}\textasciitilde 0.1: The Effect of Star Formation and Active Galactic Nuclei}",
      journal = {\apj},
         year = 2009,
        month = nov,
       volume = {705},
       number = {1},
        pages = {885-898},
          doi = {10.1088/0004-637X/705/1/885},
archivePrefix = {arXiv},
       eprint = {0909.2279},
 primaryClass = {astro-ph.CO},
       adsurl = {https://ui.adsabs.harvard.edu/abs/2009ApJ...705..885O}
}

@ARTICLE{Ogle2007ApJ...668..699O,
       author = {{Ogle}, Patrick and {Antonucci}, Robert and {Appleton}, P.~N. and {Whysong}, David},
        title = "{Shocked Molecular Hydrogen in the 3C 326 Radio Galaxy System}",
      journal = {\apj},
         year = 2007,
        month = oct,
       volume = {668},
       number = {2},
        pages = {699-707},
          doi = {10.1086/521334},
archivePrefix = {arXiv},
       eprint = {0707.0896},
 primaryClass = {astro-ph},
       adsurl = {https://ui.adsabs.harvard.edu/abs/2007ApJ...668..699O}
}

@INPROCEEDINGS{Ogle_2007,
       author = {{Ogle}, Patrick M. and {Antonucci}, R. and {Appleton}, P.~N. and {Boulanger}, F. and {Evans}, A. and {Emonts}, B.~H.~C. and {Whysong}, D.},
        title = "{Molecular Hydrogen Emission Galaxies}",
    booktitle = {American Astronomical Society Meeting Abstracts},
         year = 2007,
       series = {American Astronomical Society Meeting Abstracts},
       volume = {211},
        month = dec,
          eid = {97.21},
        pages = {97.21},
       adsurl = {https://ui.adsabs.harvard.edu/abs/2007AAS...211.9721O}
}

@ARTICLE{Ogle_2010,
       author = {{Ogle}, Patrick and {Boulanger}, Francois and {Guillard}, Pierre and {Evans}, Daniel A. and {Antonucci}, Robert and {Appleton}, P.~N. and {Nesvadba}, Nicole and {Leipski}, Christian},
        title = "{Jet-powered Molecular Hydrogen Emission from Radio Galaxies}",
      journal = {\apj},
         year = 2010,
        month = dec,
       volume = {724},
       number = {2},
        pages = {1193-1217},
          doi = {10.1088/0004-637X/724/2/1193},
archivePrefix = {arXiv},
       eprint = {1009.4533},
 primaryClass = {astro-ph.CO},
       adsurl = {https://ui.adsabs.harvard.edu/abs/2010ApJ...724.1193O}
}

@ARTICLE{Patapis2024A&A...682A..53P,
       author = {{Patapis}, Polychronis and {Argyriou}, Ioannis and {Law}, David R. and {Glauser}, Adrian M. and {Glasse}, Alistair and {Labiano}, Alvaro and {{\'A}lvarez-M{\'a}rquez}, Javier and {Kavanagh}, Patrick J. and {Gasman}, Danny and {Mueller}, Michael and {Larson}, Kirsten and {Vandenbussche}, Bart and {Lee}, David and {Klaassen}, Pamela and {Guillard}, Pierre and {Wright}, Gillian S.},
        title = "{Geometric distortion and astrometric calibration of the JWST MIRI Medium Resolution Spectrometer}",
      journal = {\aap},
         year = 2024,
        month = feb,
       volume = {682},
          eid = {A53},
        pages = {A53},
          doi = {10.1051/0004-6361/202347339},
archivePrefix = {arXiv},
       eprint = {2307.01025},
 primaryClass = {astro-ph.IM},
       adsurl = {https://ui.adsabs.harvard.edu/abs/2024A&A...682A..53P}
}

@ARTICLE{Pathak2005CP....313..133P,
       author = {{Pathak}, Amit and {Rastogi}, Shantanu},
        title = "{Computational study of neutral and cationic catacondensed polycyclic aromatic hydrocarbons}",
      journal = {Chemical Physics},
         year = 2005,
        month = jun,
       volume = {313},
       number = {1-3},
        pages = {133-150},
          doi = {10.1016/j.chemphys.2005.01.007},
       adsurl = {https://ui.adsabs.harvard.edu/abs/2005CP....313..133P}
}

@article{Peetersdoi:10.1021/acs.accounts.0c00747,
author = {Peeters, Els and Mackie, Cameron and Candian, Alessandra and Tielens, Alexander G. G. M.},
title = {A Spectroscopic View on Cosmic PAH Emission},
journal = {Accounts of Chemical Research},
volume = {54},
number = {8},
pages = {1921-1933},
year = {2021},
doi = {10.1021/acs.accounts.0c00747},
    note ={PMID: 33780617},

URL = { https://doi.org/10.1021/acs.accounts.0c00747},
eprint = { https://doi.org/10.1021/acs.accounts.0c00747}
}

@ARTICLE{Peeters2002A&A...390.1089P,
       author = {{Peeters}, E. and {Hony}, S. and {Van Kerckhoven}, C. and {Tielens}, A.~G.~G.~M. and {Allamandola}, L.~J. and {Hudgins}, D.~M. and {Bauschlicher}, C.~W.},
        title = "{The rich 6 to 9 vec mu m spectrum of interstellar PAHs}",
      journal = {\aap},
         year = 2002,
        month = aug,
       volume = {390},
        pages = {1089-1113},
          doi = {10.1051/0004-6361:20020773},
archivePrefix = {arXiv},
       eprint = {astro-ph/0205400},
 primaryClass = {astro-ph},
       adsurl = {https://ui.adsabs.harvard.edu/abs/2002A&A...390.1089P}
}

@ARTICLE{Pereira-Santaella2010ApJ...725.2270P,
       author = {{Pereira-Santaella}, Miguel and {Diamond-Stanic}, Aleksandar M. and {Alonso-Herrero}, Almudena and {Rieke}, George H.},
        title = "{The Mid-infrared High-ionization Lines from Active Galactic Nuclei and Star-forming Galaxies}",
      journal = {\apj},
         year = 2010,
        month = dec,
       volume = {725},
       number = {2},
        pages = {2270-2280},
          doi = {10.1088/0004-637X/725/2/2270},
archivePrefix = {arXiv},
       eprint = {1010.5129},
 primaryClass = {astro-ph.CO},
       adsurl = {https://ui.adsabs.harvard.edu/abs/2010ApJ...725.2270P}
}

@article{Quillen2006_disk,
doi = {10.1086/504418},
url = {https://doi.org/10.1086/504418},
year = {2006},
month = {jul},
publisher = {},
volume = {645},
number = {2},
pages = {1092},
author = {Quillen, Alice C. and Brookes, Mairi H. and Keene, Jocelyn and Stern, Daniel and Lawrence, Charles R. and Werner, Michael W.},
title = {Spitzer Observations of the Dusty Warped Disk of Centaurus A},
journal = {The Astrophysical Journal}
}

@article{Quillen2006_shell,
doi = {10.1086/503670},
url = {https://doi.org/10.1086/503670},
year = {2006},
month = {mar},
publisher = {},
volume = {641},
number = {1},
pages = {L29},
author = {Quillen, Alice C. and Bland-Hawthorn, Joss and Brookes, Mairi H. and Werner, Michael W. and Smith, J. D. and Stern, Daniel and Keene, Jocelyn and Lawrence, Charles R.},
title = {Discovery of a 500 Parsec Shell in the Nucleus of Centaurus A},
journal = {The Astrophysical Journal}
}

@article{Quillen2008,
    author = {Quillen, Alice C. and Bland-Hawthorn, Joss and Green, Joel D. and Smith, J. D. and Prasad, D. Amelia and Alonso-Herrero, Almudena and Cleary, Kieran and Brookes, Mairi H. and Lawrence, Charles R.},
    title = {Spitzer Space Telescope Infrared Spectrograph mapping of the central kpc of Centaurus A},
    journal = {Monthly Notices of the Royal Astronomical Society},
    volume = {384},
    number = {4},
    pages = {1469-1482},
    year = {2008},
    month = {02},
    issn = {0035-8711},
    doi = {10.1111/j.1365-2966.2007.12768.x},
    url = {https://doi.org/10.1111/j.1365-2966.2007.12768.x},
    eprint = {https://academic.oup.com/mnras/article-pdf/384/4/1469/2832430/mnras0384-1469.pdf},
}

@ARTICLE{RamosAlmeida2023A&A...669L...5R,
       author = {{Ramos Almeida}, C. and {Esparza-Arredondo}, D. and {Gonz{\'a}lez-Mart{\'\i}n}, O. and {Garc{\'\i}a-Bernete}, I. and {Pereira-Santaella}, M. and {Alonso-Herrero}, A. and {Acosta-Pulido}, J.~A. and {Bessiere}, P.~S. and {Levenson}, N.~A. and {Tadhunter}, C.~N. and {Rigopoulou}, D. and {Mart{\'\i}nez-Paredes}, M. and {Cazzoli}, S. and {Garc{\'\i}a-Lorenzo}, B.},
        title = "{Absence of nuclear polycyclic aromatic hydrocarbon emission from a compact starburst: The case of the type-2 quasar Mrk 477}",
      journal = {\aap},
         year = 2023,
        month = jan,
       volume = {669},
          eid = {L5},
        pages = {L5},
          doi = {10.1051/0004-6361/202245409},
archivePrefix = {arXiv},
       eprint = {2212.01258},
 primaryClass = {astro-ph.GA},
       adsurl = {https://ui.adsabs.harvard.edu/abs/2023A&A...669L...5R}
}

@ARTICLE{Riechers2014ApJ...786...31R,
       author = {{Riechers}, Dominik A. and {Pope}, Alexandra and {Daddi}, Emanuele and {Armus}, Lee and {Carilli}, Christopher L. and {Walter}, Fabian and {Hodge}, Jacqueline and {Chary}, Ranga-Ram and {Morrison}, Glenn E. and {Dickinson}, Mark and {Dannerbauer}, Helmut and {Elbaz}, David},
        title = "{Polycyclic Aromatic Hydrocarbon and Mid-Infrared Continuum Emission in a z > 4 Submillimeter Galaxy}",
      journal = {\apj},
         year = 2014,
        month = may,
       volume = {786},
       number = {1},
          eid = {31},
        pages = {31},
          doi = {10.1088/0004-637X/786/1/31},
archivePrefix = {arXiv},
       eprint = {1306.5235},
 primaryClass = {astro-ph.CO},
       adsurl = {https://ui.adsabs.harvard.edu/abs/2014ApJ...786...31R}
}

@ARTICLE{Rigopoulou1999AJ....118.2625R,
       author = {{Rigopoulou}, D. and {Spoon}, H.~W.~W. and {Genzel}, R. and {Lutz}, D. and {Moorwood}, A.~F.~M. and {Tran}, Q.~D.},
        title = "{A Large Mid-Infrared Spectroscopic and Near-Infrared Imaging Survey of Ultraluminous Infrared Galaxies: Their Nature and Evolution}",
      journal = {\aj},
         year = 1999,
        month = dec,
       volume = {118},
       number = {6},
        pages = {2625-2645},
          doi = {10.1086/301146},
archivePrefix = {arXiv},
       eprint = {astro-ph/9908300},
 primaryClass = {astro-ph},
       adsurl = {https://ui.adsabs.harvard.edu/abs/1999AJ....118.2625R}
}

@ARTICLE{Rigopoulou2021MNRAS.504.5287R,
       author = {{Rigopoulou}, D. and {Barale}, M. and {Clary}, D.~C. and {Shan}, X. and {Alonso-Herrero}, A. and {Garc{\'\i}a-Bernete}, I. and {Hunt}, L. and {Kerkeni}, B. and {Pereira-Santaella}, M. and {Roche}, P.~F.},
        title = "{The properties of polycyclic aromatic hydrocarbons in galaxies: constraints on PAH sizes, charge and radiation fields}",
      journal = {\mnras},
         year = 2021,
        month = jul,
       volume = {504},
       number = {4},
        pages = {5287-5300},
          doi = {10.1093/mnras/stab959},
archivePrefix = {arXiv},
       eprint = {2011.10114},
 primaryClass = {astro-ph.GA},
       adsurl = {https://ui.adsabs.harvard.edu/abs/2021MNRAS.504.5287R}
}

@ARTICLE{Rigopoulou2024MNRAS.532.1598R,
       author = {{Rigopoulou}, D. and {Donnan}, F.~R. and {Garc{\'\i}a-Bernete}, I. and {Pereira-Santaella}, M. and {Alonso-Herrero}, A. and {Davies}, R. and {Hunt}, L.~K. and {Roche}, P.~F. and {Shimizu}, T.},
        title = "{Polycyclic aromatic hydrocarbon emission in galaxies as seen with JWST}",
      journal = {\mnras},
         year = 2024,
        month = aug,
       volume = {532},
       number = {2},
        pages = {1598-1611},
          doi = {10.1093/mnras/stae1535},
archivePrefix = {arXiv},
       eprint = {2406.11415},
 primaryClass = {astro-ph.GA},
       adsurl = {https://ui.adsabs.harvard.edu/abs/2024MNRAS.532.1598R}
}

@ARTICLE{Robinson2026arXiv260109810R,
       author = {{Robinson}, L. and {Farrah}, D. and {Efstathiou}, A. and {Engholm}, A. and {Hatziminaoglou}, E. and {Joyce}, M. and {Lebouteiller}, V. and {Petty}, S. and {Pitchford}, L.~K. and {Afonso}, J. and {Clements}, D. and {Lacy}, M. and {Pearson}, C. and {Rigopoulou}, D. and {Rowan-Robinson}, M. and {Wang}, L.},
        title = "{Calibrating Mid-Infrared Emission Features As Diagnostics of Star Formation in Infrared-Luminous Galaxies via Radiative Transfer Modeling}",
      journal = {arXiv e-prints},
         year = 2026,
        month = jan,
          eid = {arXiv:2601.09810},
        pages = {arXiv:2601.09810},
          doi = {10.48550/arXiv.2601.09810},
archivePrefix = {arXiv},
       eprint = {2601.09810},
 primaryClass = {astro-ph.GA},
       adsurl = {https://ui.adsabs.harvard.edu/abs/2026arXiv260109810R}
}

@ARTICLE{Roche1991MNRAS.248..606R,
       author = {{Roche}, Patrick F. and {Aitken}, David K. and {Smith}, Craig H. and {Ward}, Martin J.},
        title = "{An atlas of mid-infrared spectra of galaxy nuclei.}",
      journal = {\mnras},
         year = 1991,
        month = feb,
       volume = {248},
        pages = {606},
          doi = {10.1093/mnras/248.4.606},
       adsurl = {https://ui.adsabs.harvard.edu/abs/1991MNRAS.248..606R}
}

@ARTICLE{Rothschild2011ApJ...733...23R,
       author = {{Rothschild}, R.~E. and {Markowitz}, A. and {Rivers}, E. and {Suchy}, S. and {Pottschmidt}, K. and {Kadler}, M. and {M{\"u}ller}, C. and {Wilms}, J.},
        title = "{Twelve and a Half Years of Observations of Centaurus a with the Rossi X-Ray Timing Explorer}",
      journal = {\apj},
         year = 2011,
        month = may,
       volume = {733},
       number = {1},
          eid = {23},
        pages = {23},
          doi = {10.1088/0004-637X/733/1/23},
archivePrefix = {arXiv},
       eprint = {1102.5076},
 primaryClass = {astro-ph.HE},
       adsurl = {https://ui.adsabs.harvard.edu/abs/2011ApJ...733...23R}
}

@ARTICLE{Roussel2007ApJ...669..959R,
       author = {{Roussel}, H. and {Helou}, G. and {Hollenbach}, D.~J. and {Draine}, B.~T. and {Smith}, J.~D. and {Armus}, L. and {Schinnerer}, E. and {Walter}, F. and {Engelbracht}, C.~W. and {Thornley}, M.~D. and {Kennicutt}, R.~C. and {Calzetti}, D. and {Dale}, D.~A. and {Murphy}, E.~J. and {Bot}, C.},
        title = "{Warm Molecular Hydrogen in the Spitzer SINGS Galaxy Sample}",
      journal = {\apj},
         year = 2007,
        month = nov,
       volume = {669},
       number = {2},
        pages = {959-981},
          doi = {10.1086/521667},
archivePrefix = {arXiv},
       eprint = {0707.0395},
 primaryClass = {astro-ph},
       adsurl = {https://ui.adsabs.harvard.edu/abs/2007ApJ...669..959R}
}

@ARTICLE{Sales2010ApJ...725..605S,
       author = {{Sales}, Dinalva A. and {Pastoriza}, M.~G. and {Riffel}, R.},
        title = "{Polycyclic Aromatic Hydrocarbon and Emission Line Ratios in Active Galactic Nuclei and Starburst Galaxies}",
      journal = {\apj},
         year = 2010,
        month = dec,
       volume = {725},
       number = {1},
        pages = {605-614},
          doi = {10.1088/0004-637X/725/1/605},
archivePrefix = {arXiv},
       eprint = {1010.2170},
 primaryClass = {astro-ph.CO},
       adsurl = {https://ui.adsabs.harvard.edu/abs/2010ApJ...725..605S}
}

@ARTICLE{Sandstrom2023ApJ...944L...7S,
       author = {{Sandstrom}, Karin M. and {Chastenet}, J{\'e}r{\'e}my and {Sutter}, Jessica and {Leroy}, Adam K. and {Egorov}, Oleg V. and {Williams}, Thomas G. and {Bolatto}, Alberto D. and {Boquien}, M{\'e}d{\'e}ric and {Cao}, Yixian and {Dale}, Daniel A. and {Lee}, Janice C. and {Rosolowsky}, Erik and {Schinnerer}, Eva and {Barnes}, Ashley. T. and {Belfiore}, Francesco and {Bigiel}, F. and {Chevance}, M{\'e}lanie and {Grasha}, Kathryn and {Groves}, Brent and {Hassani}, Hamid and {Hughes}, Annie and {Klessen}, Ralf S. and {Kruijssen}, J.~M. Diederik and {Larson}, Kirsten L. and {Liu}, Daizhong and {Lopez}, Laura A. and {Meidt}, Sharon E. and {Murphy}, Eric J. and {Sormani}, Mattia C. and {Thilker}, David A. and {Watkins}, Elizabeth J.},
        title = "{PHANGS-JWST First Results: Mapping the 3.3 {\ensuremath{\mu}}m Polycyclic Aromatic Hydrocarbon Vibrational Band in Nearby Galaxies with NIRCam Medium Bands}",
      journal = {\apjl},
         year = 2023,
        month = feb,
       volume = {944},
       number = {2},
          eid = {L7},
        pages = {L7},
          doi = {10.3847/2041-8213/acb0cf},
archivePrefix = {arXiv},
       eprint = {2301.00854},
 primaryClass = {astro-ph.GA},
       adsurl = {https://ui.adsabs.harvard.edu/abs/2023ApJ...944L...7S}
}

@ARTICLE{Sajina2022Univ....8..356S,
       author = {{Sajina}, Anna and {Lacy}, Mark and {Pope}, Alexandra},
        title = "{The Past and Future of Mid-Infrared Studies of AGN}",
      journal = {Universe},
         year = 2022,
        month = jun,
       volume = {8},
       number = {7},
          eid = {356},
        pages = {356},
          doi = {10.3390/universe8070356},
archivePrefix = {arXiv},
       eprint = {2210.02307},
 primaryClass = {astro-ph.GA},
       adsurl = {https://ui.adsabs.harvard.edu/abs/2022Univ....8..356S}
}

@ARTICLE{Sellgren2007ApJ...659.1338S,
       author = {{Sellgren}, K. and {Uchida}, K.~I. and {Werner}, M.~W.},
        title = "{The 15-20 {\ensuremath{\mu}}m Spitzer Spectra of Interstellar Emission Features in NGC 7023}",
      journal = {\apj},
         year = 2007,
        month = apr,
       volume = {659},
       number = {2},
        pages = {1338-1351},
          doi = {10.1086/511805},
archivePrefix = {arXiv},
       eprint = {astro-ph/0612544},
 primaryClass = {astro-ph},
       adsurl = {https://ui.adsabs.harvard.edu/abs/2007ApJ...659.1338S}
}

@ARTICLE{Sellgren2010ApJ...722L..54S,
       author = {{Sellgren}, Kris and {Werner}, Michael W. and {Ingalls}, James G. and {Smith}, J.~D.~T. and {Carleton}, T.~M. and {Joblin}, Christine},
        title = "{C$_{60}$ in Reflection Nebulae}",
      journal = {\apjl},
         year = 2010,
        month = oct,
       volume = {722},
       number = {1},
        pages = {L54-L57},
          doi = {10.1088/2041-8205/722/1/L54},
archivePrefix = {arXiv},
       eprint = {1009.0539},
 primaryClass = {astro-ph.GA},
       adsurl = {https://ui.adsabs.harvard.edu/abs/2010ApJ...722L..54S}
}

@ARTICLE{Skov2014FaDi..168..223S,
       author = {{Skov}, A.~L. and {Thrower}, J.~D. and {Hornek{\ae}r}, L.},
        title = "{Polycyclic aromatic hydrocarbons - catalysts for molecular hydrogen formation}",
      journal = {Faraday Discussions},
         year = 2014,
        month = jan,
       volume = {168},
        pages = {223},
          doi = {10.1039/C3FD00151B},
       adsurl = {https://ui.adsabs.harvard.edu/abs/2014FaDi..168..223S}
}

@ARTICLE{Smith2007ApJ...656..770S,
       author = {{Smith}, J.~D.~T. and {Draine}, B.~T. and {Dale}, D.~A. and {Moustakas}, J. and {Kennicutt}, Jr., R.~C. and {Helou}, G. and {Armus}, L. and {Roussel}, H. and {Sheth}, K. and {Bendo}, G.~J. and {Buckalew}, B.~A. and {Calzetti}, D. and {Engelbracht}, C.~W. and {Gordon}, K.~D. and {Hollenbach}, D.~J. and {Li}, A. and {Malhotra}, S. and {Murphy}, E.~J. and {Walter}, F.},
        title = "{The Mid-Infrared Spectrum of Star-forming Galaxies: Global Properties of Polycyclic Aromatic Hydrocarbon Emission}",
      journal = {\apj},
         year = 2007,
        month = feb,
       volume = {656},
       number = {2},
        pages = {770-791},
          doi = {10.1086/510549},
archivePrefix = {arXiv},
       eprint = {astro-ph/0610913},
 primaryClass = {astro-ph},
       adsurl = {https://ui.adsabs.harvard.edu/abs/2007ApJ...656..770S}
}

\begin{appendix}

\section{Moment-0 maps of the most prominent PAH features}
\begin{figure*}[h!]
\centering
\includegraphics[width=0.95\hsize]{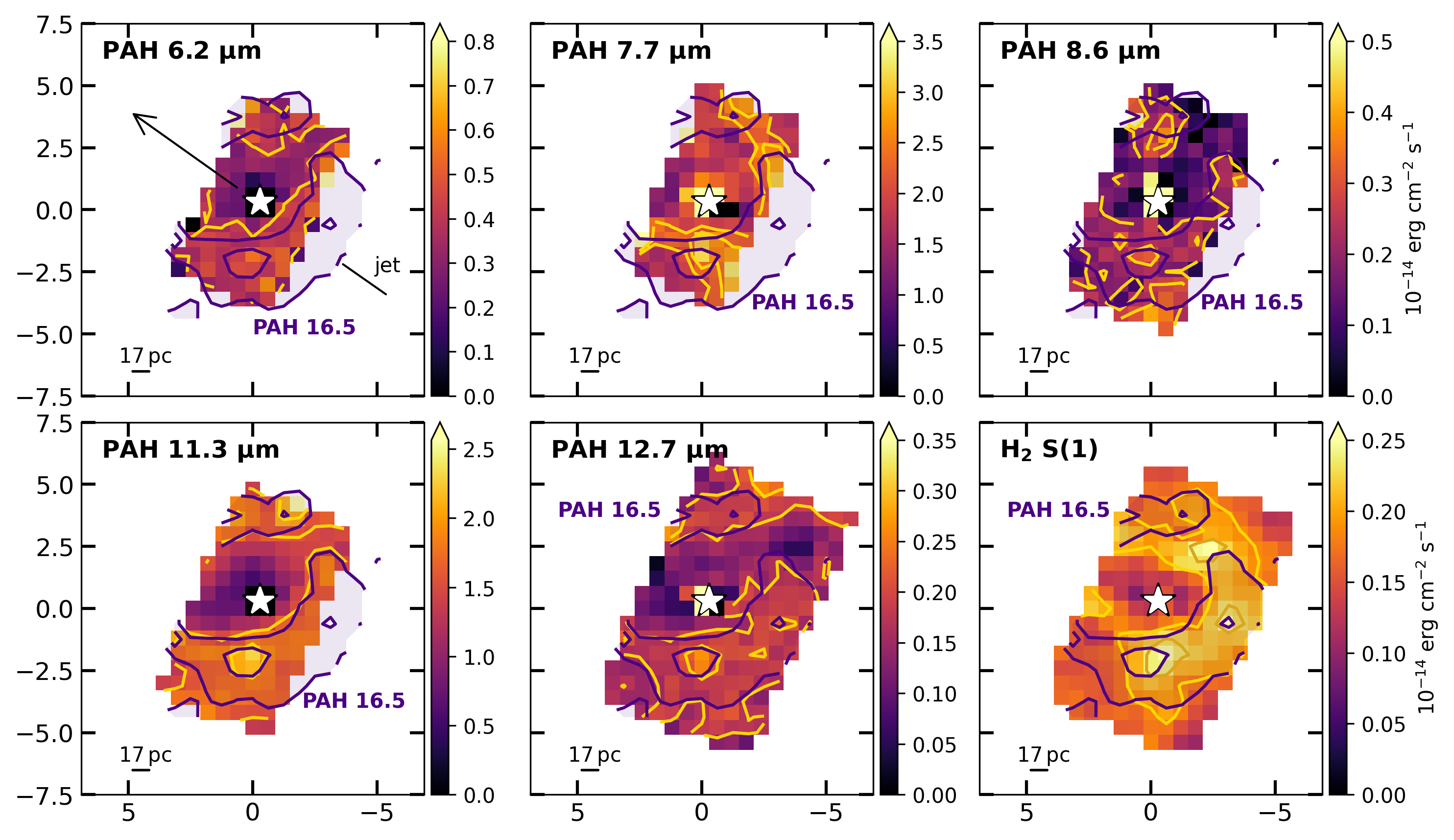}
\caption{From top left to bottom right: we show the mom0 maps of the most prominent PAH features (at 6.2, 7.7, 8.6, 11.3, 12.7~\upmicron) and the \htwo~0-0~S(1)~17.03~\upmicron line. We obtained the mom0 maps after subtraction of the local continuum, from the MIRI-MRS cube rebinned to spaxel scale of $0.6^{\prime\prime}$ (Sect. \ref{sect:2D-ext-method}).
Yellow contours trace the feature emission at levels of: $(0.4, 0.55)\times10^{-14}$~erg~cm$^{-2}$~s$^{-1}$ (PAH 6.2); $(2, 2.75)\times10^{-14}$~erg~cm$^{-2}$~s$^{-1}$ (PAH 7.7); $(0.18, 0.27)\times10^{-14}$~erg~cm$^{-2}$~s$^{-1}$ (PAH 8.6); $(1.5,2)\times10^{-14}$~erg~cm$^{-2}$~s$^{-1}$ (PAH 11.3); $(0.17, 0.22)\times10^{-14}$~erg~cm$^{-2}$~s$^{-1}$ (PAH 12.7); $(0.19, 0.23)\times10^{-14}$~erg~cm$^{-2}$~s$^{-1}$ (\htwo~S(1)).
Violet contours show the PAH 16.5~\upmicron emission at $(0.18,0.225)\times10^{-14}$~erg~cm$^{-2}$~s$^{-1}$.
In the top left panel, the black arrow indicates the position angle of the jet (i.e., PA$_{\rm jet}=51\,$deg).
The stars correspond to the position of the AGN. North is up, and East is to the left.}
\label{Fig:2D_maps}
\end{figure*}

Figure~\ref{Fig:2D_maps} shows the mom0 maps of the PAH bands at 6.2, 7.7, 8.6, 11.3, and 12.7~\upmicron, while the 16.5~\upmicron feature is displayed in the main text (Fig.~\ref{Fig:PAH16p5}). For completeness, we also include the mom0 map of the \htwo~0-0~S(1) emission line. The latter was obtained through local continuum subtraction, following the prescriptions listed in Table \ref{tab:H22D}.
\begin{table}
\caption{Setting for the mom0 extraction of \htwo~0-0~S(1) emission line.}
\label{tab:H22D}
\centering 
\begin{tabular}{cccc}
\hline\hline
\noalign{\smallskip}
$\mathbf{\lambda_{H2}}$& $\mathbf{\Delta\lambda_{H2}}$ & $\mathbf{\Delta\lambda_{cont}^{blue}}$ & $\mathbf{\Delta\lambda_{cont}^{red}}$ \\ 
\noalign{\smallskip}
[\tabupmicron] & [\tabupmicron] & [\tabupmicron] & [\tabupmicron] \\
\noalign{\smallskip}
\hline     
\noalign{\smallskip}
17.06 & $17.04-17.08$ & $17.00-17.04$ & $17.08-17.10$\\
\noalign{\smallskip}
\hline 
\end{tabular}
\tablefoot{In the order we list: the observed central wavelength of the \htwo~0-0~S(1) line ($\lambda_{\rm H2}$), the wavelength interval used for its extraction ($\Delta\lambda_{\rm H2}$), and the two continuum windows ($\Delta\lambda_{\rm cont}^{\rm blue}$, $\Delta\lambda_{\rm cont}^{\rm red}$).}
\end{table}

The most prominent PAH features follow the morphology of the 16.5~\upmicron PAH, with the exception of the 7.7~\upmicron feature that, instead, appears to follow the \htwo distribution. We caution, however, that this behaviour may be influenced by residual contamination from the four emission lines blended with the 7.7~\upmicron PAH complex (see Table~\ref{tab:PAH2D}).

Bright pixels near the AGN position are artifacts introduced by the local continuum subtraction and do not affect the PAH flux distribution at larger radii.

\label{App:PAH2Dmaps}

\section{Regions for one‐dimensional PAH extraction}
\label{App:regions}
We extract 1D spectra over the wavelength range
$4.9-21$~\upmicron from five regions of interest in the MIRI-MRS mosaic.\\
(i) The full Ch~1A mosaic, spanning $7.2^{\prime\prime}\times3.6^{\prime\prime}\sim120\times60\,\mathrm{pc}^2$, which allows comparison with studies of local Seyfert nuclei on scales of a few hundred parsecs.\\
(ii) The nucleus, treated as a point source and extracted using a conical aperture with a radius equal to twice the wavelength-dependent FWHM \citep{Law2023AJ....166...45L}:
\begin{equation}
   \mathrm{FWHM}(\lambda) = 0.033^{\prime\prime}\,\frac{\lambda}{\mu\mathrm{m}} + 0.106^{\prime\prime},
\end{equation}
with $\mathrm{FWHM}=0.35^{\prime\prime}$ for $\lambda<7$~\upmicron and $\mathrm{FWHM}=0.9^{\prime\prime}$ for Ch~4A.\\
(iii) The circum-nuclear region, that we obtained by subtracting the nuclear spectrum from that of the full Ch~1A mosaic.\\
(iv) The PAH ring, traced by the brightest pixels in the mom0 map of the 16.5\,~\upmicron PAH feature (Fig. \ref{Fig:PAH16p5}, left panel), i.e. with $S>4\sigma\approx0.18\times10^{-14}\,\mathrm{erg\,cm^{-2}\,s^{-1}}$.\\
(v) The PAH-deficient region, defined as the complement of the PAH ring area (excluding the nucleus). This region is spatially aligned with both the ionized-gas bubble identified by \citet{Alonso2025A&A...699A.334A}, which expands perpendicular to the jet, and the innermost inflowing branch of molecular gas traced by the CO(3--2) line \citep{Espada2009ApJ...695..116E}. In addition, it shows the highest \htwo/PAH ratio.

A schematic view of the regions for 1D PAH extraction is shown in the right panel of Fig. \ref{Fig:PAH16p5}. All regions cover scales larger than the FWHM at 21~\upmicron ($\sim0.9^{\prime\prime}$), which ensures reliable spectral extraction and obviates the need to convolve to a common angular resolution.

\section{On the extraction of PAH features in the nucleus}
\label{App:nucleus}
\begin{figure*}[h!]
\centering
\includegraphics[width=0.95\hsize]{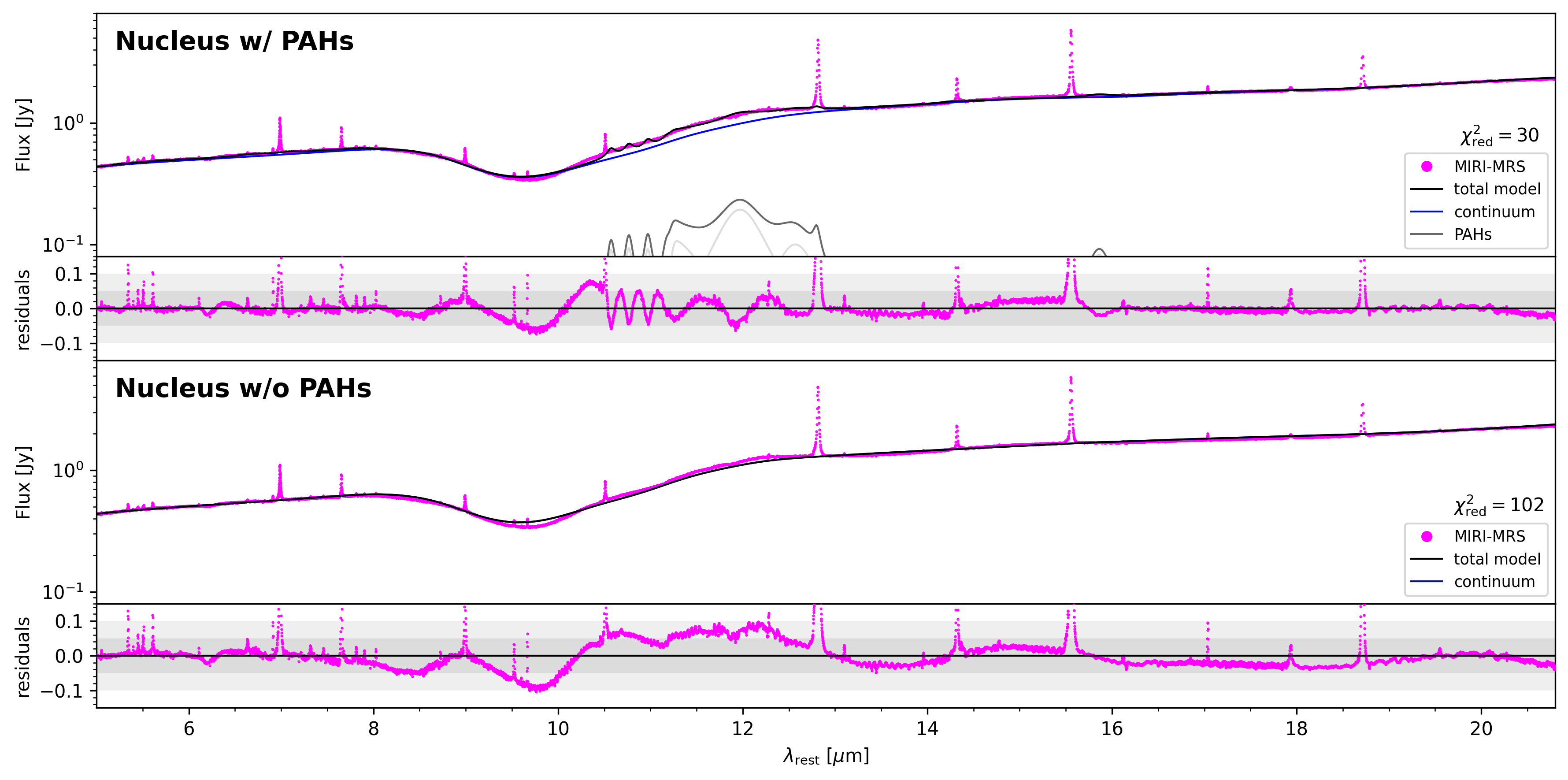}
\caption{
One-dimensional MIRI-MRS spectrum of the Cen~A nucleus (magenta). The extraction regions are shown in Appendix~\ref{App:regions}. Black curves indicate the total best-fit models from the MIR decomposition tool \citep{Donnan2024MNRAS.529.1386D}; the blue and grey curves show the fitted continua (stellar + AGN) and PAH components, respectively. Residuals are displayed beneath each panel, with the corresponding reduced $\chi^2$ reported next to the legend. The top panel includes PAH features in the fit, while the bottom panel shows the fit obtained when PAH features are excluded.}
\label{Fig:nucleus-fit}
\end{figure*}
Between all the extraction regions, the nucleus yields the poorest fit, with $\chi_{\rm red}^{2}=30$, and exhibits the largest residuals, especially across the $9-12$~\upmicron range. This clearly indicates that the MIR continuum from the central engine can significantly hinder a reliable extraction of the PAH features.

Our modelling framework includes PAH emission by default, which may not be appropriate for the nucleus, where the features do not clearly emerge above the continuum. In this Appendix, we assess the impact of excluding the PAH components from the spectral fit.

The resulting best-fit model without PAHs is shown in the bottom panel of Fig.~\ref{Fig:nucleus-fit}, together with the residuals. For comparison, the top panel reports the original fit obtained with the default configuration described in Sect.~\ref{sect:1Dextraction}, which includes PAH components.
Although both fits display residual structures with deviations of up to $8-10$\%, the reduced $\chi^{2}$ is substantially lower when PAHs are included ($\chi^{2}_{\rm red}=30$ compared to $\chi^{2}_{\rm red}=\mathbf{102}$ for the continuum-only model). An F-test indicates that the fit including the PAH features provides a statistically significant improvement over the alternative model, with a p-value $\mathbf{\ll0.05}$. This indicates that some PAH emission is likely present even in the nucleus of Cen~A; however, the strong MIR continuum prevents these features from being reliably constrained. 

\section{PAH modelling with \cafe}
\label{App:cafe}
\begin{figure*}[h!]
\centering
\includegraphics[width=0.95\columnwidth]{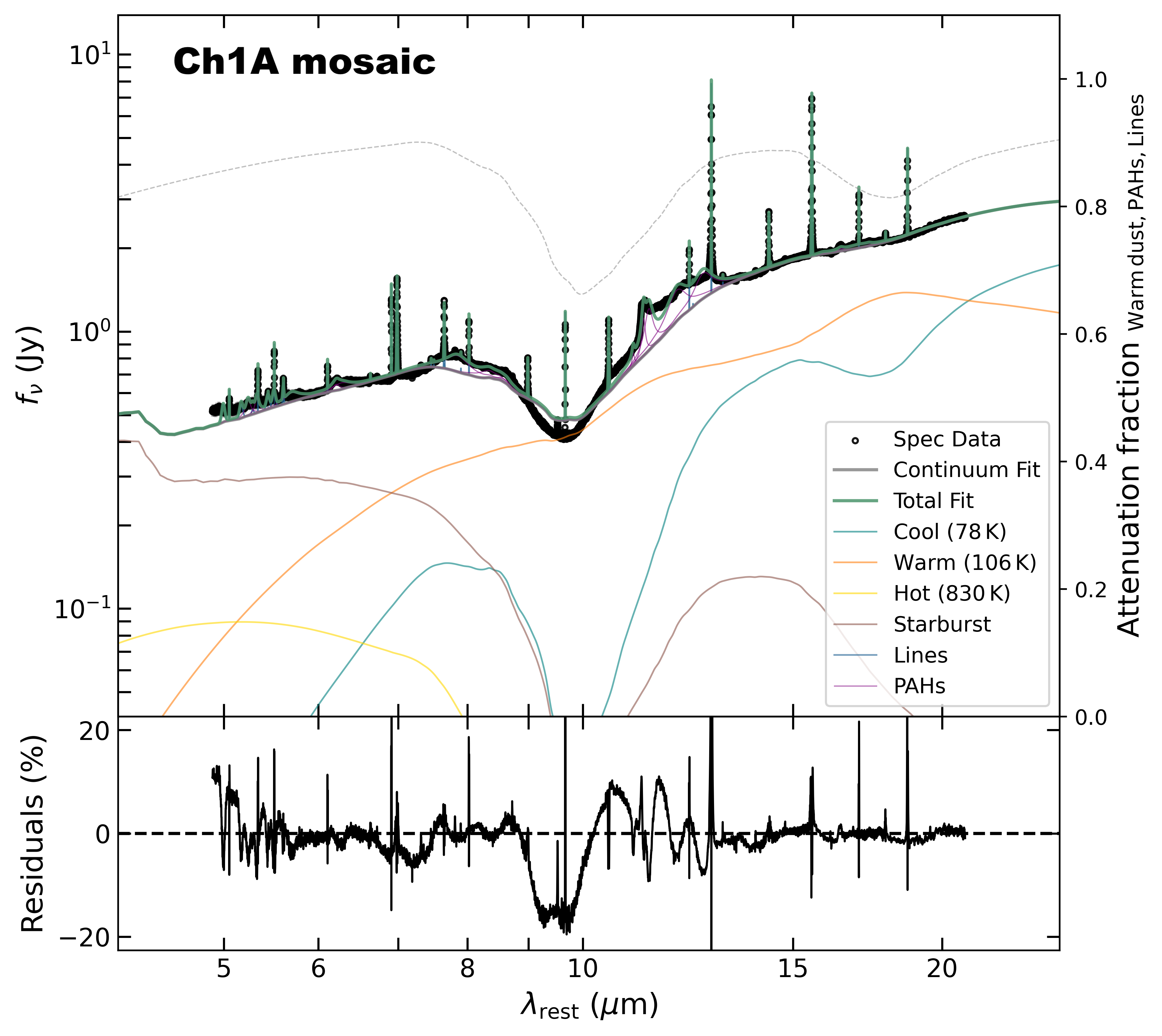}
\includegraphics[width=0.95\columnwidth]{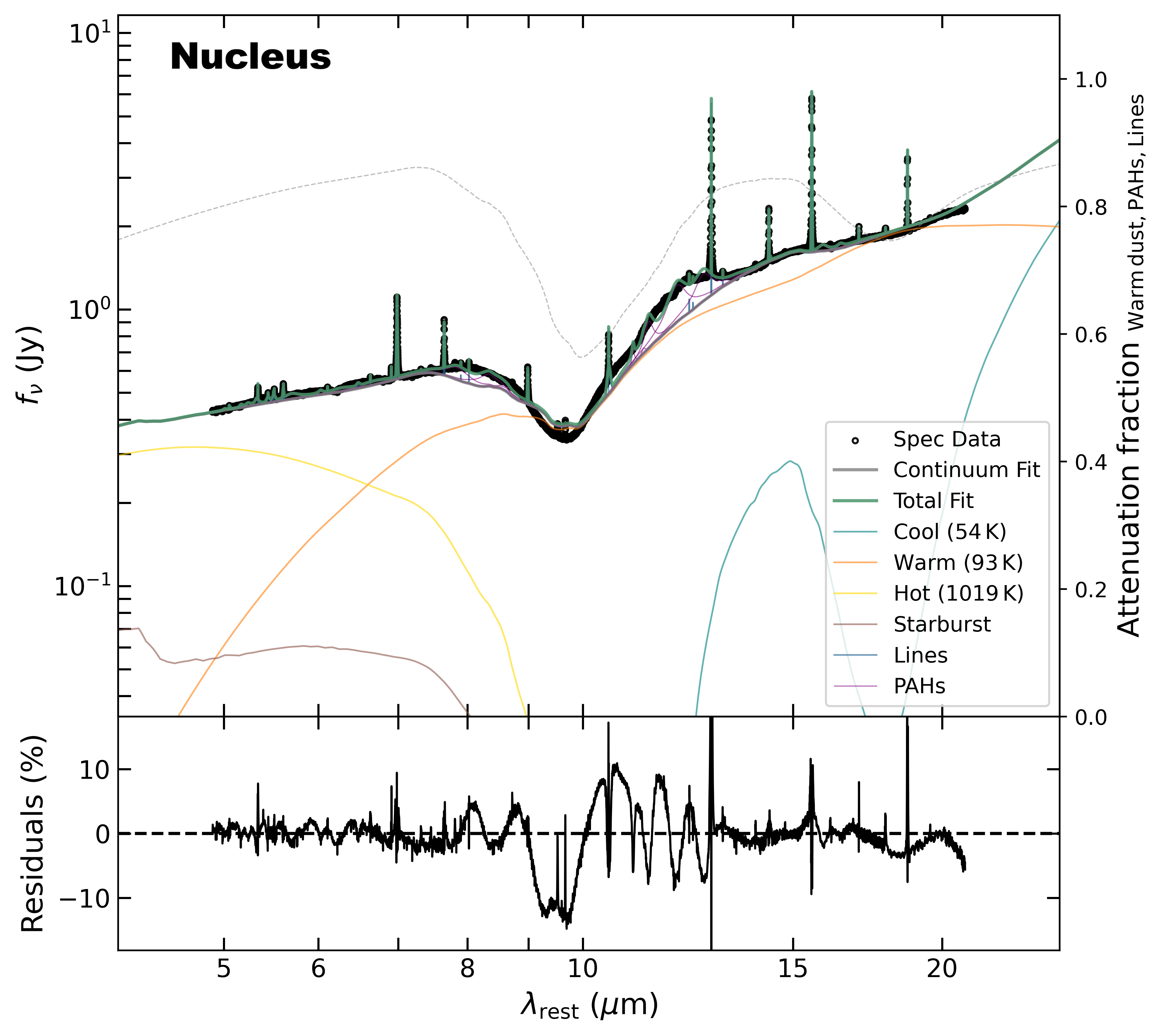}\\
\includegraphics[width=0.95\columnwidth]{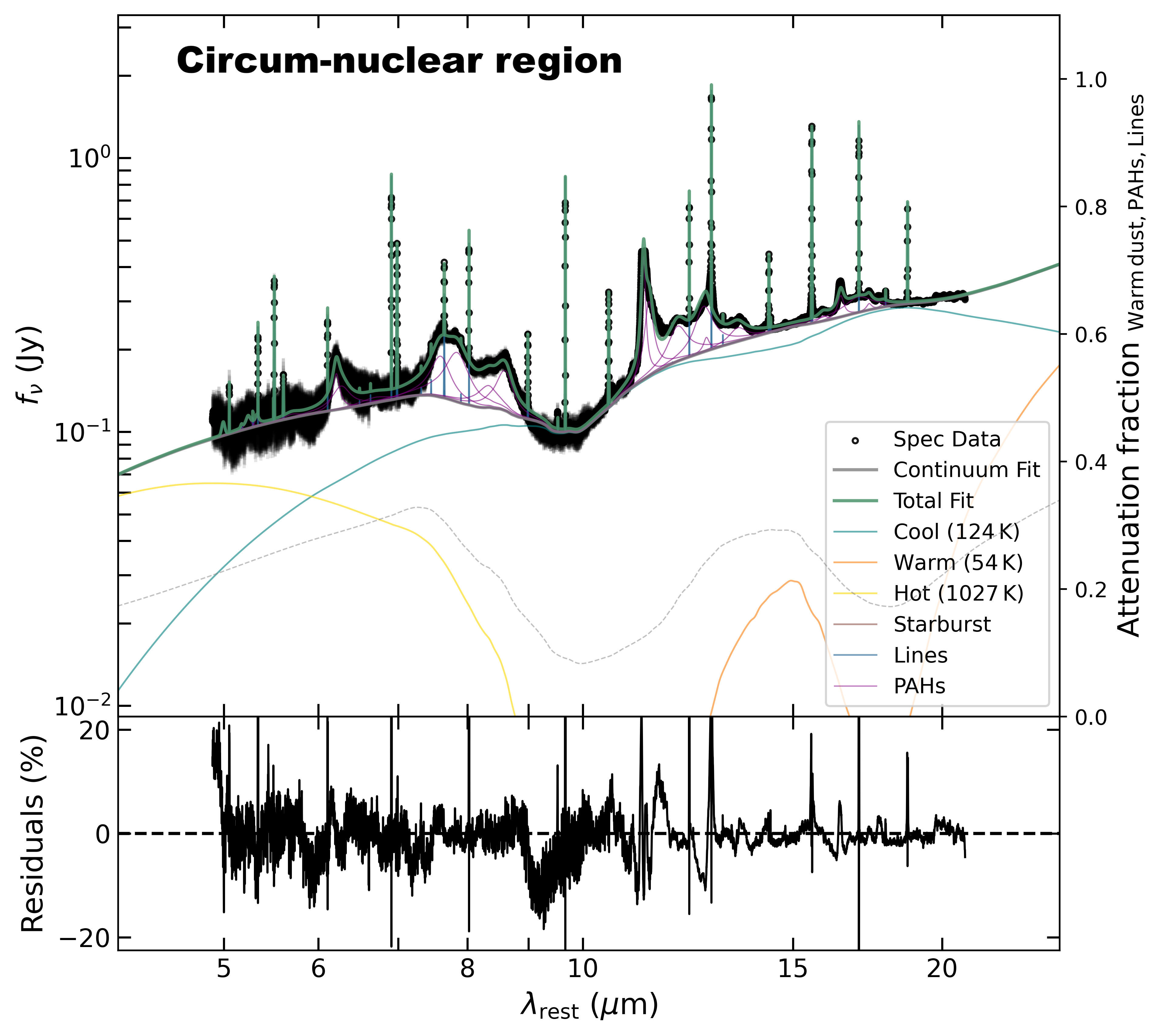}
\includegraphics[width=0.95\columnwidth]{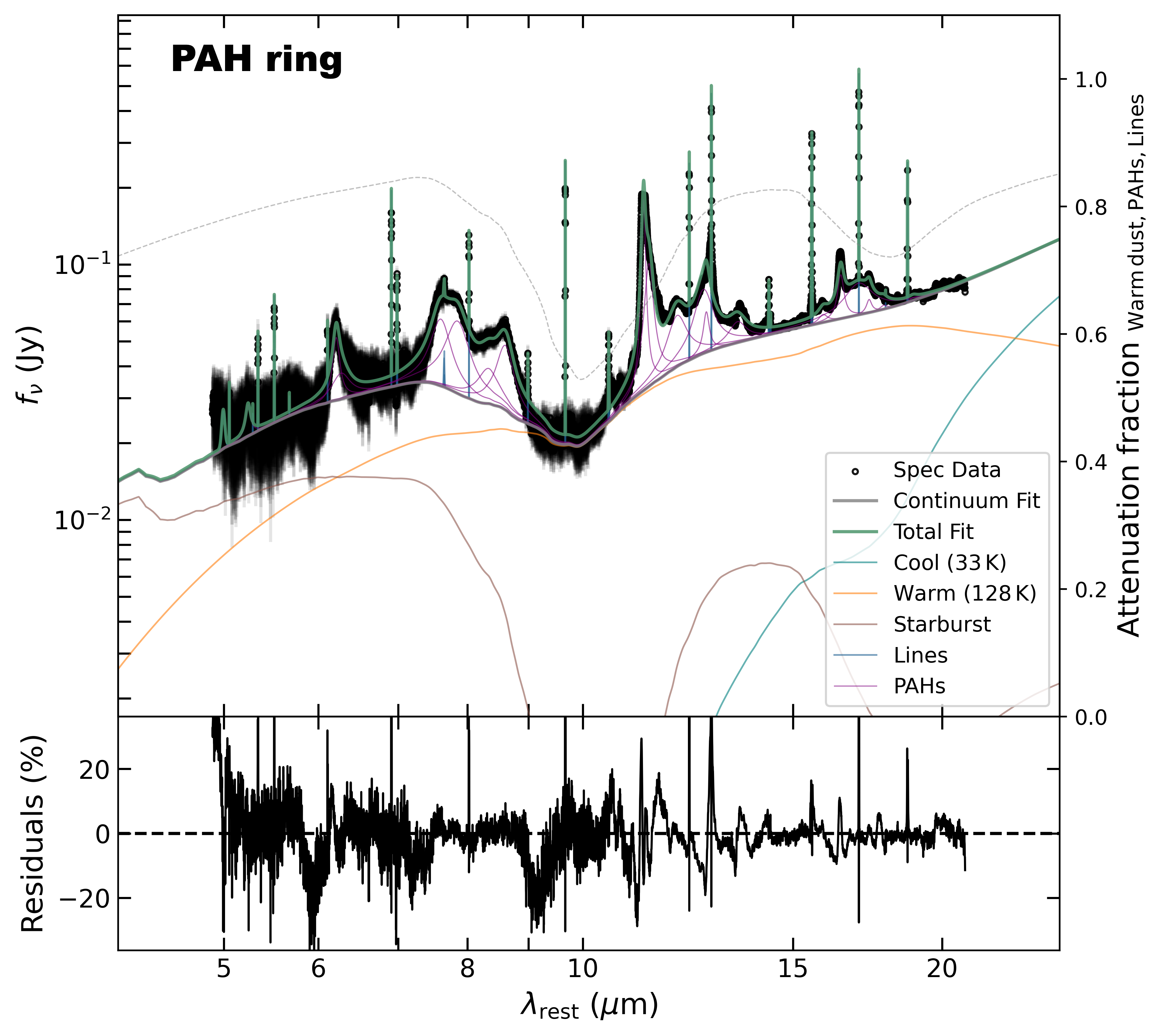}\\
\includegraphics[width=0.95\columnwidth]{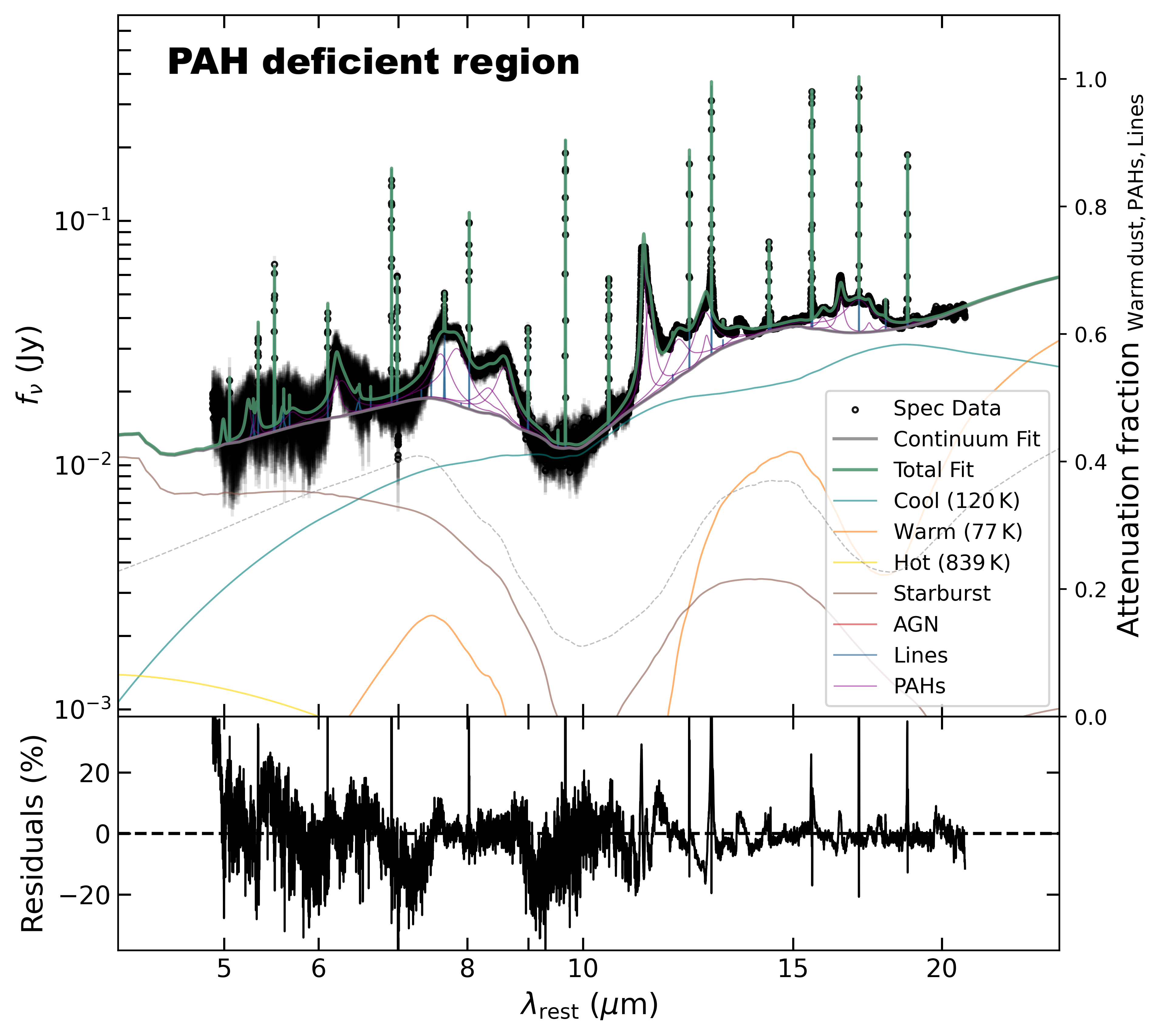}
\caption{One-dimensional MIRI–MRS spectra extracted from the full Ch~1A mosaic, the nucleus, the circum-nuclear region, the PAH ring, and the PAH-deficient region (see Appendix~\ref{App:regions}), modelled with \cafe. The observed spectra are shown in black, the fitted continuum in grey, the PAH components in violet, and the total model in green. The continuum comprises a starburst component (brown) and cool, warm, and hot dust continua (light blue, orange, and yellow, respectively). The best-fit temperature of each component is indicated in brackets on the panels. The dashed gray line shows the dust attenuation. The lower insets display the residuals.}
\label{fig:cafe_fits}
\end{figure*}
\begin{table*}[h!]
\caption{PAH, warm molecular hydrogen and ionized gas emission line intensities.} 
\label{tab:cafe_results}
\centering
\begin{tabular}{lcccccc} 
\hline\hline
\noalign{\smallskip}
\textbf{Line~/~feature}& $\mathbf{\lambda_c}$ & \textbf{Ch~1A mosaic} & \textbf{Nucleus} & \textbf{Circumn-nucl. reg.} & \textbf{PAH ring} & \textbf{PAH def. reg.}\\ 
\noalign{\smallskip}
& [\tabupmicron] & $[10^{-17}$ W/m$^{2}]$ & $[10^{-17}$ W/m$^{2}]$ & $[10^{-17}$ W/m$^{2}]$ & $[10^{-17}$ W/m$^{2}]$ & $[10^{-17}$ W/m$^{2}]$\\ 
\noalign{\smallskip}
\hline     
\noalign{\smallskip}
PAH & 6.2 & $160\pm140$ & $41\pm17$ & $443\pm130$ & $68\pm15$ & $77\pm29$\\ 
PAH complex & 7.7 &  $370\pm90$ & $220\pm20$ & $1230\pm280$ & $210\pm10$ & $180\pm40$ \\ 
PAH & 8.61 & $44\pm45$ & $40\pm10$ & $500\pm170$ & $60\pm5$ & $80\pm30$ \\ 
PAH complex & 11.3 & $500\pm50$ & $225\pm9$ & $1600\pm370$ & $162\pm6$ & $235\pm63$ \\ 
PAH & 12 & $680\pm70$ & $750\pm20$ & $600\pm180$ & $58\pm2$  & $73\pm25$ \\
PAH complex & 12.7 & $560\pm55$ & $524\pm16$ & $650\pm190$ & $85\pm3$ & $90\pm30$ \\ 
PAH complex & 17 & $140\pm40$ & $67\pm20$ & $400\pm120$ & $70\pm3$ &  $100\pm30$ \\
\htwo S(3) & 9.664 & $11\pm2$ & $-$ & $90\pm30$ & $5.2\pm0.4$ &  $19\pm8$ \\
\htwo S(2) &12.279 & $8\pm1$ & $1.7\pm0.5$ & $26\pm8$ & $2.8\pm0.1$ & $6\pm2$ \\
\htwo S(1) & 17.035 & $12\pm1$ & $2.0\pm0.5$ & $44\pm14$ & $5.8\pm0.2$ & $11\pm4$ \\
\neii & 12.814 & $69\pm5$ & $55\pm2$ & $70\pm20$ & $5.3\pm0.2$ &  $9\pm3$ \\
\nev & 14.3217 & $26\pm3$ & $22\pm3$ & $14\pm5$ & $0.9\pm0.2$ & $1.9\pm0.6$ \\
\neiii & 15.555& $130\pm10$ & $118\pm4$ & $76\pm23$ & $7.2\pm0.3$ &  $16\pm5$ \\
\noalign{\smallskip}
\hline
\end{tabular}
\tablefoot{Intensities are derived using \cafe. The second column lists the central wavelength of each feature. PAH complexes include the following subcomponents: 7.42, 7.60, 7.85~\upmicron features (PAH 7.7); 11, 11.23 and 11.33~\upmicron (PAH 11.3); 12.62 and 12.69~\upmicron (PAH 12.7); and 16.45, 17.04, 17.375, and 17.87~\upmicron (PAH 17).}
\end{table*}
\begin{table*}[h!]
\caption{PAH equivalent widths (EWs).}
\label{tab:PAH_1DEW_cafe}
\centering
\begin{tabular}{lccccc} 
\hline\hline
\noalign{\smallskip}
\textbf{PAH}& \textbf{Ch~1A mosaic} & \textbf{Nucleus} & \textbf{Circumn-nucl. reg.} & \textbf{PAH ring} & \textbf{PAH def. reg.}\\
\noalign{\smallskip}
$\lambda_c$ [\tabupmicron] & [nm] & [nm] & [nm] & [nm] & [nm] \\
\noalign{\smallskip}
\hline     
\noalign{\smallskip}
6.2 & 28 & 9 & 415 & 253 & 546\\   
7.7 complex& 100 & 76 & 1777 & 1185 & 1916 \\ 
8.61 & 15 & 18 & 935 & 499 & 1134 \\
11.3 complex & 222 & 115 & 3847 & 1781 & 4834\\ 
12 & 276 & 364 & 1437 & 602 & 1395 \\
12.7 complex & 188 & 215 & 1448 & 796 & 1459 \\
17 complex &  61 & 41 & 1455 & 1036 & 2649\\
\noalign{\smallskip}
\hline 
\end{tabular}
\tablefoot{Same as Table \ref{tab:cafe_results}. Uncertainties are not available.
}
\end{table*}
To assess the dependence of our results on the adopted spectral-decomposition method, we repeated the analysis using \cafe \citep{Marshall2007ApJ...670..129M, CAFE2025ascl.soft01001D}. This tool is widely employed in the literature to model the MIR spectra of starburst galaxies and AGNs \citep[e.g.,][]{U2022ApJ...940L...5U, Lai2022ApJ...941L..36L, Lai2023ApJ...957L..26L, Armus2023ApJ...942L..37A, Bohn2024ApJ...977...36B}, and differs from \spirit primarily in its treatment of extinction.
Other commonly used decomposition tools, such as \pahfit \citep{Smith2007ApJ...656..770S, VanDePutte2025A&A...701A.111V}, are optimized, instead, for modelling emission from the diffuse ISM \citep[e.g.][]{Chastenet2026}. Notice that differences of about 7\% in the neutral PAH fraction were reported by \citet{Maragkoudakis2025ApJ...979...90M}, using different decomposition codes. 

As in \spirit, \cafe models PAH features with Drude profiles and allows variations in PAH central wavelength and width, though using fewer PAH subcomponents. Emission lines from warm molecular hydrogen and ionized gas are fitted with Gaussian profiles.
The MIR continuum is described by:
(i) dust re-emission, represented by cool, warm, and hot MBB components (initial temperatures $T= 75,\, 135,\, 500$ K) each with different opacity at 9.8~\upmicron; 
(ii) direct continua, including a component mimicking the ISRF in the solar neighbourhood; starburst templates of age 2, 10, 100 Myr; and an AGN accretion-disc component, parameterized by multiple power laws.
For modelling the integrated Cen~A spectra, we adopted the default MIRI-AGN input parameter file.

Table~\ref{tab:cafe_results} lists the PAH feature intensities and associated uncertainties derived with \cafe, together with the fluxes of other relevant spectral lines. For reference, Table~\ref{tab:PAH_1DEW_cafe} reports the PAH EWs, although we do not use them for our analysis. Fig.~\ref{fig:cafe_fits} shows the corresponding best-fit models and residuals.

We find that the \cafe fits exhibit systematically larger residuals than those obtained with \spirit (cf. Fig.~\ref{Fig:PAH-SED}). In particular, residuals reach up to $\sim20$\% around 9.8~\upmicron and in the vicinity of several PAH complexes, including the 6.2, 11.3, and 12.7~\upmicron features. Moreover, approximately half of the PAH fluxes derived with \cafe are systematically higher by a factor of $\sim2$.

\section{UV radiation field intensity in the centre of Cen~A}\label{App:ISRF}

UV radiation from the AGN accretion disc can efficiently excite PAH molecules within the central $\sim500$ pc \citep{Jensen2017MNRAS.470.3071J, Li2020NatAs...4..339L}. In this Appendix we estimate an upper limit to the UV radiation field illuminating the circum-nuclear region of Cen~A. The UV luminosity of an AGN can be approximated as:
\begin{equation}
    L_{\rm UV} = L_{\rm bol} - L_{\rm 2-10\,keV},
\end{equation}
where the bolometric and X-ray luminosities are taken from \citet{Rothschild2011ApJ...733...23R} and \citet{Beckmann2011A&A...531A..70B}. 
These values yield:
\begin{equation}
    L_{\rm UV} \simeq (10 - 3)\times10^{42}\ {\rm erg\ s^{-1}} \approx 7\times10^{42}\ {\rm erg\ s^{-1}}.
\end{equation}

Since the PAH-emitting region lies at an average distance $R=50$~pc (\(\sim1.54\times10^{18}\,{\rm m}\)) from the central black hole, the corresponding ultraviolet flux is: 
\begin{equation}
    F_{\rm UV} = \frac{L_{\rm UV}}{4\pi R^2} 
               = \frac{7\times10^{42}}{4\pi\times(1.54\times10^{18})^2} 
               \sim 2\times10^{5}\ {\rm erg\ s^{-1}\ m^{-2}}.
\end{equation}

For comparison, the interstellar radiation field in the solar neighbourhood is 
\({\rm ISRF} = 2.2\times10^{-5}\ {\rm W\ m^{-2}} = 220\ {\rm erg\ s^{-1}\ m^{-2}}\) 
\citep{Mathis1983A&A...128..212M, Galliano2022HabT.........1G}. 
Hence, the radiation field impinging on the PAH ring in Cen~A must be lower than 
\(\sim10^{3}\,{\rm ISRF}\).

\section{Neon fine structure line ratios and AGN models}\label{App:AGNmodels}
 
\citet{Alonso2025A&A...699A.334A}, based on an analysis of fine-structure line intensity ratios, argued that the gas in the MIRI-MRS mosaic of Cen~A is at least partially excited and ionized by shocks. In this Appendix, we compare the Neon fine-structure line ratios measured in the five regions analysed in this work (Appendix \ref{App:regions}) with the predictions of AGN radiative models and combined AGN+shock models \citep{Feltre2023A&A...675A..74F}.

\begin{figure*}
\sidecaption
  \includegraphics[width=12cm]{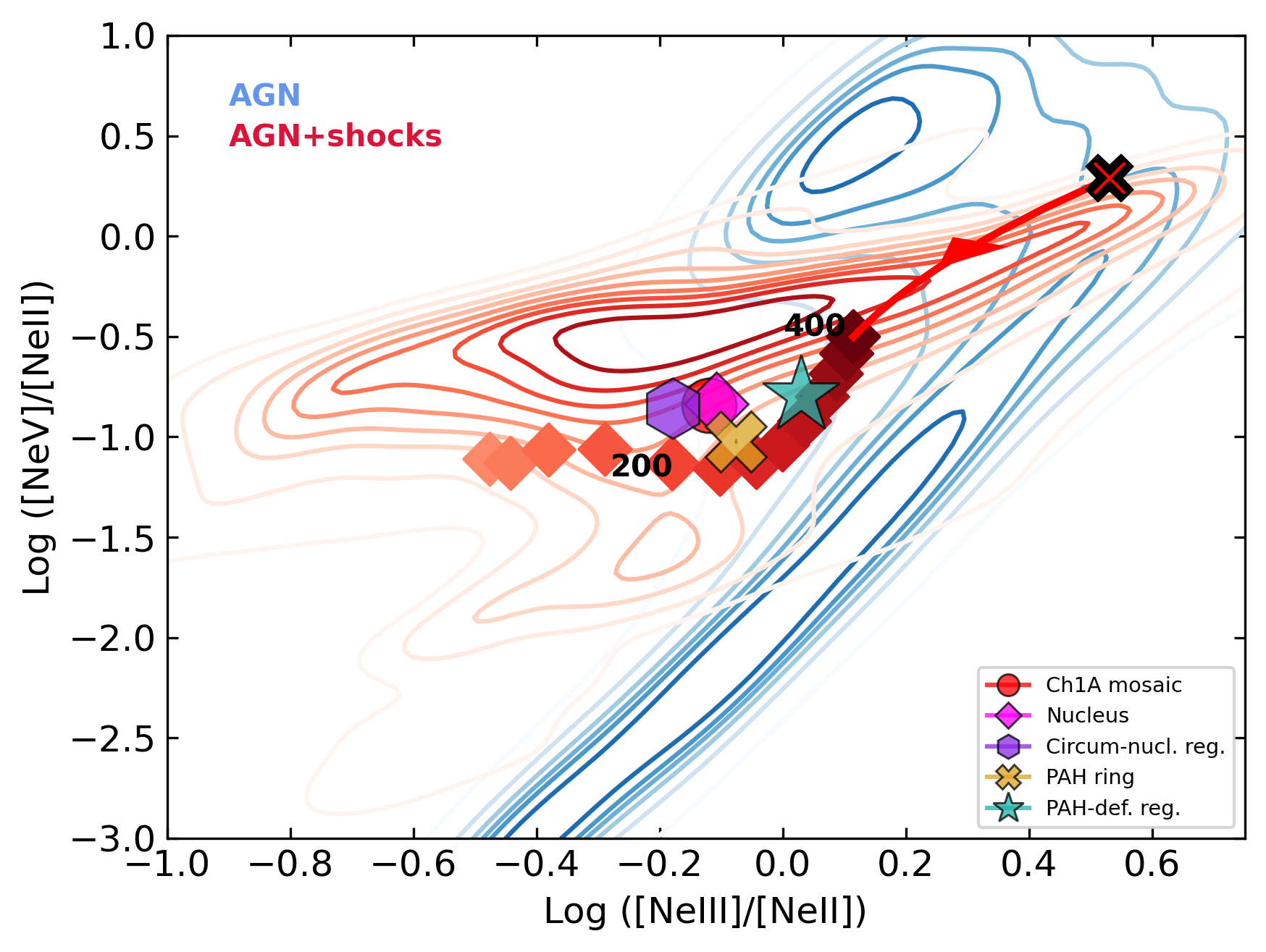}
     \caption{Neon line-ratio diagnostic diagram, showing \nev$_{14}\,/\,$\neii$_{12}$ versus \neiii$_{15}\,/\,$\neii$_{12}$ for the five regions analysed: full Ch~1A mosaic (red circle), nucleus (magenta diamond), circum-nuclear region (violet hexagon), PAH ring (yellow cross), and PAH-deficient region (green star). Blue density contours indicate the pure AGN models of \citet{Feltre2016MNRAS.456.3354F}, while red contours show the AGN+shock models of \citet{Feltre2023A&A...675A..74F}. The red arrow illustrates the effect of adding a $0-90$\% shock contribution to H$\beta$ to an AGN model with $Z = 0.017$, $\log(\langle U \rangle) = -2.5$, n$_H = 103\, {\rm cm}^{-3}$, $\xi_d = 0.3$, and $\alpha = -1.7$ \citep[black cross;][]{Feltre2016MNRAS.456.3354F}. Orange diamonds mark AGN+shock models with a 90\% shock contribution, with shock velocities increasing from 100 to 400 km s$^{-1}$ (light to dark).}
     \label{fig:AGNmod_Feltre}
\end{figure*}

In Fig. \ref{fig:AGNmod_Feltre} we show the \neiii/\neii$\times$\nev/\neii diagnostic plot for the five region of interest. Their position on the plot is inconsistent with purely AGN radiative models, requiring a non-negligible contribution from shocks to excite the lines.

\end{appendix}

\end{document}